\documentclass[oldversion,longauth]{aa}

\usepackage{graphicx}
\usepackage{txfonts}

\usepackage{epsfig}
\usepackage{arydshln}
\usepackage{amsmath}
\usepackage{wasysym}
\usepackage{marvosym}
\usepackage{txfonts}
\usepackage{graphicx}
\pdfoutput=1
\usepackage[pdftex]{color,xcolor}
\usepackage[pdfauthor=Paul Molliere, backref=true, pagebackref=true, hyperindex=true, breaklinks=true,colorlinks=true,urlcolor=blue, linkcolor=blue,citecolor=blue,pagecolor=red, bookmarks=true, bookmarksopen=true]{hyperref}
\usepackage{epstopdf}
\usepackage{lipsum}
\usepackage{natbib}
\usepackage{mathtools}

\bibpunct{(}{)}{;}{a}{}{,} 
\usepackage{mhchem}
\usepackage{inputenc}
\usepackage{rotating}
\usepackage[flushleft]{threeparttable}
\usepackage{adjustbox}

\usepackage{booktabs}

\def\pso{PSO~J318}

\def\apjl{ApJ}                
\def\apjs{ApJS}               
\def\ao{Appl.Optics}          
\def\aap{A\&A}                

\newcommand{\rch}[1]{{{\color{black}#1}}}

\usepackage[colorinlistoftodos]{todonotes}
\newcounter{todocounter}
\let\oldtodo\todo
\renewcommand{\todo}[1]{\stepcounter{todocounter}\oldtodo[inline]{\thetodocounter: #1}}

\begin{document}

   \title{Evidence for SiO cloud  nucleation in the rogue planet PSO~J318}


   \author{P. Molli\`{e}re
          \inst{1}
          \and
          H. K\"uhnle\inst{2}
          \and
          E.~C. Matthews\inst{1}
          \and
          Th. Henning\inst{1}
          \and
          M. Min\inst{3}
          \and
          P. Patapis\inst{2}
          \and
          P.-O. Lagage\inst{4}
          \and
          L.~B.~F.~M. Waters\inst{5,3}
          \and
          M. G\"udel\inst{6,2}
          \and
          Cornelia J\"ager\inst{1,7}
          \and
          Z. Zhang\inst{8,9}\thanks{NASA Sagan Fellow}
          \and
          L. Decin\inst{10}
          \and
          B.~A. Biller\inst{11,12}
          \and
          O. Absil\inst{13}
          \and
          I. Argyriou\inst{10}
          \and
          D. Barrado\inst{14}
          \and
          C. Cossou\inst{15}
          \and
          A. Glasse\inst{16}
          \and
          G. Olofsson\inst{17}
          \and
          J. P. Pye\inst{18}
          \and
          D. Rouan\inst{19}
          \and
          M. Samland\inst{1}
          \and
          S. Scheithauer\inst{1}
          \and
          P. Tremblin\inst{4,20}
          \and
          N. Whiteford\inst{21}
          \and
          E. F. van Dishoeck\inst{22}
          \and
          {G. \"Ostlin\inst{17}}
          \and
          T. Ray\inst{23}
          }

   \institute{Max-Planck-Institut f\"ur Astronomie, K\"onigstuhl 17, 69117 Heidelberg, Germany\\
              \email{molliere@mpia.de}
         \and
             Institute for Particle Physics and Astrophysics, ETH Zurich, Wolfgang-Pauli-Str 27, 8093 Z\"urich, Switzerland
         \and
             SRON Space Research Organisation Netherlands, Niels Bohrweg 4, 2333 CA Leiden, The Netherlands
         \and
             Université Paris Cité, Université Paris-Saclay, CEA, CNRS, AIM, F-91191 Gif-sur-Yvette, France
         \and
             Department of Astrophysics/IMAPP, Radboud University, PO Box 9010, 6500 GL The Netherlands
         \and
             Dept. of Astrophysics, University of Vienna, T\"urkenschanzstr. 17, A-1180 Vienna, Austria
         \and
             Institute of Solid State Physics, Friedrich Schiller University, Jena
         \and
             Department of Astronomy \& Astrophysics, University of California, Santa Cruz, CA 95064, USA
         \and
             Department of Physics \& Astronomy, University of Rochester, Rochester, NY 14627, USA
         \and
             Institute of Astronomy, KU Leuven, Celestijnenlaan 200D, B-3001 Leuven, Belgium
         \and
             Institute for Astronomy, University of Edinburgh, Royal Observatory, Edinburgh EH9 3HJ, UK
         \and
             Centre for Exoplanet Science, University of Edinburgh, Edinburgh, EH9 3FD, UK 
         \and
             STAR Institute, Universit\'e de Li\`ege, All\'ee du Six Ao\^ut, 19C, B-4000 Li\`ege, Belgium
         \and
             Centro de Astrobiolog\'{\i}a (CAB), CSIC-INTA, ESAC Campus, Camino bajo del Castillo s/n, E-28692 Villanueva de la Ca\~nada, Madrid, Spain
         \and
             Université Paris-Saclay, CEA, IRFU, 91191, Gif-sur-Yvette, France
         \and
             UK Astronomy Technology Centre, Royal Observatory Edinburgh, Blackford Hill, Edinburgh EH9 3HJ, UK
         \and
             Department of Astronomy, Stockholm University, AlbaNova University Center, 10691 Stockholm, Sweden
         \and
             School of Physics \& Astronomy, Space Park Leicester, University of Leicester, 92 Corporation Road, Leicester, LE4 5SP, UK
         \and
             LIRA, Observatoire de Paris, Universit{\'e} PSL, Sorbonne Universit{\'e}, Sorbonne Paris Cit{\'e}, CY Cergy Paris Universit{\'e}, CNRS, 5 place Jules Janssen, 92195 Meudon, France
         \and
             Université Paris-Saclay, UVSQ, CNRS, CEA, Maison de la Simulation, 91191, Gif-sur-Yvette, France
         \and
             Department of Astrophysics, American Museum of Natural History, Central Park West at 79th Street, New York, NY 10024, USA
         \and
            Leiden Observatory, Leiden University, PO Box 9513, 2300 RA Leiden, The Netherlands
         \and
             School of Cosmic Physics, Dublin Institute for Advanced Studies, 31 Fitzwilliam Place, Dublin D02 XF86, Ireland
}

   \date{Received May 29, 2025; accepted --}

 \abstract{Silicate clouds have long been known to significantly impact the spectra of late L-type brown dwarfs, with observable absorption features at $\sim 10$~µm. The \emph{James Webb Space Telescope} (\emph{JWST}) has reopened our window to the mid-infrared with unprecedented sensitivity, bringing the characterization of silicate clouds into focus again. 
 Using \emph{JWST}, we aim to characterize the planetary-mass brown dwarf PSO J318.5338-22.8603, concentrating on any silicate cloud absorption the object may exhibit. \pso's spectrum is extremely red, and its flux is variable, both of which are thought to be hallmarks of cloud absorption. 
 We present \emph{JWST} \emph{NIRSpec PRISM}, \emph{G395H}, and \emph{MIRI MRS} observations of \pso\ from 1 to 18~µm. We introduce a method based on \pso's brightness temperature to generate a list of cloud species that are likely present in its atmosphere. We then test for the species' presence with \texttt{petitRADTRANS} retrievals. Using retrievals and grids from various climate models, we derive bulk parameters from \pso's spectra, which are mutually compatible. 
 Our retrieval results point to a solar to slightly super-solar atmospheric C/O, a slightly super-solar metallicity, and a $\rm ^{12}C/^{13}C$ below ISM values. The atmospheric gravity proves difficult to constrain for both retrievals and grid models. Retrievals describing the flux of \pso \ by mixing two 1-D models (``two-column models'') appear favored over single-column models; this is consistent with \pso's variability. The \emph{JWST} observations also reveal a pronounced absorption feature at $10$~µm. This absorption is best reproduced by introducing a high-altitude cloud layer of small ($<$0.1~µm), amorphous \ce{SiO} grains. 
 The retrieved particle size and location of the cloud is consistent with \rch{SiO condensing as cloud seeding} nuclei. High-altitude clouds comprised of small SiO particles have been suggested in previous studies, therefore the SiO nucleation we potentially observe in \pso \ could be a more wide-spread phenomenon.}
  


   \maketitle
%

\section{Introduction}
Brown dwarfs can be thought of as occupying the mass space between the most massive gas giant planets and the lowest mass stars. While their formation likely corresponds to the low-mass end of the star formation process \citep[e.g.,][]{chabrierjohansen2014}, they differ from stars by having masses too low ($M\lesssim75 \ {\rm M_{Jup}}$) to sustain prolonged hydrogen fusion on the main sequence \citep[e.g.,][]{burrows1997}. The main observable difference between the physical properties of brown dwarfs and gas giant planets is that the latter tend to be less massive and may have different metal enrichment patterns in their \ce{H2}/He-dominated envelopes and atmospheres; since planet formation is strongly dependent on initial conditions and driven by many complex, interlocked and stochastic processes, a greater diversity in compositions is expected for planets \citep[e.g.,][]{oeberg2011,madhusudhan2014,mordasinivanboekel2016,mollieremolyarova2022}. The brief period of deuterium burning, often used to define the lower mass limit of brown dwarfs, has little effect on the properties of a brown dwarf (except to slow down cooling somewhat), and both massive planets and low-mass brown dwarfs with masses around $13 \ {\rm M_{Jup}}$ can burn a fraction of their deuterium over time \citep[e.g.,][]{spiegelburrows2011,mollieremordasini2012,bodenheimerdangelo2013}.

The similarity between low-mass brown dwarfs and exoplanets has led to brown dwarfs being called ``free-floating''  or ``rogue'' planets, or "isolated planetary-mass objects". This similarity is reflected in the spectra of particularly young, low-mass brown dwarfs, which are closely related to those of young gas giant exoplanets. Studying brown dwarfs and planets together will thus reveal a more complete picture of cold substellar atmospheres \citep[e.g.,][]{faherty2018}, while isolated brown dwarfs have the obvious advantage of not suffering from the photon noise of an overwhelmingly bright host star.

The pressure-temperature conditions in the atmospheres of exoplanets and brown dwarfs include regions where silicates are thermodynamically stable. Therefore, we expect to detect their presence via spectroscopy. Indirect evidence for clouds in brown dwarfs has long been claimed from the ``reddened'' appearance of their spectra \citep[i.e., near-IR emission is suppressed when compared to theoretical cloud-free predictions, see, e.g.,][]{tsuji2996,allardhauschildt2001}. Mid-IR wavelengths were first accessible with \emph{Spitzer} and revealed the absorption due to Si-O stretching modes of silicate grains with sizes $\lesssim 1$~µm at wavelengths of $\sim$10~µm \citep{cushing2006}. This signature was subsequently detected in a number of L-type brown dwarfs \citep{suarezmetchev2022}. Work by \citet{lunamorley2021,suarezmetchev2023b} indicated that the corresponding clouds in some objects may be located at higher altitudes than classically expected, are consistent with small particle sizes $\lesssim 0.1$~µm, and may consist of amorphous \ce{SiO} or \ce{MgSiO3}. \rch{Relatedly, the work by \citet{camposestrada2025} indicated that small (SiO)$_N$ clusters forming as cloud seeding nuclei high in the atmospheres of brown dwarfs and directly imaged planets may describe the subsequent cloud formation better than the typically assumed \ce{TiO2} seeds when comparing models spectra with observations (but we note that their models did not produce a 10~µm feature itself).} Going forward, it was clear that the increased signal-to-noise (S/N) of the James Webb Space Telescope \citep[\emph{JWST},][]{gardnermather2023}, especially its \emph{MIRI} instrument, would enable a detailed characterization of clouds. \emph{JWST} has now detected evidence for 10~µm absorption in a number of brown dwarf companions and exoplanets \citep[e.g.,][]{milesbiller2023,grantlewis2023,dyrekmin2024,inglisbatalha2024,hochrowland2025}.

Here we study the atmosphere of the young, low-mass brown dwarf PSO J318.5338-22.8603 (called PSO~J318 in the following). PSO~J318's discovery with Pan-STARRS1 was reported by \citet{liumagnier2013}, who emphasized the brown dwarf's extremely red color and planet-like faintness. Together with the weak alkali absorption and the triangular shape of PSO~J318's H-band spectrum, two signs of low gravity, the picture of a low-mass brown dwarf with a likely very cloudy atmosphere arises. PSO~J318 is an assigned member of the $\beta$~Pic moving group (\citealp[]{Barrado1999_BPMG}), so it should be young, and \citet{liumagnier2013} estimated its mass to be $6.5_{-1.0}^{+1.3} \ {\rm M_{Jup}}$ using evolutionary models, assuming a uniform age prior of $12_{-4}^{+8}$~Myr, which is consistent with the low gravity inferred from its spectrum \rch{\citep[we note that the currently accepted age is around 24 Myr, however, see][]{mamajekbell2014}}. An updated value of $8.3\pm0.5 \ {\rm M_{Jup}}$ was determined by \citet{allersgallimore2016}. The near-infrared (NIR) spectra presented in \citet{liumagnier2013} lacked any sign of methane absorption, and the authors derived a spectral type of L7$\pm$1. This makes PSO~J318 about 400~K cooler (they derived $T_{\rm eff}=1160_{-40}^{+30}$~K) than field brown dwarfs of the same spectral type.

Another noteworthy property of \pso\ is its flux variability \citep{billervos2015}. It is thought to stem from brightness inhomogeneities across its top-of-atmosphere structure that rotate in and out of view and evolve. Multiple properties could conceivably vary and thus produce these inhomogeneities \citep[clouds, temperature, or chemical composition, see][]{biller2017}. The contemporaneous \emph{HST} and \emph{Spitzer} light curves of \pso \ reported in \citet{billervos2018} resulted in a constraint on the rotational period of $8.6\pm 0.1$~hr, which led to an inclination of $i=56.2\pm 8.1^\circ$ when combined with the $v \ {\rm sin}(i)=17.5_{-2.8}^{+2.3}$~km/s of \citet{allersgallimore2016}. \citet{billervos2018} derived peak-to-trough variabilities  $3.4\pm 0.1$ \ \% for \emph{Spitzer} and 4.4–5.8~\% across the spectral points of \emph{HST WFC3}. For \emph{WFC3} the authors found the magnitude of  variability to be approximately independent of wavelength. Since cloud cross-sections for absorption and scattering can be gray (i.e., constant with wavelength), this hints at cloud heterogeneity as a possible cause. Intriguingly, the authors also find a phase offset of $\sim 200^\circ$ between the variability of \emph{Spitzer} and \emph{HST} measurements, indicating at a pressure-dependent heterogeneity structure.

Recent evidence for the presence of clouds and the variability of brown dwarfs being correlated was established by \citet{vosallers2017} and \citet{suarezvos2023}. \citet{vosallers2017} found that the most variable brown dwarfs, that is, those seen equator-on, exhibit redder spectral energy distributions than brown dwarfs seen pole-on. Additionally, \citet{suarezvos2023} showed that silicate cloud absorption features in the mid-infrared are more prominent for brown dwarfs viewed equator-on. Thus variability correlates with redness and cloud absorption, and equatorial regions appear to be more cloudy. Another noteworthy finding has been presented in \citet{vosburningham2023}, where the authors showed with archival \emph{Spitzer IRS} data that the atmospheres of the two variable, early-T brown dwarfs SIMP-0136 and 2M~2139 are best described as having patchy silicate clouds. Similarly, \citet{zhangmolliere2025} find that the atmosphere of the known variable companion 2M1207b is best described by patchy silicate clouds, based on \emph{JWST NIRSpec} data. This may point to clouds as a likely variability driver. We note, however, that spectra could be reddened by two effects, in principle: (i) clouds could ``hide'' the deep hot atmosphere otherwise probed in the NIR; (ii) the deep atmosphere could be colder than expected, so it no longer needs to be hidden \citep{tremblinamundsen2015,tremblinamundsen2016,tremblinchabrier2017,tremblinpadioleau2019}. Numerous ongoing \emph{JWST} programs are investigating the cause of variability in brown dwarf atmospheres, pointing to a complex picture of cloud, temperature, and compositional heterogeneities \citep[see][and many more ongoing \emph{JWST} programs]{billervos2024,mccarthyvos2025,chenbiller2025}. \rch{A noteworthy recent study is the work by \citet{nasedkinschrader2025}, who ran the first time-dependent retrievals of a variable brown dwarf observed with \emph{JWST} (the T2 dwarf SIMP-0136), and found that its variability is likely caused by temperature and chemical fluctuations.}

In this paper, we aim to study the panchromatic \emph{JWST} spectrum of \pso\ obtained with the GTO program 1275 (PI Lagage) with the goal of characterizing \pso's atmosphere and studying its silicate cloud absorption. Our focus lies on identifying the clouds' constituent species (e.g., \ce{MgSiO3} vs. \ce{Mg2SiO4}), constraining the particle structure (i.e., crystalline vs. amorphous) \rch{and size}, and on assessing their spatial distribution (are clouds homogeneous or ``patchy''?) and altitude in the atmosphere. While our observations were taken over multiple hours, so a significant fraction of \pso's rotational period, we will not attempt to constrain whether they contain a variability signal.

In Section~\ref{sect:observations} we describe the new and archival observations used in this work. Section~\ref{sect:silicate_characterization} contains an overview over the relevant factors that shape the wavelength-dependent opacity of silicates, and then focuses on the identification of the species that causes \pso's silicate feature with a method based on brightness temperatures and with retrievals. In Section~\ref{sect:bulk_properties} we discuss the bulk atmospheric properties of \pso, as obtained from comparisons to grid models in radiative-convective equilibrium and retrievals, and further characterize \pso's silicate cloud. Our findings are summarized and discussed in Section~\ref{sect:discussion}.


\section{Observations}
\label{sect:observations}

\begin{figure*}
    \centering
    \includegraphics[width=0.86\textwidth]{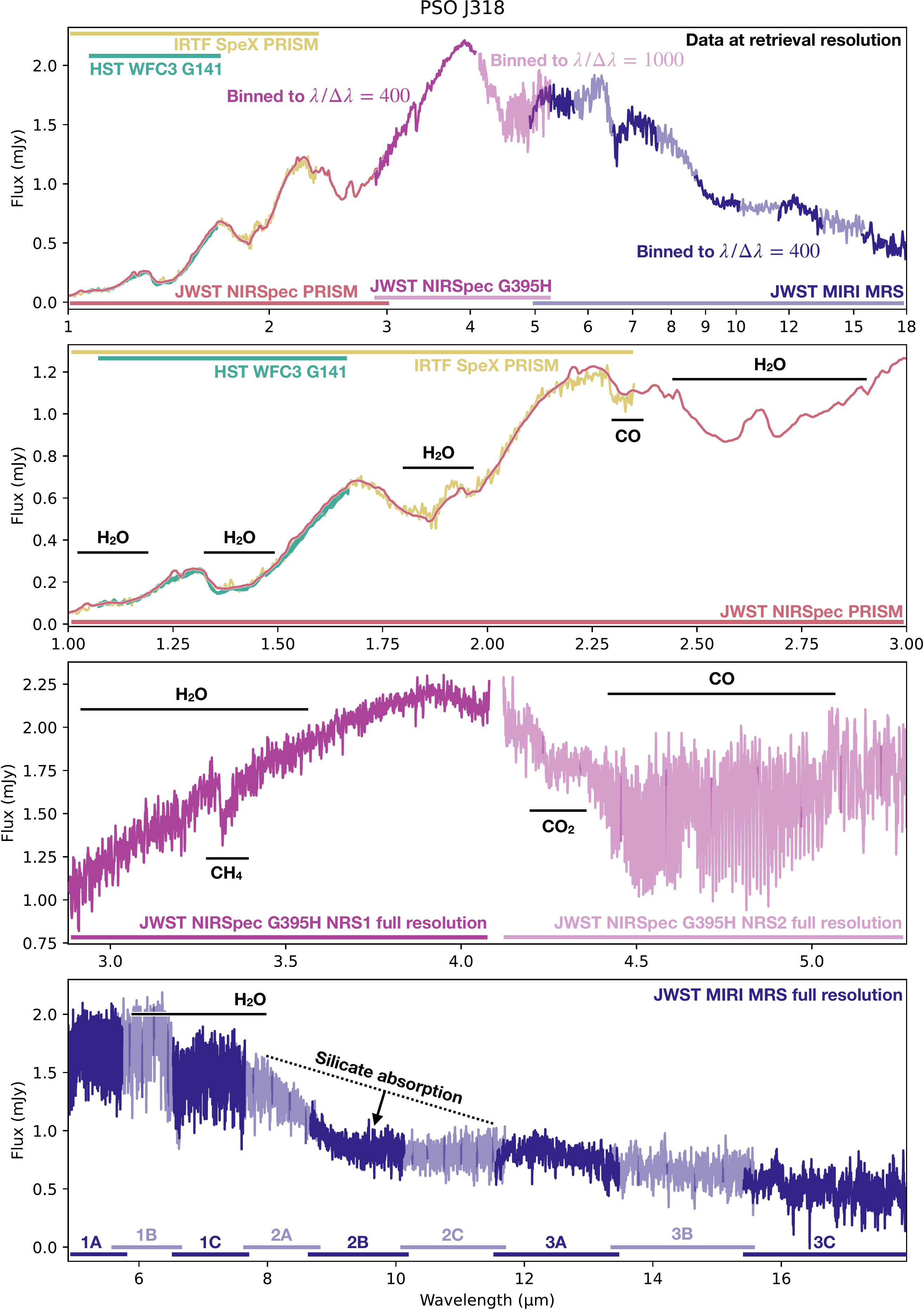}
    \caption{\rch{Observations considered for the atmospheric characterization of \pso.} {\it Uppermost panel:} all data considered for the spectral characterization of \pso, at the wavelength binning fed into the retrieval and self-consistent grid model fits. Specifically, \emph{MIRI MRS} data was binned down to $\lambda/\Delta\lambda=400$, while NIRSpec G395H was binned to $\lambda/\Delta\lambda=400$ and $\lambda/\Delta\lambda=1000$ for NRS1 and NRS2, respectively. {\it Second panel:} the \emph{HST WFC3}, \emph{IRTF SpeX} and \emph{JWST NIRSpec PRISM} data over the 1-3~µm wavelength range. {\it Third panel:} the \emph{JWST NIRSpec G395H} data at the full spectral resolution. {\it Lowermost panel:} the \emph{JWST MIRI MRS} data at the full spectral resolution.}
    \label{fig:jwstdata}
\end{figure*}

The full spectrum of \pso, combining all the observations as used in our atmospheric analysis, is shown in Figure \ref{fig:jwstdata}. Below we describe the JWST observations and data reduction, as well as the provenance of the archival data.

\subsection{JWST observations}
We observed \pso~with \emph{JWST} as part of the ExOMIRI GTO consortium, GTO program \#1275 (PI Lagage). We collected data with \emph{NIRSpec PRISM}, \emph{NIRSpec G395H} \citep{jakobsenferruit2022}, all three \emph{MIRI MRS} gratings, and with the \emph{MIRI} imager \citep{riekewright2015}. Together, this provides a spectrum of \pso~with complete coverage from 0.6~µm to 27.9~µm, with a resolution $R>100$ throughout and $R>1500$ for all wavelengths longer than 2.87~µm.

Our observations were collected on 12 June 2023, between UTC 03:45:34 and 06:19:44. The whole sequence lasted 2~h and 34~min, corresponding to roughly 30\% of a full rotation of \pso. Our observing sequence was as follows: we first collected \emph{MIRI MRS} observations, using all three gratings (\emph{LONG}, \emph{MEDIUM}, \emph{SHORT}) sequentially. Combined with all four channels (observed simultaneously), this provides data from 4.9~µm to 27.9~µm at $R\sim$~1500-3500. However, the thermal background becomes increasingly dominant at the longest wavelengths while \pso \ becomes increasingly faint. In the current work we consider only channels 1, 2, and 3, and ignore data beyond 18~µm (i.e., data from Channel 4). 
For these \emph{MIRI MRS} observations we used a 2-point dither pattern, with 86 groups/integration and 1 integration/exposure, for a total integration time of 477.3~s per sub-channel. While integrating with \emph{MIRI MRS}, we also collected simultaneous \emph{MIRI} images in the \emph{F1280W} filter, but those are not used in the current work. We then collected MIRI imaging in four photometric filters: \emph{F1800W}, \emph{F2100W}, \emph{F1280W} and \emph{F1500W}; also these images are not used in the current work, since they add little additional information compared to the spectra.

Next we collected \emph{NIRSpec} data using the \emph{PRISM} grating and the \emph{CLEAR} filter, providing data from 0.6~µm to 5.3~µm at a mean resolution of $R\sim$~100. We used a 4-point nod pattern, with 2 groups/integration and 2 integrations/exposure, for a total of 350.1~s of integration. Finally, we collected observations with \emph{NIRSpec} using the \emph{G395H} grating and the \emph{F290LP} filter, providing data from 2.87~µm to 5.14~µm at $R\sim$~2700. We used a 4-point-nod pattern, with 6 groups/integration and 1 integration/exposure, for a total of 408.5~s of integration.

\subsection{JWST data reduction}

For the \emph{NIRSpec} data reductions (both \emph{PRISM} and \emph{G395H}) we directly used the \texttt{*x1d.fits} files from the MAST data archive. We include the \emph{PRISM} data from 1~µm onwards, and disregard it for wavelengths longer than 3~µm, since they exhibit a flux excess and unexpected systematic behavior when compared to the \emph{G395H} data. Within our adopted range the \emph{NIRSpec} data of \emph{PRISM} and \emph{G395H} are in good mutual agreement. \emph{PRISM} also lines up with archival \emph{HST} and \emph{IRTF} observations at short, while \emph{G395H} lines up with the bluest \emph{MIRI MRS} channel (\emph{1A}) at long wavelengths (see Fig.~\ref{fig:jwstdata}).

We performed our own reduction for the \emph{MIRI MRS} data, using the public \texttt{jwst} pipeline\footnote{\url{ https://jwst-pipeline.readthedocs.io/en/latest/}} (version 1.12.5, CRDS reference files jwst\_1149.pmap, \citealt{Bushouse2023}). We started from the \texttt{*uncal.fits} files, and applied all three stages of the standard data reduction. Stage 1 produces \texttt{*rate.fits} files, that is, the rate of photoelectric charge accumulation on the detector. In Stage 2 the world coordinate system (wcs) is applied, and several calibrations are applied to produce photometrically calibrated \texttt{*cal.fits} files. The calibrations applied here include a flat field, stray light, residual fringe and photometric correction. Between Stage 2 and 3 we perform a nod subtraction between dither positions to reduce the residual background. Finally, Stage 3 projects the 2D spectra into 3D cubes (x/y/$\lambda$, \texttt{*s3d.fits}) using the ``drizzle'' weighting algorithm \citep{Law2023}. Here, we include an outlier detection, flagging remaining outliers due to, for example, cosmic rays. A one-dimensional spectrum (\texttt{*x1d.fits}) can then be extracted from this cube by placing an aperture of one Full Width Half Maximum (FWHM) of the corresponding Point Spread Function (PSF) around the source and measuring the contained flux at each wavelength. The mid-point of the source is detected using the \texttt{ifu\_autocen()} built-in function.

We also applied an additional correction to the \emph{MRS} data to account for large-scale systematics, which were apparent in the spectrum produced by the \texttt{jwst} pipeline. Examination of the IFU cubes revealed significant, wavelength-dependent structure in the background, even after the dither subtraction; this leads to wavelength-dependent systematics in the extracted spectrum when using the default pipeline. To correct for this systematic we follow the process described in \citet{matthewsmolliere2025}. Briefly, we calculated the 3D cubes in the \texttt{ifualign} orientation, where the striping is largely horizontal. We then masked out the source and modelled the background structure for each individual $\lambda$ slice in a data-driven fashion. This background can then be subtracted, and the spectrum extracted from the cube using the standard \texttt{jwst} pipeline methods. With this correction, our final spectra are smooth in the mid-IR, well-explained with physical models, and show good agreement in the regions where consecutive sub-channels overlap.

\subsection{Archival data}

In addition to \emph{JWST} observations we considered the archival near-infrared (NIR) spectra taken with \emph{HST WFC3} by \cite{billervos2018}. We included all spectra taken during five consecutive \emph{HST} orbits in our analysis, spanning 7~hr, so almost one full rotation of the object. We also included ground-based archival NIR spectroscopy from \emph{IRTF SpeX}, taken by \citet{liumagnier2013}. This data set was included from wavelengths larger than 1~µm, out to wavelengths of 2.35~µm. Wavelengths shorter than 1~µm were excluded for numerical efficiency, and because \pso \ becomes increasingly faint (leading to low-$S/N$ data). Wavelengths longer than 2.3~µm were excluded because of low $S/N$. These archival data sets show good mutual agreement (neglecting the fact that \emph{HST} is actually precise enough to show \pso's variability), and agree well with \emph{JWST NIRSpec PRISM}, see Fig.~\ref{fig:jwstdata}.

\section{Characterizing \pso's 10 µm silicate feature}
\label{sect:silicate_characterization}

The data we present in Fig.~\ref{fig:jwstdata} exhibits a prominent absorption feature at $\sim$10~µm, which we attribute to silicates, that is, absorption by condensed silicon-oxide-rich material in the atmosphere of \pso. Silicon is one of the most abundant refractory elements in the universe \citep[e.g.,][]{asplund2009}. In its condensed form it constitutes a major opacity source, especially at 10~µm. In addition to the aforementioned brown dwarfs and planets, silicate absorption is evident in outflows of evolved AGB stars, ejecta of supernovae, proto-planetary and debris disks, the interstellar medium and AGN dust tori \citep[for a detailed review see][]{henning2010}.

Since we posit that silicates cause the 10~µm absorption, we will explore in the following what type of silicate material may be present in the atmosphere of \pso. For planets or brown dwarfs silicates of olivine-type stoichiometry (${\rm Mg}_{2-x}{\rm Fe}_{x}{\rm SiO}_4$), pyroxene-type (${\rm Mg}_{1-x}{\rm Fe}_{x}{\rm SiO}_3$) and quartz (\ce{SiO2}) are expected species, with their relative importance likely determined by atomic abundance ratios such as Mg/Si \citep{calamarifaherty2024} and potentially C/O \citep{wetzel2013}. The internal structure of silicates can either be crystalline, that is, the elemental building blocks are arranged on a regularly repeating lattice, or amorphous. We note that the amorphous state is not well defined. It can range from configurations of small crystalline domains that do not properly connect to disarray on a smaller level, where crystalline ordering is missing altogether \citep{henning2010}. In studies of silicates forming around evolved stars and in proto-planetary disks it was found that crystalline silicates tend to be Fe-poor \citep[e.g.,][]{jaeger1998,olofssonaugereau2009,juhaszbouwman2010}, which is consistent with theoretical models predicting that Fe should be more abundant in amorphous silicates, that form at lower temperatures \citep{gail2010}.

Recent studies that investigated the broad silicate feature visible in \emph{Spitzer IRS} spectra found that the 10~µm feature is best explained by amorphous silicate absorption. \cite{lunamorley2021} constrained particle sizes to be small ($\lesssim$0.1-1~µm), and consisting of \ce{SiO}, \ce{MgSiO3}, or \ce{Mg2SiO4}. \citet{suarezmetchev2023b} found that the visible clouds consist of amorphous pyroxene-type (${\rm Mg}_{1-x}{\rm Fe}_{x}{\rm SiO}_3$) material for low-gravity brown dwarfs (with particle sizes around 1~µm), while high-gravity brown dwarfs may be dominated by small ($\lesssim$0.1~µm) grains composed of amorphous \ce{SiO} and \ce{MgSiO3}.

Since a 10 µm-feature of a putative silicate cloud appears to be present in \pso's spectrum, we will first attempt to study its grain properties based on the spectral shape from 7-13~µm alone in Section~\ref{sect:brightness_temperature_method}. This will be done in an agnostic manner (i.e., neglecting the constraints on composition from \citealt{lunamorley2021,suarezmetchev2023b}). This can be seen as a precursor step before turning to the numerically costly retrievals, which are presented in Section~\ref{sect:10mu_retrievals}. We expect that the shape of the 10~µm feature is sensitive to the following particle properties:

\begin{itemize}
    \item {\bf Particle structure}: silicate crystals have sharp absorption features stemming from Si-O stretching transitions. In amorphous particles the crystal structure is lost. The corresponding distribution of bond lengths and angles in the resulting solid leads to a much wider, less structured absorption feature at 10~µm \citep{dorschnerbegemann1995,henning2010}. If condensation nuclei \rch{that form the seed particles for cloud particle growth} are prevalent and mixing is slow, silicates in the atmospheres of exoplanets and brown dwarfs should condense at high temperatures, as soon as gas is mixed into regions where condensates are thermodynamically stable. This would result in crystalline particles, because amorphous grains are very efficiently and quickly converted into crystals at high temperatures \citep[so-called ``annealing'', see, e.g.,][]{fabianjaeger2000,gail2001,harkerdesch2002,gail2004}. The broad, presumably amorphous absorption features seen in VHS~1256b \citep{milesbiller2023}, \pso \ and the \emph{Spitzer} spectra of many brown dwarfs \citep{lunamorley2021,suarezmetchev2023b} are therefore unexpected and point to a gap in our understanding of cloud physics. \rch{We also note that, for some stoichiometries, multiple crystalline forms may exist, with distinct spectral features. The stability regimes of these so-called polymorphs depend on temperature. Conversion timescales between polymorphic phases at temperatures relevant for atmospheres are in the range of hours. This may also make the occurrence of mixed polymorph grains possible, which could further transition through an amorphous stage during polymorph transformation \citep{moranmarley2024}.}
    \item {\bf Particle composition}: particle composition has a large effect on the shape and the exact location of the 10~µm feature. For example, as one moves from quartz (\ce{SiO2}) to enstatite (\ce{MgSiO3}) to forsterite (\ce{Mg2SiO4}) the degree of polymerization of \ce{SiO4} tetrahedra drops, leading to a shifting of the 10~µm feature to redder wavelengths \citep{henning2010}. The location and number of the narrow absorption maxima observed for crystalline silicates likewise changes as the composition of the grains is varied. Adding iron to the condensates described above (e.g., \ce{MgFeSiO4} instead of \ce{Mg2SiO4} when considering particles of olivine-type stoichiometry) further reddens the onset of the 10~µm band, because the bond lengths between Fe and O are longer than between Mg and O \citep{jaeger1998}. This effect is clearly discernible in crystalline silicates. In amorphous silicates, however, it can be obscured by other factors, including differences in disorder, shape and size. In addition, it is not necessarily the case that cloud particles are composed of just one species. In principle, particles could be layered (one species condensing on top of another) or more strongly mixed. For mixed particles, the characteristic absorption features (e.g., crystalline absorption features) could be muted \citep{kiefersamra2024}. The degree to which this is important is not clear, and we note that the silicate absorption seen in circumstellar disks can show both amorphous and crystalline features \rch{of uniquely identifiable species simultaneously} \citep[e.g.,][]{boekelmin2005,juhaszbouwman2010}.
    \item {\bf Particle shape}: the most common assumption for the shape of a cloud particle is a sphere, which enables the use of Mie theory to calculate the cross-sections of particles \citep[e.g.,][]{bohrenhuffman1983}. This approximation is likely incorrect, especially for solid particles (e.g., consider the shape of a snowflake). Several treatment options exist, all of which are more numerically costly than a simple Mie treatment, and all of them require additional parameters to describe the deviation of a particle from a sphere. Examples are treating particles as distributions of hollow spheres \citep[DHS, see][]{minhovenier2005}, continous distributions of ellipsoids \citep[CDE, see][]{bohrenhuffman1983} or to fully specify the shape of a grain and estimating its cross-section using the discrete dipole approximation \citep[DDA, see, e.g.,][]{purcell1973,draine1988}. The wavelength location of the crystalline absorption features can move significantly if deviations from spherical grains are considered, and indeed such a departure has been inferred for most of the crystalline silicate absorption in protoplanetary disks \citep[e.g.,][]{bouwman2001,juhaszbouwman2010}. An example of how the particle shapes affect exoplanet transmission spectra with crystalline clouds can be found in \citet{mollierevanboekel2016}. Amorphous silicate absorption is less strongly influenced by the assumed particle shape; it mainly changes the extent of the 10~µm-feature towards red wavelengths, while the blue onset of the feature is largely unaffected \citep[e.g.,][]{henningsognienko1993,Min2015}.
    \item {\bf Particle size (distribution)}: the particle size significantly affects the wavelength-dependent cross-sections of clouds. For example -- as follows from Mie theory -- scattering is strongest for wavelengths $\lambda \ll 2 \pi r$, where $r$ is the particle radius, and decreases with $\lambda^{-4}$ for wavelengths $\gg 2 \pi r$. In addition, the 10~µm silicate absorption feature begins to disappear for grains of sizes $\gtrsim 1$~µm, and its shape likewise depends on the particle size \citep[e.g.,][]{Min2015}. It is therefore not surprising that also the particle size distribution plays an important role for determining the shape of the 10~µm absorption feature. Log-normal particle size distributions, which appear symmetric around a characteristic size in logged particle size are a common assumption \citep[e.g.,][to name just a few studies]{ackermanmarley2001,morleyfortney2012,nasedkinmolliere2024,morleymukherjee2024}. Another example is the Hansen distribution \citep{hansen1971,burninghamfaherty2021,vosburningham2023} which can be asymmetric, depending on the choice of parameters, and may exhibit broad shoulders. Finally, there are fully microphysical models that solve for the particle size distribution as a function of altitude by considering processes such as condensation, settling, and coagulation, etc. \citep[e.g.,][]{hellingwoitke2008,gaomarley2018,powellzhang2018}. These studies often point to complex, multi-modal distributions, although it should be explored which particle sizes inferred from this full treatment actually matter when calculating spectra. In addition, it is not likely that cloud properties are constant across the atmosphere of a planet or brown dwarf, which can affect the aggregate shape of the 10~µm feature that arises from averaging fluxes across the visible hemisphere of the object.
\end{itemize}

Even if the above properties of the cloud species were perfectly known, biases are likely to impact analyses due to the challenges of deriving optical constants in laboratories. For instance, samples of the species of interest may be hard to synthesize. One way of producing amorphous silicates is through melts in a high-temperature furnace and subsequent cooling (with rates of $\sim 1000$~K/s) to prevent crystallization \citep{jaegermutschke1994,dorschnerbegemann1995}. Such samples are called ``glassy'' in the following. However, while \ce{MgFeSiO4} has a melting point of manageable 1900~K, \ce{Mg2SiO4} only melts at $\sim 2200$~K, which is hard to achieve in a laboratory setting. For such species the so-called ``sol-gel'' method is used, where metal organic compounds are dissolved in mixtures of water and alcohol (methanol or ethanol), hydrolized and finally condensed as a three-dimensional magnesium silicate network, the ``silicate gel'' \citep[see, e.g.,][]{jaegerdorschner2003}. The condensed gel is then distilled in order to remove the water and alcohol present, and subsequently annealed at elevated temperatures in order to remove the porosity and to densify the material (but not enough to let it crystallize). A challenge associated with silicate samples produced either by melting or by the sol-gel method is the derivation of optical constants at wavelengths corresponding to low absorption. The best solution would be transmission measurements of thick samples to determine the absorption coefficient directly from the transmission. If this is not possible, extrapolation is a commonly employed solution. However, this can result in discrepancies in the optical data close to the blue onset of the 10~µm feature between the silicates generated by melting or by sol-gel.

A last complication we want to mention here is that optical properties of a material sample depend on its temperature. This is most important for the absorption features of crystalline silicates where peaks become broader and shift location for higher temperatures \citep[e.g.,][]{zeidlermutschke2015}. Interestingly, this may affect the 10~µm feature somewhat less than the longer wavelength silicate absorption features towards the far-infrared \citep[$\lambda\gtrsim30$~µm, see][]{koikemutschke2006}. In all analyses in this work we neglect the temperature dependence of the silicate opacities. This is a common, but not necessarily justified, assumption in the community.

\subsection{Silicate feature analysis with the brightness temperature method}
\label{sect:brightness_temperature_method}

Given the aforementioned complexities that influence what the 10~µm absorption of a cloud will look like, it would be preferable to have an efficient method for identifying cloud candidate species in a first-look approach, that is, before running numerically costly retrievals. We present and apply a potentially useful method for this in the following. Since we treat this demonstration as a proof of principle, only a subset of the cloud complexities was explored in this study.

\subsubsection{Fitting procedure}
In an optically thin slab of gas and condensates, information about the silicates could be extracted by directly comparing the observed flux $F(\nu)$, where $\nu$ is the spectral frequency, to predicted opacities $\kappa(\nu)$, where opacity is defined as cross-section per unit mass. This is because in the optically thin limit the flux is described by
\begin{equation}
    F(\nu) \propto \kappa(\nu) B(\nu,T),
\end{equation}
where $B(\nu,T)$ is the Planck function, and $T$ is the silicate dust temperature. The atmosphere of a giant planet or brown dwarf is never optically thin. Instead, at every frequency one can only probe into the atmosphere until it becomes optically thick, which defines the so-called photosphere. Commonly the optical depth at the photosphere is assumed to be $\tau \approx 2/3$, although we note that emission is a continuous process, and the observed flux arises from regions at both lower and higher $\tau$. The above relation then no longer holds, but can be replaced by an expression that relates the atmospheric brightness temperature $T_{\rm bright}$ to the opacity, if making the simplifying assumption that the flux is emitted from $\tau = 2/3$ (or another fixed point), exactly:
\begin{equation}
    {\rm ln}[\kappa(\nu)] \approx  - \frac{{\rm ln}[T_{\rm bright}(\nu)]}{\nabla} + {\rm cst} ,
    \label{equ:brigthness_temp_master_equation}
\end{equation}
where $\nabla = d{\rm ln}(T)/d{\rm ln}(P)$ is a power law index approximating the average temperature gradient of the atmosphere in the regions probed by the observations, and $\rm cst$ is a place holder for a constant which is irrelevant to the problem. The derivation of this expression, a list of assumptions that go into obtaining it \rch{and a discussion of the validity of these assumptions are presented} in Appendix~\ref{app:opacity_properties}. As long as the cloud is the dominant opacity source over the frequency range of interest, an approximated cloud opacity can thus be extracted, up to a scaling constant, from the measured brightness temperature.

In what follows, we use Eq.~\ref{equ:brigthness_temp_master_equation} to fit the opacities predicted for various silicate species to the observed brightness temperature. In practice, we assume that the silicate cloud resides in a cloudy column of the atmosphere, and add a correction to the brightness temperature to account for the emission of another atmospheric column with a constant brightness temperature. In the derivation this column is called ``clear'', but its defining propertiy is that its spectrum is featureless (i.e., at constant brightness temperature) across the 10~µm region. We treat $\nabla$, the cloud column coverage  $f$ and the clear column photospheric temperature (again, see Appendix~\ref{app:opacity_properties} for more details) as free parameters to extract ${\rm ln}(\kappa)$. This ${\rm ln}(\kappa)$ is then compared to predicted silicate cloud (scattering+absorption) opacities, assuming a log-normal particle size distribution (in principle, any distribution may be adopted). For this, the mean particle size and width of the distribution are free parameters. The maximum of the log-opacity of the resulting cloud description is scaled to the maximum of the log-opacity derived from the brightness temperature, with an additional scaling (additive term in log space) as a free parameter. We note that all free parameters are varied simultaneously during the fit.  \rch{We then investigate which species best describes the 10~µm feature over a wavelength range from 8-12.5~µm, which we determined to be the range over which the silicate cloud is the dominant opacity contributor.} The priors adopted for the free parameters are given in Table~\ref{table:fit_priors}. The fits were run by estimating the posterior probability distribution of the free parameters, given the observed brightness temperatures. For this we used PyMultiNest \citep{buchnergeorgakakis2014}, which is a Python wrapper of MultiNest \citep{ferrozhobson2008,ferrozhobson2009,ferozhobson2019}. For our runs we assumed 200 live points, and the standard parameter values of \verb|pymultinest.run()|, that is \verb|evidence_tolerance = 0.5| and \verb|sampling_efficiency = 0.8|. The error bars on the brightness temperature observations were derived from the flux uncertainties (see Section~\ref{appendix:kappa_uncertainties}). For the latter we used the error bar inflation from our best-fitting retrieval model ($b=-8.589$ at $\lambda/\Delta\lambda=400$, see Section~\ref{sect:10mu_retrievals} for more information), adjusted to our fitting resolution $\lambda/\Delta\lambda=100$. 

\begin{table}[t!]
    \centering
    \begin{threeparttable}
    \caption{Priors adopted for the cloud feature fit.}
    \label{table:fit_priors}
    \begin{tabular}{l l}
        \hline\hline
        Parameter & Prior  \\
        \hline
$\nabla$ (powerlaw $T$-gradient) & $\mathcal{U}(0.01,0.5)$ \\
$s$  (log-opacity scaling) & $\mathcal{U}(-1,1)$ \\
${\rm log}_{10}(a/{\rm 1 \ cm})$  (log-mean particle radius) & $\mathcal{U}(-6,0)$ \\
$\Delta {\rm log}_{10}\sigma$  (particle distribution width$^{\rm (a)}$) & $\mathcal{U}(-2,0)$ \\
$f$ (cloud coverage fraction) & $\mathcal{U}(0,1)$ \\
$T_{\rm clear}$ (clear column temperature in K) & $\mathcal{U}(500,2000)$ \\
        \hline
    \end{tabular}
    \begin{tablenotes}
\footnotesize
\item Notes: $\mathcal{U}(x_1,x_2)$ denotes a uniform distribution from $x_1$ to $x_2$. (a): the width of the log-normal particle size distribution is constructed as $\sigma=1+2\times 10^{\Delta {\rm log}_{10}\sigma}$, with the log-normal particle size distribution defined as in \citet{ackermanmarley2001}. A value of $\sigma=1$ corresponds to a $\delta$ function.
\end{tablenotes}
\end{threeparttable}

\vspace{4mm}

\begin{threeparttable}
    \caption{\rch{Ranked list of cloud opacity fits using the brightness temperature method for the wavelength range of 8-12.5~µm (species considered in the full retrievals in Sect.~\ref{sect:10mu_retrievals} are highlighted in bold).}}
    
    \label{table:cloud_fit_full}
    {\fontsize{8.3}{\baselineskip}\selectfont
    \begin{tabular}{l c c c}

        \hline\hline
        Species, structure, shape & ${\rm log}_{10}(a/{\rm 1 cm})$ & $\sigma$ & $\chi_{\rm red}^2$ \\
        \hline
        {\bf \ce{SiO}, amorph., sph.} & $-3.99_{-0.14}^{+0.06}$ & $1.22_{-0.15}^{+0.19}$ & 0.65 \\
{\bf \ce{MgSiO3}$^{\rm (s)}$, amorph., sph.} & $-3.8_{-0.13}^{+0.04}$ & $1.2_{-0.14}^{+0.23}$ & 0.75 \\
{\bf \ce{SiO}, amorph., irr.} & $-4.28_{-0.85}^{+0.22}$ & $1.26_{-0.2}^{+0.79}$ & 0.82 \\
{\bf \ce{MgSiO3}$^{\rm (s)}$, amorph., irr.} & $-3.66_{-0.13}^{+0.05}$ & $1.1_{-0.07}^{+0.27}$ & 0.85 \\
{\bf \ce{Mg2SiO4}$^{\rm (s, \nabla)}$, amorph., sph.} & $-5.34_{-0.42}^{+0.51}$ & $1.11_{-0.08}^{+0.28}$ & 0.96 \\
{\bf \ce{MgSiO3}$^{\rm (g)}$, amorph., sph.} & $-3.92_{-0.06}^{+0.03}$ & $1.09_{-0.05}^{+0.15}$ & 1.20 \\
{\bf \ce{MgSiO3}$^{\rm (g)}$, amorph., irr.} & $-4.5_{-0.94}^{+0.44}$ & $1.16_{-0.12}^{+0.42}$ & 1.46 \\
{\bf Mg$_{0.5}$Fe$_{0.5}$SiO$_3$$^{\rm (s)}$, amorph., sph.} & $-3.98_{-0.04}^{+0.03}$ & $1.05_{-0.02}^{+0.07}$ & 1.59 \\
\ce{SiO2}$^{\rm (\nabla)}$, cryst., irr. & $-3.45_{-0.03}^{+0.04}$ & $1.13_{-0.09}^{+0.14}$ & 1.84 \\
Mg$_{0.5}$Fe$_{0.5}$SiO$_3$$^{\rm (s)}$, amorph., irr. & $-5.29_{-0.5}^{+0.61}$ & $1.13_{-0.09}^{+0.36}$ & 1.87 \\
{\bf \ce{MgSiO3}$^{\rm (\nabla)}$, cryst., sph.} & $-5.06_{-0.55}^{+0.49}$ & $1.12_{-0.08}^{+0.36}$ & 2.09 \\
\ce{SiO2}$^{\rm (\nabla)}$, cryst., sph. & $-3.64_{-0.01}^{+0.01}$ & $1.15_{-0.04}^{+0.03}$ & 2.55 \\
{\bf \ce{Mg2SiO4}$^{\rm (s, \nabla)}$, amorph., irr.} & $-5.23_{-0.47}^{+0.38}$ & $1.13_{-0.08}^{+0.25}$ & 2.62 \\
\ce{MgSiO3}, cryst., irr. & $-5.37_{-0.39}^{+0.42}$ & $1.13_{-0.1}^{+0.33}$ & 2.85 \\
{\bf \ce{MgFeSiO4}$^{\rm (g)}$, amorph., sph.} & $-5.36_{-0.44}^{+0.51}$ & $1.13_{-0.09}^{+0.29}$ & 3.96 \\
\ce{SiO2}$^{\rm (s)}$, amorph., sph. & $-3.5_{-0.02}^{+0.02}$ & $1.25_{-0.19}^{+0.16}$ & 5.19 \\
\ce{MgFeSiO4}$^{\rm (g)}$, amorph., irr. & $-5.35_{-0.43}^{+0.44}$ & $1.11_{-0.08}^{+0.27}$ & 6.44 \\
\ce{Mg2SiO4}, cryst., sph. & $-5.52_{-0.28}^{+0.39}$ & $1.11_{-0.07}^{+0.24}$ & 6.60 \\
\ce{SiO2}$^{\rm (s)}$, amorph., irr. & $-0.31_{-0.56}^{+0.23}$ & $1.11_{-0.08}^{+0.34}$ & 8.88 \\
\ce{SiC}$^{\rm (\nabla)}$, cryst., sph. & $-3.5_{-0.01}^{+0.06}$ & $1.06_{-0.01}^{+0.04}$ & 9.04 \\
\ce{Mg2SiO4}, cryst., irr. & $-5.5_{-0.33}^{+0.4}$ & $1.12_{-0.09}^{+0.3}$ & 9.66 \\
\ce{Fe2SiO4}$^{\rm (\nabla)}$, cryst., sph. & $-0.89_{-0.32}^{+0.25}$ & $1.35_{-0.21}^{+0.51}$ & 10.13 \\
\ce{SiC}$^{\rm (\nabla)}$, cryst., irr. & $-0.9_{-0.34}^{+0.25}$ & $1.28_{-0.09}^{+0.2}$ & 10.35 \\
\ce{Fe2SiO4}$^{\rm (\nabla)}$, cryst., irr. & $-2.0_{-0.03}^{+0.02}$ & $1.07_{-0.03}^{+0.05}$ & 10.75 \\
        \hline
    \end{tabular}}
        \begin{tablenotes}
            \footnotesize
            \item Notes: (s): sol-gel. (g): glassy. \rch{($\nabla$): fits with a resulting ${\rm median}(\nabla)>0.3$. This may be problematic because convection drives the atmosphere towards $\nabla={\rm min}(\nabla_{\rm rad},\nabla_{\rm ad})$, where $\nabla_{\rm rad}$ and $\nabla_{\rm ad}$ are the radiative and adiabatic temperature gradients, respectively, and it holds that $\nabla_{\rm ad}\approx 0.3$.}
        \end{tablenotes}
    \end{threeparttable}

\end{table}

\begin{figure*}[t!]
    \centering
    \includegraphics[width=0.95\textwidth]{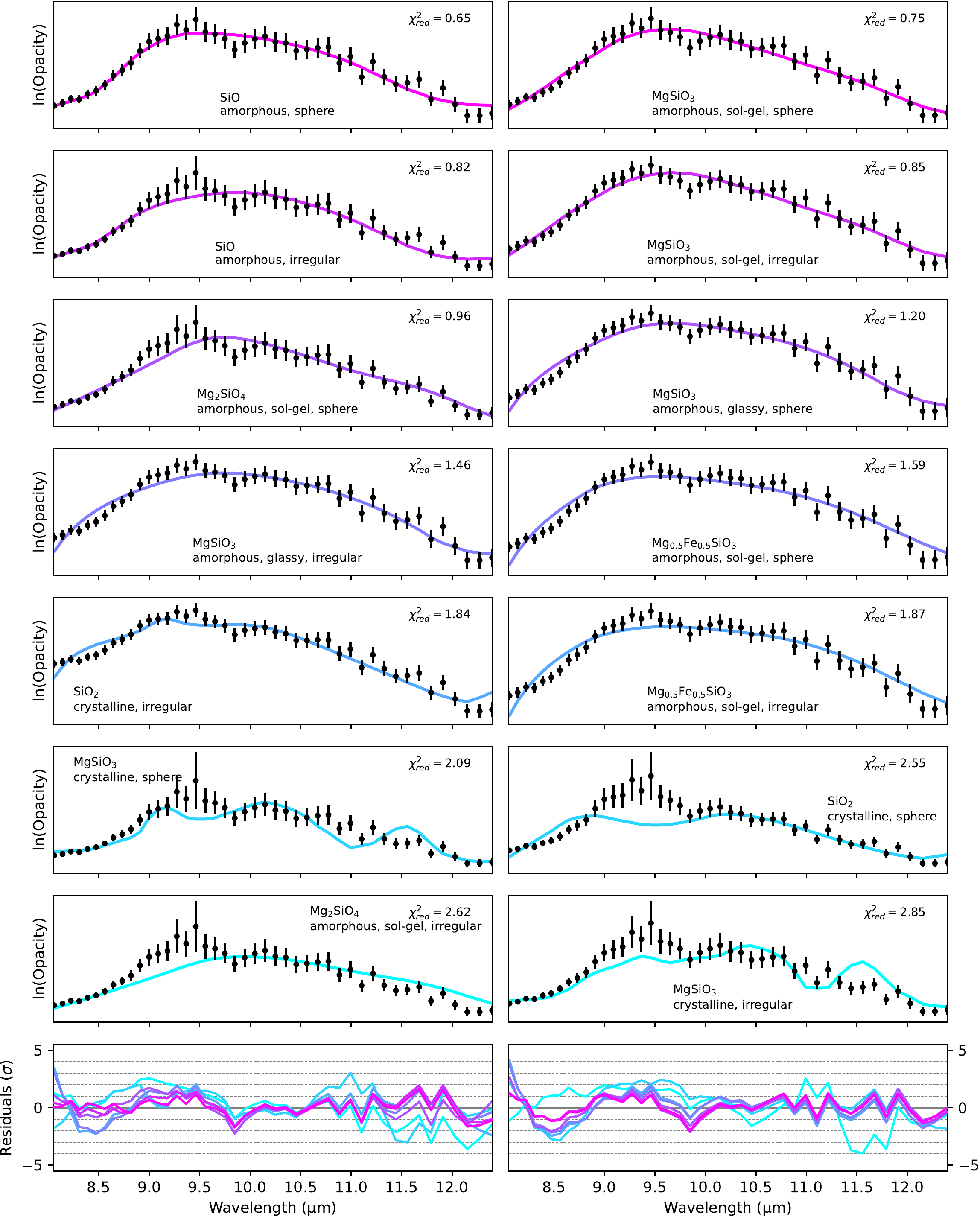}
    \caption{\rch{Opacity fits of the $10$~µm feature for the ``Top-14'' silicate cloud species, using the brightness temperature method. The remaining fits for the less well fitting species can be found in Fig.~\ref{fig:brightness_temperature_all_appendix_1}}}
    \label{fig:cloud_shape_fits}
\end{figure*}

\rch{The $\chi^2_{\rm red}$ values for all considered silicate species are given in Table~\ref{table:cloud_fit_full} (see Table~\ref{tab:si_cloud_species_for_brigthness_temp_fitting} for a list of references of the optical constants). The fits of the ``Top-14'' species can be seen in Fig.~\ref{fig:cloud_shape_fits} and fits of all remaining species are shown in figure \ref{fig:brightness_temperature_all_appendix_1}. We find that the best-fit $\chi^2_{\rm red}$ are smaller than one, which could indicate overfitting. We note, however, that the flux error bars we are using here have been scaled up, based on values found in the full retrievals (see Sect.~\ref{sect:10mu_retrievals}), and may account for both underestimated uncertainties and retrieval model insufficiencies.}

In addition to the nominal single-species fits described here, we also attempted to fit the silicate feature by combining two cloud species. For this we considered all $N_{\rm c}(N_{\rm c}-1)/2=276$ possible combinations of species, where $N_{\rm c}=24$ is the number of Si-bearing cloud species in our database. To keep the number of free parameters low, we only added a relative weighting between the two species as an additional free parameter ($\kappa_{\rm tot}=\kappa_1+w\kappa_2$, with $w$ going from $10^{-5}$ to $10^5$ on a log-uniform prior). The mean particle size and width of the size distribution was therefore the same for both species. We robustly find that SiO is the species leading to the best fits: \rch{it is one of the two combined species in 21 out of the top 25 combinations, and also is present in the best combination. We find that adding an additional species can increase the fit quality, pushing $\chi^2_{\rm red}$ to values as low as 0.52, when compared to the single-species best-fit (also SiO) of 0.65.} We refrained from a more detailed exploration with multiple species, because we considered the brightness temperature technique as a tool for finding likely cloud candidates for in-depth analyses via retrievals here. The underlying assumptions (see Appendix \ref{app:opacity_properties}) may be satisfactorily met for our intended purposes, but may not be good enough to replace an in-depth retrieval analysis over the full wavelength range, with proper radiative transfer. For crystalline silicate features, for which cloud species have more distinct appearances for different species, the multi-species approach may be worth revisiting.

\subsubsection{Results of the brightness temperature analysis}

The most likely cloud species, if taking the results of the brightness temperature method at face value, is the absorption of amorphous SiO. This is surprising, because SiO would oxidize to form \ce{SiO2} very quickly. If correct, this could mean that the SiO forms in a quite strongly O-depleted (relative to carbon) environment \citep{wetzel2013}, or that we trace the formation of SiO nuclei \rch{that form the seeds for further cloud condensation \citep[e.g.,][]{gailwetzel2013} (this option is discussed further in Sect.~\ref{sect:sio_evidence_discussion}, since we deem it a possible scenario).} One may argue that while SiO fits the spectrum well over the narrow wavelength range considered here (8-12.5~µm), the required particle properties may lead to imperfect fits over the full \emph{JWST} wavelength range. We thus progressed by using the brightness temperature method as a way to inform our decision on which condensate species should be tested in retrievals. However, as we describe in Section~\ref{sect:10mu_retrievals}, the retrievals also point to SiO being the most likely species. We also note again that SiO was identified as a possible species in \citep{lunamorley2021,suarezmetchev2023b}. The implications of this finding, if correct, are discussed in Sections~\ref{sect:sio_evidence_discussion} and \ref{sect:discussion}. Lastly we note that even for the best-fitting species a residual wavelength-dependent structure is visible in the opacities inferred from brightness temperatures, from $\sim$9.5-10~µm, \rch{see Fig.~\ref{fig:cloud_shape_fits}}. This will likewise be discussed in Section~\ref{sect:sio_evidence_discussion}.


\subsection{Retrievals}
\label{sect:10mu_retrievals}

Armed with a \rch{ranked} list of likely cloud species, we study how well we can reconstruct the spectrum of \pso \ through retrievals below, and how robustly we can constrain the underlying properties of the atmosphere. Particular focus is placed on the identity of the clouds causing the 10 µm feature. \rch{More specifically, we will test the top 8 species identified with the brightness temperature method. In addition, we will look at three more species from further down in the list, resulting in 11 species with $\chi_{\rm red}^2$ values from 0.65 to 4 in the brightness temperature method (these species are also highlighted in bold in Table~\ref{table:cloud_fit_full}).}

Retrievals generally aim to constrain the probability distribution of parameters thought to describe the atmosphere (e.g., temperature, composition, cloud properties) while leveraging any prior beliefs we have about the distribution of said parameters. A fairly recent review on retrievals can be found in \citet{madhusudhan2018b}. A number of retrieval codes exist now, many of them open source; we refer the reader to \citet{macdonaldbatalha2023}\footnote{\url{https://zenodo.org/records/7859170}} for an up-to-date list. In the work presented below we use \texttt{petitRADTRANS} \citep[\texttt{pRT}, see][]{mollierewardenier2019,mollierestolker2020,blainmolliereJOSS2024}; retrievals were run with \texttt{pRT}'s retrieval package described in \citet{nasedkinmolliere2024}.

\subsubsection{Forward model description}
The forward model is the function that returns flux predictions given a set of atmospheric input parameter values. A retrieval inverts the forward model and returns the parameter distribution, given an observation. The forward model is thus the central ingredient of any retrieval. We start by defining a single-column forward model which takes the atmospheric temperature, cloud and compositional structure as input parameters for calculating the flux emerging at the top of the atmosphere (including multiple scattering). A single-column model can also be used as a building block to approximate horizontal atmospheric heterogeneities. For this we assume that the atmosphere is well described by the flux predicted from a set of 1-dimensional columns, weighted by the fractional area they occupy in the atmosphere. Below we assume at most two columns.

\begin{itemize}
    \item {\bf Atmospheric temperature structure}: we described the atmospheric temperature structure using the approach reported in \citet{zhangmolliere2023}, that is, the power law dependence of the temperature with pressure $d{\rm ln}T/d{\rm ln}P$ was retrieved at 10 points in the atmosphere (equidistantly spaced in log-pressure), and quadratically interpolated between these layers. The priors for the seven lowest (i.e., highest pressure) points were determined from constructing the distribution of $d{\rm ln}T/d{\rm ln}P$ for the temperature profiles reported in \citet{morleymukherjee2024}, as has been reported in \citet{zhangmolliere2025}. The corresponding 1-$\sigma$ ranges of these distributions defined our Gaussian priors. In the three uppermost layers the power law index could vary freely (excluding inversions) because our pressure range extends to $10^{-6}$ bar, but the \citet{morleymukherjee2024} models end at $10^{-4}$ bar. The adopted priors for all retrieval parameters and the pressure coordinates of the 10 points are given in Table \ref{table:retrieval_priors}.
    \item {\bf Temperature excursion}: we additionally allow for the temperature profile to deviate from the above treatment. This is only used for some multi-column setups. Nominally, the columns share the same temperature profile, but with the excursion treatment the temperature structure of a given column is allowed to deviate. The temperature excursion is defined by multiplying the nominal $d{\rm ln}T/d{\rm ln}P$ structure by
    \begin{equation}
        1+f_{\rm exc}\left(1-\frac{2|{\rm log}_{10}(P)-{\rm log}_{10}(P_{\rm exc})|}{\Delta{\rm log}_{10}(P_{\rm exc})}\right),
    \end{equation}
    for all $|{\rm log}_{10}(P/P_{\rm exc})|\leq \Delta{\rm log}_{10}(P_{\rm exc})/2$. The prior for $f_{\rm exc}$ is chosen such that the resulting behavior of the excursion spans from $P$-$T$ curves exhibiting (weak) inversions to curves with strong boosts of the $d{\rm ln}T/d{\rm ln}P$ gradient. Also ${\rm log}_{10}(P_{\rm exc})$ and $\Delta{\rm log}_{10}(P_{\rm exc})$ are free parameters.
    \item {\bf Chemical composition}: we determined the chemical composition by prescribing chemical equilibrium for all absorber species we expect to play a minor role. For this we retrieved the atmospheric metallicity, [M/H], and C/O, where the latter was changed by scaling the oxygen abundance after all metals have been scaled by $10^{\rm [M/H]}$. The composition is obtained by interpolating in the chemical equilibrium table that is part of \emph{pRT}, which itself has been prepared with \emph{easyCHEM} \citep{leimolliere2024}\footnote{\url{https://easychem.readthedocs.io/en/latest/index.html}}. \rch{No rainout is included in these calculations, while it is implicitly taken into account for some species by limiting the selection of condensates. For example, feldspars such as orthoclase are not included in our chemical calculations, with the goal of preventing a sequestration of alkalis into these species at the temperatures of L/T transition objects -- which is not observed \citep{linemarley2017,zaleskyline2019}.} Since we wanted to treat the major absorbing species more flexibly, we retrieved the mass fractions of $\rm ^{12}CO$, $\rm ^{13}CO$, \ce{CO2}, \ce{CH4}, \ce{CrH} and \ce{H2O} independently, assuming them to be vertically constant. \ce{CrH} was retrieved independently because it is not included in the pre-computed equilibrium table of \emph{pRT}. The retrieved metallicity and C/O values reported in Table~\ref{tab:all_retrieval_posteriors} are obtained from considering all metal gas phase abundances, irrespective of whether they were obtained from the chemical interpolation or retrieved independently.
    \item {\bf Gas opacity sources}: we included the following line opacities in our analysis: \ce{CH4} \citep{hargreaves2020}, $^{12}$CO \citep{rothman2010}, $^{13}$CO \citep{rothman2010}, \ce{CO2} \citep{yurchenkomellor2020}, CrH \citep{burrowsram2002}, FeH \citep{wendereiners2010}, HCN \citep{barberstrange2013}, \ce{H2O} \citep{polanskykyuberis2018}, \ce{H2S} \citep{azzamtennyson2016}, K \citep[line profiles by N. Allard, see][]{mollierewardenier2019}, Na \citep[][]{allardspiegelman2019}, \ce{NH3} \citep{colesyurchenko2019}, \ce{PH3} \citep{sousasilva2014}, SiO \citep{yurchenkotennyson2021}, TiO \citep{mckemmishmasseron2019}. Where available, correlated-k opacities in the \texttt{pRT} format were taken from the ExoMolOP database \citep{chubb2021exomolop-1d1} or computed with the method described in \citet{mollierevanboekel2015} otherwise. Rebinning to lower resolution opacities was done using \texttt{Exo\_k} \citep{leconte2021}.
    \item {\bf Clouds}: the parameterized behavior of the clouds is motivated by the semi-analytical model presented in \citet{ackermanmarley2001}, but is more flexible. For any given cloud species we freely retrieve the position of the cloud base, $P_{i{\rm ,base}}$. The cloud mass fraction at the cloud base $X_{i{\rm, base}}$, where $i$ stands for a cloud species like \ce{MgSiO3}, is found by scaling the elemental mass budget by a free parameter $10^{s_i}$. The elemental mass budget is determined by locking all elemental building blocks into the condensate in question, until the first species is depleted (e.g., for \ce{MgSiO3} the limiting element is Si, when considering scaled solar composition). The cloud mass fraction in the atmosphere is then $X_i(P) = X_{i{\rm, base}}(P/P_{i{\rm ,base}})^{f_{{\rm sed}, i}}$ for $P\leq P_{i{\rm ,base}}$ and 0 for $P>P_{i{\rm ,base}}$. The power law index $f_{\rm sed}$ is likewise a free parameter. The particle size distribution is assumed to be log-normal, as defined in \citet{ackermanmarley2001}, with the mean particle size $a$ and the width $\sigma$ as free parameters in our standard approach. We set the prior for $\sigma$ up in the same way as in the brightness temperature fitting, see Section~\ref{sect:brightness_temperature_method}. 
    \begin{table}
    \centering
    \begin{threeparttable}        
    \caption{Retrieval priors adopted for the single-column forward model.}
    \label{table:retrieval_priors}
    {\fontsize{9.0}{\baselineskip}\selectfont
    \begin{tabular}{l l}
        \hline\hline
        Parameter & Prior  \\
        \hline
        $P$-$T$ profile & \\
        $T_{\rm bottom}$ (K) & $\mathcal{U}(100,8900)$ \\
        $(d {\rm ln}T/d {\rm ln}P)_1$ & $\mathcal{N}(0.25,0.025)$ \\
        $(d {\rm ln}T/d {\rm ln}P)_2$ & $\mathcal{N}(0.15, 0.03)$ \\
        $(d {\rm ln}T/d {\rm ln}P)_3$ & $\mathcal{N}(0.18, 0.045)$ \\
        $(d {\rm ln}T/d {\rm ln}P)_4$ & $\mathcal{N}(0.21, 0.06)$ \\
        $(d {\rm ln}T/d {\rm ln}P)_5$ & $\mathcal{N}(0.16, 0.05)$ \\
        $(d {\rm ln}T/d {\rm ln}P)_6$ & $\mathcal{N}(0.08, 0.025)$ \\
        $(d {\rm ln}T/d {\rm ln}P)_7$ & $\mathcal{N}(0.07, 0.02)$ \\
        $(d {\rm ln}T/d {\rm ln}P)_8$ & $\mathcal{U}(0,0.1)$ \\
        $(d {\rm ln}T/d {\rm ln}P)_9$ & $\mathcal{U}(0,0.1)$ \\
        $(d {\rm ln}T/d {\rm ln}P)_{10}$ & $\mathcal{U}(0,0.1)$ \\ \hdashline
        \multicolumn{2}{l}{\underline{If using a temperature excursion:}} \\
        Location ${\rm log}_{10}(P_{\rm exc}/1 {\rm bar})$ & $\mathcal{U}(-6, 3)$ \\
        Excursion width $\Delta {\rm log}_{10}(P_{\rm exc})$ & $\mathcal{U}(0, 5)$ \\
        Slope factor $f_{\rm exc}$ & $\mathcal{U}(-1.1, 2)$ \\ \hline
        Composition & \\
        $\rm [M/H]$ & $\mathcal{U}(-1,1.5)$ \\
        C/O & $\mathcal{U}(0.1,1.6)$ \\
        ${\rm log}_{10}(X_i)^{\rm (a)}$ & $\mathcal{U}(-10,-1)$ \\ \hline
        Clouds & \\
        $s_i^{\rm (b)}$ & $\mathcal{U}(-6,2)$ \\
        ${\rm log}_{10}(P_{i{\rm ,base}}/{\rm 1 \ bar})$ & $\mathcal{U}(-6,3)$ \\
        $\Delta {\rm log}_{10}\sigma^{\rm (c)}$ & $\mathcal{U}(-2,0)$ \\
        $f_{{\rm sed}, i}^{\rm (d)}$ & $\mathcal{U}(0,10)$  \\ 
        ${\rm log}_{10}(a_i/{\rm 1 \ cm})$ & $\mathcal{U}(-6,0)$ \\ \hline 
        \underline{If using evolutionary priors:} \\ 
        $R$ ($\rm R_{Jup}$) & $\mathcal{N}(1.36,0.09)$ \\
        ${\rm log}_{10}(g/{\rm 1 \ cm \ s^{-2}})$ & $\mathcal{N}(4,0.04)$ \\ \hdashline
        \underline{If not using evolutionary priors:} \\
        $R$ ($\rm R_{Jup}$) & $\mathcal{U}(0.8,2.5)$ \\
        ${\rm log}_{10}(g/{\rm 1 \ cm \ s^{-2}})$ & $\mathcal{U}(3,6)$ \\ \hline
        Column coverage \\
        $B_{\rm HST}^{\rm (e)}$ & $\mathcal{U}(0,1)$ \\
        $B_{\rm SpeX}$ & $\mathcal{U}(0,1)$ \\
        $B_{\rm JWST}$ & $\mathcal{U}(0,1)$ \\ \hline
        Uncertainty scaling & \\
        ${10^b}^{\rm (f)}$ for \emph{HST WFC3} & $\mathcal{U}[0.01 \ {\rm min}(\sigma^2),100 \ {\rm max}(\sigma^2)]$ \\
        ${10^b}$ for \emph{SpeX} & $\mathcal{U}[0.01 \ {\rm min}(\sigma^2),100 \ {\rm max}(\sigma^2)]$ \\
        ${10^b}$ for \emph{NIRSpec PRISM} & $\mathcal{U}[0.01 \ {\rm min}(\sigma^2),100 \ {\rm max}(\sigma^2)]$ \\
        ${10^b}$ for \emph{NIRSpec G395H NRS 1} & $\mathcal{U}[0.01 \ {\rm min}(\sigma^2),100 \ {\rm max}(\sigma^2)]$ \\
        ${10^b}$ for \emph{NIRSpec G395H NRS 2} & $\mathcal{U}[0.01 \ {\rm min}(\sigma^2),100 \ {\rm max}(\sigma^2)]$ \\
        ${10^b}$ for \emph{MIRI MRS} & $\mathcal{U}[0.01 \ {\rm min}(\sigma^2),100 \ {\rm max}(\sigma^2)]$ \\ \hline
    \end{tabular}
    }
    \begin{tablenotes}
        \item Notes: $\mathcal{U}(x_1,x_2)$ denotes a uniform distribution from $x_1$ to $x_2$. $\mathcal{N}(\mu,\sigma)$ denotes a normal distribution with mean $\mu$ and standard deviation $\sigma$. Notes: the power law indices $d {\rm ln}T/d {\rm ln}P$ are separated by one dex in pressure, going from $10^3$ for $(d {\rm ln}T/d {\rm ln}P)_1$ to $10^{-6}$ bar for $(d {\rm ln}T/d {\rm ln}P)_{10}$. (a): $X_i$ stands for the mass fractions of $\rm ^{12}CO$, $\rm ^{13}CO$, \ce{CO2}, \ce{CH4}, \ce{CrH} and \ce{H2O}, respectively. (b): $s_i$ stands for the mass fraction {\it scaling} (see text) applied for cloud species $i$ (e.g., \ce{MgSiO3}). (c): same meaning as in Table~\ref{table:fit_priors}. (d): power law factor controlling the cloud scale height. (e): weight associated to a column in a multicolumn fit. For a two-column fit the other column has a weight of $1-B$. (f): factor controlling the uncertainty scaling, where $\sigma_{\rm scaled} = (\sigma^2+10^b)^{1/2}$, following \citet{lineteske2015}.
    \end{tablenotes}
    \end{threeparttable}
    \vspace{5mm}
\end{table}
    \item {\bf Evolutionary priors:} Despite excellent data we noticed the tendency of our retrievals to approach unphysical values for bulk parameters (e.g., ${\rm log}(g) \rightarrow 3$ and lower, depending on the prior range). We therefore decided to prescribe priors on the radius and ${\rm log}(g)$, using the values derived in \citet{zhangliu2020}. For these values the authors assumed an age for \pso \ consistent with the $\beta$~Pic moving group ($24\pm3$~Myr), used \pso's inferred bolometric luminosity, and the cooling curves by \cite{saumonmarley2008}. Alternatively we adopted free priors on the radius and gravity to explore their effect on our best-fitting model.
    \item {\bf Column coverage:} our forward model is set up in such a way that it can handle multiple 1-d columns, with the total flux being
    \begin{equation}
        F = \sum_{i = 1}^{N_{\rm column}}b_i F_i,
    \end{equation}
    where $F_i$ is the top-of-atmosphere flux of column $i$, and $b_i\geq 0$ its weight, respectively. In principle, the different columns may be fully independent, but in practice they share most of their parameters and only the cloud parameters are varied on a per-column basis, for example. Because the retrievals presented here assumed at most two columns we defined $b_1 = B$, $b_2 = 1-B$, $B\leq 1$. Since the various data sets we consider were taken at different epochs we retrieve three different values: $B_{\rm HST}$, $B_{\rm SpeX}$, $B_{\rm JWST}$, thought to express different average climate states at the respective times of observation with the three observatories. That is, if $\tilde{B}(t)$ is the actual time-dependent change due to rotation of the object then $B$ is the time average of $\tilde{B}(t)$ during the observation. We neglect that the JWST observations with the \emph{NIRSpec PRISM}, \emph{G395H} and \emph{MIRI MRS} instruments were taken sequentially, thus recorded time-averaged fluxes from different rotational phase intervals.
    \item {\bf Uncertainty scaling:} the flux uncertainties of individual instruments were scaled by setting $\sigma_{\rm scaled}(\lambda)=[\sigma^2(\lambda)+10^b]^{1/2}$, where $\sigma$ is the reported flux uncertainty of the observation. This treatment allows to correct for underestimated uncertainties or shortcoming in the model that would lead to systematic biases. It serves to give a more conservative estimate of the parameter distribution width as it widens the posteriors \citep{lineteske2015}. The $b$ values were retrieved on a per-data-set basis, where \emph{G395H}'s \emph{NRS~1} and \emph{NRS~2} detectors were treated separately, because we binned \emph{NRS~1} to $\lambda/\Delta\lambda=400$ and \emph{NRS~2} to $\lambda/\Delta\lambda=1000$ during the retrievals. The priors for the $b$ values were taken from \citet{lineteske2015}.
\end{itemize}

\subsubsection{Retrieval runs}

We tested a number of different forward model setups in which the atmosphere was approximated by two columns. For example, we let the temperature structure vary between the columns, or the free parameters associated with the cloud, or the parameters associated with the chemical composition, or mixtures of these setups. The motivation for these options are the various processes (cloud cover, temperature, compositional variations) that have been suggested as drivers for atmospheric variability \citep[see, e.g.,][]{radiganlafreniere2014,robinsonmarley2014,tremblinphillips2020}. For studying whether we can constrain the most likely cloud species with retrievals we finally adopted a model in which both columns shared most of their properties: $P$-$T$ structure, gas phase abundances, and the parameters describing the iron cloud. A global iron cloud (in the lower atmosphere) is a common finding in retrieval studies \citep{burninghamfaherty2021,vosburningham2023}. Then, for Column 1, a silicate cloud with its associated parameters was retrieved, while for Column 2 the silicate cloud was neglected (by setting $s_{\rm silicate}=-50$), motivated by the patchy silicate cloud findings of, for example, \citet{apairadigan2013,vosburningham2023,zhangmolliere2025}. We thus stress that ``patchy silicate clouds'' does not mean that the atmosphere is described by a cloudy and a cloud-free column, since the iron cloud is present in both columns. The two Columns 1 and 2 also retrieved separate $\Delta {\rm log}\sigma$, allowing for different widths of the particle size distributions between Columns 1 and 2. Results obtained with retrieval models that differ from our standard setup, for example assuming a single-column atmosphere, or keeping the cloud parameters fixed in both columns and varying the temperature structure instead, or turning off the evolutionary prior, are listed in Table~\ref{tab:all_retrieval_posteriors} and discussed in Section~\ref{sect:bulk_retrievals}.
    
We note that among all our explored two-column setups, the fiducial model definition described above consistently fit the data with the least bias upon visual inspection, resulted in the smallest required error bar scalings (i.e., the smallest $b$-factors), and formally converged in retrievals most reliably. The best-practice approach would be to carry out model selection via a Bayes factor analysis, but with our 40+ free parameters, wide wavelength coverage, and high S/N-data, MultiNest starts to break down \citep[also see][for a discussion of how nested sampling can fail]{buchner2021,himes2022retrievalfail,dittmann2024}. For example, we found that models with too many free parameters failed to converge, or observed that widening the prior ranges (when turning off the evolutionary priors for ${\rm log}_{10}(g)$ and $R$) resulted in worse fits for a given model. Given the fact that we are in a high-dimensional parameter space (with signs of the sampler missing the global likelihood maximum, also see \citealt{himes2022retrievalfail}), and given that we must run MultiNest in constant sampling efficiency to enable convergence \citep[which can lead to over-confident posteriors, see][]{chubbmin2022}, we refrain from carrying out MultiNest-based Bayes factor analyses; we cannot guarantee that the posterior sampling leads to a reliable integration for the evidence $Z$. This observation and its implications for the use of nested sampling in the era of JWST (and ELT in the future) will be discussed in Section~\ref{sect:discussion}.

In practice we set up MultiNest with 3000 live points, using \verb|const_efficiency_mode=True|, \verb|sampling_efficiency=0.05| and \verb|evidence_tolerance| set to 0.5. While MultiNest results must be approached cautiously for the reasons stated above, it is the nested sampling algorithm that converges most efficiently \citep{himes2022retrievalfail} and enables us to run retrievals for this study in the first place. To speed up the retrievals further we binned the \emph{MIRI MRS} data to \texttt{pRT}'s wavelengths for a $\lambda/\Delta\lambda=400$ wavelength spacing, exactly. The same was done for \emph{NIRSpec G395H NRS1}. For \emph{NRS2} we used the spacing of \texttt{pRT} wavelength tables at $\lambda/\Delta\lambda=1000$, to retain sensitivity to secondary CO isotopologues. For \emph{HST}, \emph{SpeX} and \emph{JWST NIRSpec PRISM} the models were calculated at $\lambda/\Delta\lambda=300$ (\emph{HST}) and 140 (\emph{SpeX} and \emph{NIRSpec PRISM}) before being convolved to $R=130$ (\emph{HST}), 75 (\emph{SpeX}) and 65 (\emph{NIRSpec PRISM}) and then binned to the data's respective wavelength spacing. Retrievals were run on the Viper cluster of the Max Planck Computing \& Data Facility (MPCDF), each requiring on the order of $10^5$~core hours to finish on AMD EPYC Genoa 9554 CPUs; each retrieval ran for about a week on 1,000 cores.

\begin{table}[t!]
    \centering        
    \begin{threeparttable}
    \caption{Ranked list of likely cloud species, obtained from full retrievals.}
    \label{tab:cloud_species_significance_table}
    \begin{tabular}{l l l}
        \hline\hline
        Species, structure, shape & $\Delta {\rm BIC}$ & ${\rm log}_{10}(B)$\\
        \hline
\ce{SiO}, amorph., sph. & 0.00 & \rch{0} \\
\ce{SiO}, amorph., irr. & 56.55 & \rch{12.28} \\
\ce{MgSiO3}, amorph.$\rm ^{(g)}$, irr. & 148.21 & \rch{32.18} \\
\ce{MgSiO3}, cryst., sph. & 185.43 & \rch{40.27} \\
\ce{MgSiO3}, amorph.$\rm ^{(g)}$, sph. & 197.17 & \rch{42.81} \\
\rch{Mg$_{0.5}$Si$_{0.5}$O$_3$, amorph.$\rm ^{(s)}$, irr.} & 211.87 & \rch{46.01} \\
\ce{MgSiO3}, amorph.$\rm ^{(s)}$, irr. & 355.48 & \rch{77.19} \\
\ce{Mg2SiO4}, amorph.$\rm ^{(s)}$, irr. & 432.38 & \rch{93.89} \\
\ce{MgSiO3}, amorph.$\rm ^{(s)}$, sph. & 533.39 & \rch{115.82} \\
\rch{\ce{Mg2SiO4}, amorph.$\rm ^{(s)}$, sph.} & 746.05 & \rch{162.00} \\
\ce{MgFeSiO4}, amorph.$\rm ^{(g)}$, sph. & 856.21 & \rch{185.92} \\
\hline
    \end{tabular}
    \begin{tablenotes}
        \item Notes: \rch{\texttt{pRT} retrievals were run over the full wavelength range of the considered data (1-18~µm). The table lists $\Delta {\rm BIC}$s and \rch{corresponding Bayes factors} for rejection with respect to the retrieval model with the best fitting cloud species in the first line. (g): glassy. (s): sol-gel.}
    \end{tablenotes}
    \end{threeparttable}
\end{table}

\subsubsection{Retrieval results}

\begin{figure*}
    \centering
    \includegraphics[width=0.93\textwidth]{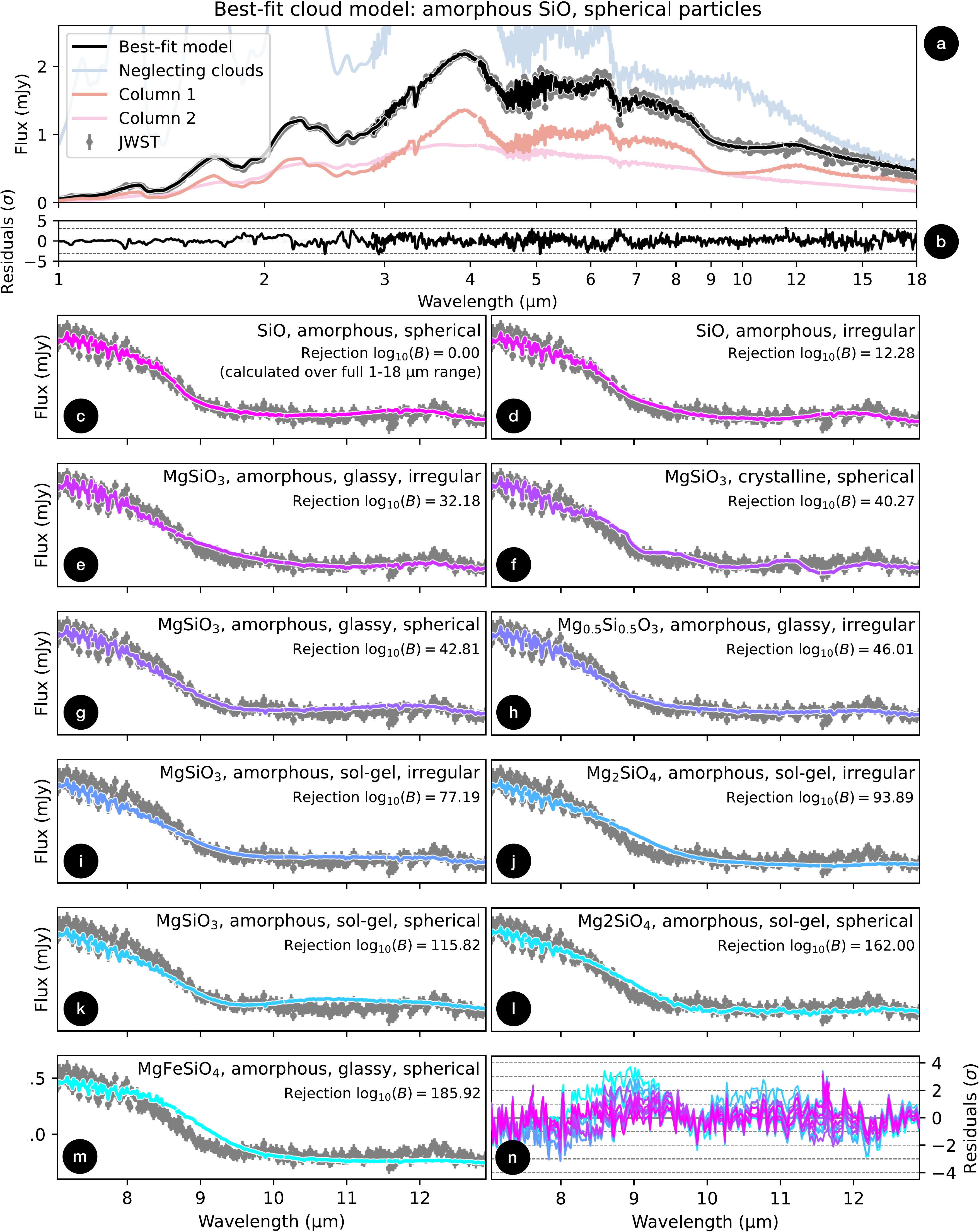}
    \caption{Model fits of PSO~J318 obtained with our fiducial two-column retrieval setup, considering different types of silicate clouds. {\it Panel a} shows the best-fit spectrum of the overall winning model (black solid line) plotted on top of the JWST data (gray circles, with $10^b$ error scaling). The winning model assumes amorphous spherical SiO particles. The flux contribution of the two individual columns is also shown, at their best-fit relative scaling (salmon and rose colored lines, respectively). The light blue line shows the result of recalculating the best-fit model, but turning off the cloud opacities. {\it Panel b} shows the residuals between the best fit model and the data (with $10^b$ error scaling). {\it Panel c} is a version of Panel~a that zooms in on the silicate feature, but with the best-fit model shown in pink instead. {\it Panels d-\rch{m}} show the same zoomed-in view of the silicate feature, but for the other tested silicate feature candidates. {\it Panel \rch{n}} shows the residuals between model and data for the various silicate cloud species, using the same colors as in panels c-\rch{m}.}
    \label{fig:retrieval_summary}
\end{figure*}

Fig.~\ref{fig:retrieval_summary} shows the best-fitting spectra of the two-column retrieval models. Panel~a shows our ``winning model'' \rch{(i.e., the model with the lowest BIC -- see discussion below)}, an atmosphere with a global iron cloud and a patchy SiO cloud with amorphous spherical particles, plotted on top of the JWST observations. Residuals (Panel b) are mostly flat but exhibit systematic behavior at some locations, for example in the water band at 6.6~µm. In Panels c-\rch{m} we show retrievals with all tested cloud candidate species, zoomed in on the silicate feature. \rch{In these zoomed-in views it} can be seen that the predicted line depth appears to be shallower than the data from 7-8~µm for some of the models. In general, the winning model retrieval finds that the flux of \pso \ is significantly reddened by clouds, since neglecting the cloud opacity for the best-fit model leads to a dramatic increase of flux in the near-infrared. What is more, the flux separates into a column component that is very red and blackbody-like, and one that retains molecular features much more strongly, similar to the results presented for 2M1207b in \citet{zhangmolliere2025}.

We cannot trust evidences from nested sampling to select the most likely model from our set of tested models, but we still want to tentatively vet the retrievals with different cloud species for fit quality (assuming that the global log-likelihood maximum, or a maximum of similar quality to the global maximum, was identified). For this we make use of the so-called Bayesian Information Criterion: ${\rm BIC}=N_{\rm param}{\rm ln}(N_\lambda)-2\rm ln(L_{\rm max})$ which allows for model selection. Here, $N_{\rm param}$ is the number of free parameters of the model, $N_{\lambda}$ is the number of wavelength points in the spectrum, and ${\rm ln}L_{\rm max}=-0.5\sum_{i=1}^{N_{\rm param}}\{(f_i-m^{\rm best}_i)^2/(\sigma_i^2+10^b)+{\rm ln}[2\pi(\sigma_i^2+10^b)]\}$ is the best-fit log-likelihood, where $f_i$, $m^{\rm best}_i$ and $\sigma_i$ are the observed flux, best-fit model, and observational uncertainties at $N_\lambda$ wavelengths $\lambda_i$, respectively, and $b$ is the error bar scaling parameter. The BIC makes the underlying assumption that posteriors are multivariate Gaussians. This is not necessarily the case, but tends to be better satisfied in high-$S/N$ regimes, over which the retrieval model can be better linearized over the width of the posteriors. Yet, this is a limitation that needs to be kept in mind, especially if a posterior exhibits parameter correlations over wide value ranges, or is multi-modal. \rch{We note that} from visual inspection, most of our retrieved parameters are tightly constrained, with mono-modal posteriors. If the a-priori probability for a given model is unknown (i.e., all models are considered to be equally likely before being applied to the data), then the probability ratio of two models $\mathcal{M}_1$ and $\mathcal{M}_2$, given the data vector $\mathbf{f}$, $P(\mathcal{M}_1|\mathbf{f})/P(\mathcal{M}_2|\mathbf{f})$, is given by the Bayes factor $B_{12}=Z_1/Z_2$. $Z$ is the evidence here, which is typically returned by nested sampling. For a posterior with a shape described by a multivariate Gaussian one can write $Z={\rm exp}(-{\rm BIC}/2)$ \citep{raftery1995}. Since the models that test different silicate cloud species have same number of free parameters and wavelength points it then holds that
\begin{equation}
    {\rm ln}B_{12}\approx\frac{1}{2}(\chi^2_2-\chi^2_1)+\frac{1}{2}\sum_{i=1}^{N_\lambda}{\rm ln}\frac{\sigma_i^2+10^{b_2}}{\sigma_i^2+10^{b_1}},    
\end{equation}
where $\chi^2=\sum_{i=1}^{N_{\rm param}}(f_i-m^{\rm best}_i)^2/(\sigma_i^2+10^b)$. We note that our error bar scaling via $10^b$ causes a $\chi^2\rightarrow N_\lambda$ convergence in retrievals \citep{lineteske2015}, unless $b$ priors are hit, which we observe for the worst cloud candidates. However, the above expression takes this into account: in a situation where all $\chi$ would indeed be identical, the model with the least upscaled error bars would win. \rch{Given a Bayes factor between a given model and the one with the hightest evidence, it is customary to calculate a rejection significance using the formalism described in \citet{bennekeseager2013}. However, \citet{kippingbenneke2025} recently argued that this is misleading, since the so-derived significance values are upper limits only. We thus refrain from a significance conversion and simply report Bayes factors.}

Panels c-\rch{m} of Fig.~\ref{fig:retrieval_summary} \rch{show} the best-fit models of all tested silicate species on top of the data, zoomed in on the 10~µm region. In agreement with the brightness temperature method, the winning silicate species of the two-column retrievals is amorphous SiO in the form of spherical particles. The panels also list the rejection \rch{Bayes factor} of the respective cloud species, when compared to the winning SiO model. This \rch{Bayes factor} was calculated using the best-fit log-likelihood from the full wavelength range (1-18~µm). After irregularly shaped amorphous SiO, \rch{the next most likely cloud consists of irregularly shaped glassy (thus amorphous) \ce{MgSiO3} particles, followed by crystalline spherical \ce{MgSiO3} particles.} We note that crystalline \ce{MgSiO3} appears unlikely, given the smoothness of the 10~µm feature visible in the data. Indeed, \rch{any} amorphous \rch{material with pyroxene-type} (Mg$_{1-x}$Si$_{x}$O$_3$) \rch{stoichiometry } appears to be fitting the data better across the 10~µm region plotted in Panels c-\rch{m}, but we note again that the rejection \rch{Bayes factors} are calculated over the full wavelength range. The best fit spectra of all tested retrieval models, over the full wavelength range from 1-18~µm, are shown in Fig.~\ref{fig:best_fits_all_retrievals}.

We list the $\Delta \rm BIC$ and corresponding rejection \rch{Bayes factors} of less favored cloud species in Table~\ref{tab:cloud_species_significance_table}. In comparison, the Bayes factors returned by MultiNest did not lead to believable results. For example, the BIC-based values indicate that spherical SiO is \rch{clearly} favored over irregularly shaped SiO (\rch{${\rm log}_{10}(B)=12.28$}). In contrast, the corresponding Bayes factor from MultiNest results in a numerical overflow.

\rch{It is worth noting that the second most likely species identified by the retrievals is glassy amorphous \ce{MgSiO3}}. \ce{MgSiO3} is potentially expected to form in a \pso-like atmosphere, while the presence of SiO is surprising (\citealt{calamarifaherty2024}, but also note \citealt{lunamorley2021,suarezmetchev2023b}). As can be seen in Panels \rch{e} and \rch{g} in Fig.~\ref{fig:retrieval_summary}, glassy \ce{MgSiO3} in spherical or irregular form actually leads to a good fit across the 10~µm range, albeit worse than the still favored SiO. In addition, both glassy \ce{MgSiO3} clouds lead to a too strong flux decrease in the predicted 4~µm flux peak of \pso \ (see Fig.~\ref{fig:best_fits_all_retrievals}), such that in total glassy \ce{MgSiO3} is disfavored with \rch{${\rm log}_{10}(B)=32.18$} and \rch{${\rm log}_{10}(B)=42.81$} respectively, when compared to SiO (see Table~\ref{tab:cloud_species_significance_table}).

\rch{Comparing the ranked lists of likely cloud species in Tables~\ref{table:cloud_fit_full} and \ref{tab:cloud_species_significance_table}} we conclude that the brightness temperature method can be a useful pointer for the most likely silicate absorbers \rch{affecting a spectrum. More specifically, the best and worst species in both lists agree, while there is some reordering present for the species in between (we note that only 11 of the 24 silicate species were tested in full retrievals due to the significant numerical cost). The brightness temperature method therefore must not be trusted blindly, but appears to lead to reasonably accurate predictions for \pso.}

\rch{It is interesting that} both the brightness temperature method and the retrievals come to the conclusion that the winning species is SiO. Making this assertion based on the shape of the amorphous 10~µm feature may still appear to be risky. But we note that the high altitude of the silicate cloud deck and the small particle sizes we retrieve, similar to the results presented in \citet{lunamorley2021}, also point to SiO, which will be discussed in Section \ref{sect:sio_evidence_discussion}. The implication of this likely SiO detection will also be further discussed in Section~\ref{sect:discussion}. 

\section{\pso's bulk and atmospheric properties}
\label{sect:bulk_properties}

In addition to constraining the identity of \pso's silicate cloud species, its spectrum should also allow us to constrain the objects bulk and atmospheric properties. An established way to judge the accuracy of such constraints is to compare results obtained with retrievals to those obtained from interpolating in so-called self-consistent model grids. Such models determine the atmospheric structure by assuming radiative-convective equilibrium, coupled to chemical schemes that determine the atmospheric composition for a given atmospheric temperature structure and elemental abundances \citep[e.g.,][]{marleyrobinson2015,hubeny2017}. For the comparison below, we derived posterior distributions of the self-consistent model parameters with PyMultiNest, using interpolated model grid spectra as the forward model. This analysis was carried out with \verb|species|\footnote{\url{https://species.readthedocs.io}} \citep{stolkerquanz2020}.

\subsection{Radiative-convective equilibrium models}
\label{sect:rce_gridtrievals}

Self-consistent models have far fewer free parameters than retrievals, because they implement atmospheric physics to determine the atmospheric state to a much greater degree. A common free parameter is the composition. It is usually expressed through atmospheric metallicity $\rm [M/H]$ and through C/O. Other free parameters are the atmospheric gravity ${\rm log}_{10}(g)$, the planetary effective temperature $T_{\rm eff}$, and parameters that describe the setup of the atmospheric clouds and departures from chemical equilibrium. In the following we describe the parameter grids we consider for our study. These are identical to the grids recently used in the Early Release Science (ERS) team's paper on VHS~1256b, an object similar to \pso \ \citep{petruswhiteford2024}. A table of the free parameters, the explored parameter range, grid spacing and fit results is given in Table~\ref{tab:self_cst_grids}. \\

\noindent {\bf Exo-REM} \citep{baudinobezard2015,charnaybezard2018,blaincharnay2021} solves for the atmospheric structure in radiative-convective equilibrium and implements \rch{chemical disequilibrium using the quench approximation} \citep{zahnlemarley2014}. Clouds are described following a combination of the time scale approach of \cite{rossow1978} with the \cite{ackermanmarley2001} approach. The atmospheric mixing strength needed for the cloud and chemical composition models is determined from mixing length theory in the convective regions, and from assuming a convective overshoot decay in the radiative regions above. Among other species (\ce{Na2S}, KCl, Fe) Exo-REM considers spherical amorphous \ce{Mg2SiO4} clouds, adopting the optical constants from \cite{jaegerilin2003}. \\

\noindent {\bf ATMO} \citep[][]{tremblinamundsen2015} likewise models the atmosphere in radiative-convective equilibrium, and implements disequilibrium chemistry for the reactions controlling the \ce{CH4}-CO and \ce{NH3}-\ce{N2} conversions. For this the vertical eddy diffusion coefficients is varied as $K_{zz} = 10^{5+2(5-{\rm log}_{10}g)} \ {\rm cm^2 \ s^{-1}}$. ATMO models are cloud-free, such that the reddening of the spectra, often considered to be caused by clouds, is achieved through a decrease of the adiabatic index $\gamma$ between $2 \times 10^{{\rm log}_{10}(g)-5}$ and $500 \times 10^{{\rm log}_{10}(g)-5}$~bar. This leads to an earlier (lower pressure) onset of convection, and convective regions with a shallower dependence of temperature on pressure, since $(d{\rm ln}T/d{\rm ln}P)_{\rm ad}=(\gamma-1)/\gamma$. This reddens the spectra -- which is very similar to the expected effect of clouds; instead of a cloud hiding the hot lower altitudes of the atmosphere, the lower altitudes are simply less hot with ATMO's modified $\gamma$ treatment. The decrease of $\gamma$, when compared to classical dry convection, can occur for multiple reasons: either due to a conversion to a higher mean molecular weight atmosphere at high altitudes (e.g., increase in \ce{CH4} and \ce{NH3}) or due to the release of latent heat if there is a condensible species in the atmosphere (moist convection). More details on the occurrence of diabatic convection in \ce{H2}/He dominated atmospheres near the L-T transition can be found in \cite{tremblinamundsen2015,tremblinamundsen2016,tremblinchabrier2017,tremblinpadioleau2019}, where it is introduced as the process that drives that transition. \\

\begin{table}[t!]
    \centering
    \begin{threeparttable}        
    \caption{Properties of the parameter grids explored for the self-consistent grid retrievals of the \pso \ data.}
    \label{tab:self_cst_grids}
    {\fontsize{9.0}{\baselineskip}\selectfont
    \begin{tabular}{l l l l}
        \hline\hline
        Parameter & Grid extent & Step & 16-84-\%-Range  \\
        \hline
        Exo-REM \\
        $T_{\rm eff}$ (K) & 800-2000 & 50 & ${1249.9991^{+0.0007}_{-0.0014}}^{\rm (a)}$ \\
        ${\rm log}_{10}(g)$ & 3.0-5.0 & 0.5 & ${3.0000004^{+0.0000007}_{-0.0000003}}^{\rm (b)}$ \\
        $\rm [M/H]$ & $-0.5$-1.0 & 0.5 & $0.1724^{+0.0001}_{-0.0001}$ \\
        C/O & 0.1-0.8 & 0.05 & ${0.6^{+0.0000001}_{-0.0000001}}^{\rm (a)}$ \\
        $R$ ($R_{\rm Jup}$) & 0.5-20 & -- &  $1.26^{+0.02}_{-0.02}$ \\ 
        \hline
        ATMO \\
        $T_{\rm eff}$ (K) & 800-2000 & 100 & $1208.93^{+0.05}_{-0.05}$ \\
        ${\rm log}_{10}(g)$ & 2.5-5.5 & 0.5 & ${2.5000006^{+0.0000009}_{-0.0000004}}^{\rm (b)}$  \\
        $\rm [M/H]$ & $-0.6$-0.6 & 0.3 & $0.0316^{+0.0004}_{-0.0004}$ \\
        C/O & [0.3, 0.55, 0.7] & -- & ${0.3000007^{+0.000001}_{-0.0000006}}^{\rm (b)}$ \\
        $\gamma$ & [1.01, 1.03, 1.05] & -- & $1.01145^{+0.00003}_{-0.00003}$ \\
        $R$ ($R_{\rm Jup}$) & 0.5-20 & -- & $1.28^{+0.02}_{-0.02}$ \\
        \hline
        Diamondback \\
        $T_{\rm eff}$ (K) & 900-2000 & 100 & $1109.08^{+0.04}_{-0.04}$ \\
        ${\rm log}_{10}(g)$ & 3.5-5.5 & 0.5 & ${3.5001^{+0.0002}_{-0.0001}}^{\rm (b)}$  \\
        $\rm [M/H]$ & $-0.5$-0.5 & 0.5 & ${0.4999998^{+0.0000002}_{-0.0000002}}^{\rm (b)}$ \\
        $f_{\rm sed}$ & [1, 2, 3, 4, 8] & -- & ${1.0000003^{+0.0000005}_{-0.0000002}}^{\rm (b)}$ \\
        $R$ ($R_{\rm Jup}$) & 0.5-20 & -- & $1.47^{+0.03}_{-0.03}$ \\
        \hline        
    \end{tabular}
    }
    \begin{tablenotes}
        \item Notes: the last column lists the range of the 16-84~percentile regions from the retrieval posterior, as returned by \texttt{species}. (a): median value at grid point coordinates. (b): posterior converged to prior boundary (i.e., grid boundary).
    \end{tablenotes}
    \end{threeparttable}
\end{table}

\noindent {\bf Sonora Diamondback} \citep{morleymukherjee2024} determines the atmospheric structure in radiative-convective and chemical equilibrium. In addition, it implements the cloud modeling approach of \cite{ackermanmarley2001}. The required atmospheric mixing for the cloud model is determined from a mixing length and convective overshoot description, see \citet{morleymukherjee2024} for more details. The following amorphous high-temperature cloud species are considered: \ce{MgSiO3} \citep{dorschnerbegemann1995}, \ce{Mg2SiO4} \citep{jaegerilin2003}, as well as Fe \citep{palik1985,kitzmannheng2018b} and \ce{Al2O3} \citep{koikekaito1995}. In addition to the vertical mixing strength $K_{zz}$, Sonora Diamondback controls the cloud properties through the $f_{\rm sed}$ parameter \citep{ackermanmarley2001}, which is defined as the mass-averaged ratio between the particles' settling and mixing velocities. At a given $K_{zz}$, $f_{\rm sed}$ controls both the mean particle size and vertical extent of the cloud. We note that the \texttt{pRT}'s forward model setup conceptually borrows from this description, defining its cloud mass fraction profile as $\propto P^{f_{\rm sed}}$, while retrieving the cloud particle size freely; this is related, but not identical, to the Sonora Diamondback approach.

\begin{figure*}
    \centering
    \includegraphics[width=0.95\textwidth]{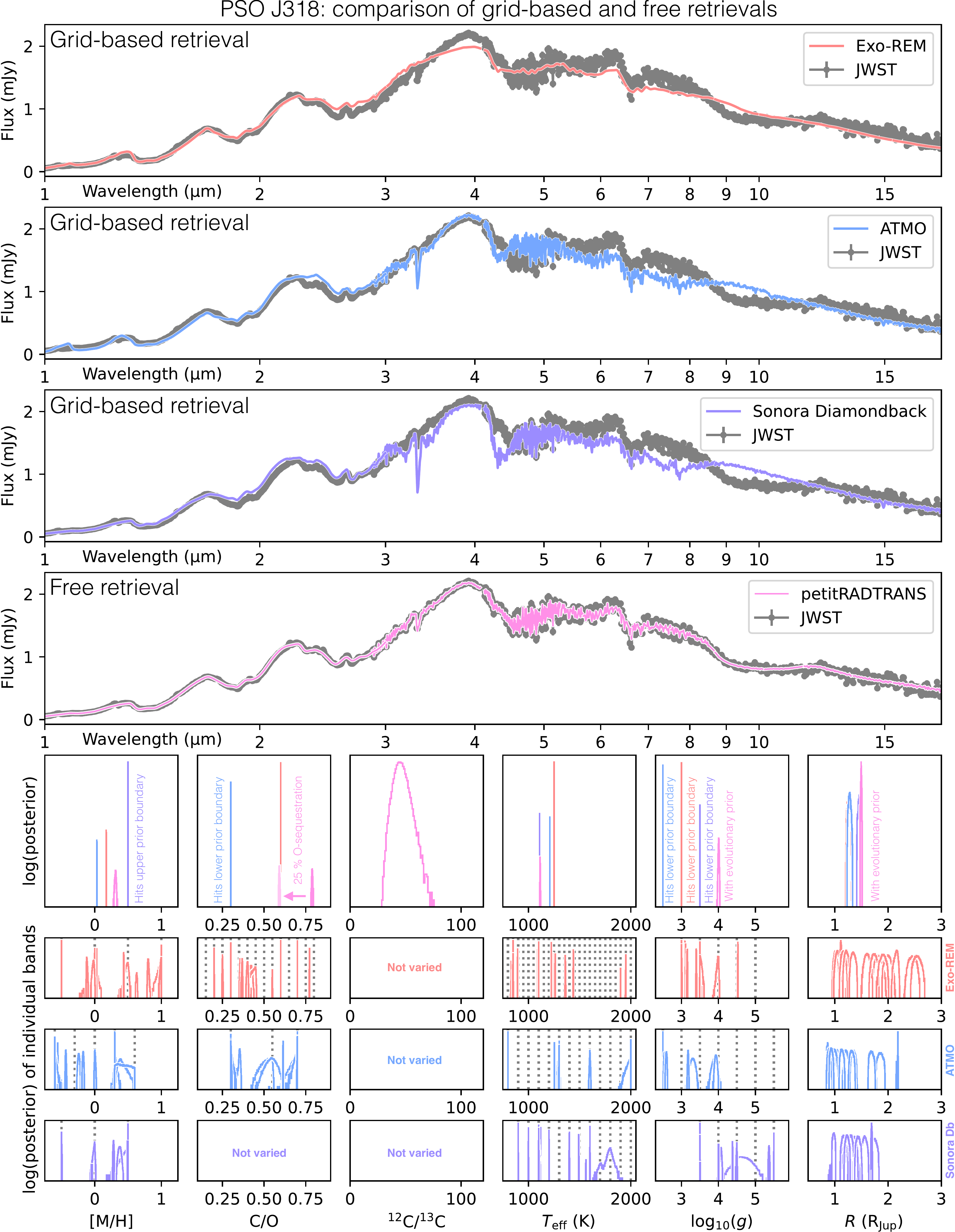}
    \caption{Best-fit spectra and marginalized posterior distributions of the grid interpolation and free retrievals. {\it First row:} Exo-REM fit; {\it second row:} ATMO fit; {\it third row:} Sonora Diamondback fit;  {\it fourth row:} \texttt{petitRADTRANS} fit. The fifth row shows the projected 1-d posteriors for [M/H], C/O, $\rm ^{12}C/^{13}C$, $T_{\rm eff}$, ${\rm log}_{10}(g)$ and the radius $R$ in the first to sixth column, respectively. For C/O, the posterior of the \texttt{petitRADTRANS} retrieval is shown twice, because it was nominally calculated from the gas phase abundances only. For solar C/O, $\sim$25~\% reduction of gas phase oxygen is expected due to condensation \citep{sanchezlopezlandman2022}, which we applied to the second, less opaque C/O posterior shown for \texttt{petitRADTRANS}. The last three rows show the posteriors obtained from considering the data's 14 wavelength bands separately for the self-consistent grids, with the vertical dashed lines indicating the \rch{parameter grid values}.}
    \label{fig:gridtrieval_vs_retrieval}
\end{figure*}

\subsection{Results in comparison to free retrievals}
\label{sect:bulk_free_comparison}

In Fig.~\ref{fig:gridtrieval_vs_retrieval} we show the best-fitting spectra of the various self-consistent grids together with the best-fit model of the winning (amorphous spherical SiO) \texttt{pRT} retrieval. For obtaining the grid fits we followed the approach presented in \citet{petruswhiteford2024}, that is, comparing all the data (\emph{HST}, \emph{SpeX}, \emph{JWST}) to the model grids at once, without accounting for a $10^{b}$ error bar scaling (such a treatment is currently not implemented in \verb|species|). The fit was run using the same data treatment as for the retrieval, such that the grid models were directly binned from their intrinsic wavelength spacings ($\lambda/\Delta\lambda = 3000$ for ATMO and Sonora Diamondback, 500 for Exo-REM) to the spacings of 400 and 1000 used for the \emph{JWST MIRI} and \emph{NIRSpec G395H} data, and were convolved and binned to match the \emph{HST}, \emph{SpeX} and \emph{JWST NIRSpec PRISM} data accordingly. We note that the $\lambda/\Delta\lambda = 500$ spacing of Exo-REM is close to the $\lambda/\Delta\lambda = 400$ spacing used for most of the \emph{JWST} data, such that aliasing effects may be present. Exo-REM's $\lambda/\Delta\lambda$ spacing is also a factor of two lower than our adopted spacing for \emph{NIRSpec G395H}'s \emph{NRS2} detector, such that we switched to $\lambda/\Delta\lambda=400$ for \emph{NIRSpec G395H NRS2}, for Exo-REM only.

Due to the low number of free parameters it is not surprising that the self-consistent models lead to a worse overall fit. While this may appear to put them at a disadvantage at first, herein lies also their strength, when compared to the free retrieval results. These self-consistent models test how well we can reproduce the data, given our state-of-the-art understanding of one-dimensional atmospheric physics and chemistry. Any departure between model and data uncovers processes which require an improved description in the era of JWST, going forward (assuming that biases from the data reduction are less important than model biases). In addition, studying how inferred parameter values agree between the self-consistent grids, and when comparing to the retrievals, allows us to make statements about the robustness of derived atmospheric properties.

We begin by noting that reproducing the shape of the 4~µm flux peak appears to be challenging for the cloudy self-consistent forward models (Exo-REM and Sonora Diamondback). In various retrievals with \texttt{pRT} we observed that the shape of this feature depends on the treatment of the deep Fe cloud: turning off the Fe cloud in the best-fit model changed the peak shape, while turning off the silicate cloud had less of an effect on its shape (while the presence of both clouds results in an flux decrease). The putative effect of the Fe cloud appears to be comparatively well reproducible by the more isothermal temperature profile of the ATMO model. It also appears as if an over-prediction of the \ce{CO2} abundance is responsible for the difference between model and data seen at $\sim 4.3$~µm for all self-consistent grids. Likewise, an overabundance of \ce{CH4}, especially in the Diamondback grid assuming chemical equilibrium, and to a lesser degree in ATMO, causes visible differences at $\sim3.3$~µm, and in the wavelength range from 6 and 8~µm. We note that, conversely, Exo-REM fully misses the 3.3~µm feature of \ce{CH4}. These comparisons highlight the importance of describing the deep atmospheric cloud or temperature gradient well, as well as the necessity of correctly describing the vertical mixing and disequilibrium chemistry in the atmosphere. In addition, none of the self-consistent models reproduces the silicate absorption feature at 10~µm. This is obviously the case for the cloud-free ATMO model, while the cloud parameterizations in Exo-REM and Sonora Diamondback may result in particle sizes too large for producing an appreciable silicate feature. As noted already in \citet{lunamorley2021}, an ad-hoc cloud of small silicate particles high in the atmosphere may be necessary for producing this feature, which is consistent with our retrieval findings. In any case, the absence of a 10~µm absorption feature in the grid models has also been discussed in \citet{petruswhiteford2024}.

The fact that \texttt{pRT}, with ``more knobs to turn'' leads to a better fit does not necessarily mean that its results for the bulk atmospheric parameters are more trustworthy (what if we have too many, or the wrong knobs?). Therefore, in Fig.~\ref{fig:gridtrieval_vs_retrieval} we also compare the posterior distributions of the bulk parameters, namely [M/H], C/O, $T_{\rm eff}$, ${\rm log}_{10}(g)$ and the object's radius $R$, of the grid models and the \texttt{pRT} retrievals. We will focus on \texttt{pRT}'s winning model here (two-column model, spherical cloud particles of amorphous SiO, evolutionary priors). However, given the spread in grid retrieval results, none of findings below significantly change when considering the results of the other retrievals presented in Table~\ref{tab:all_retrieval_posteriors}.

Similar to the findings of \citet{petruswhiteford2024} for VHS~1256b, we find that the posteriors of the grid model fits are mostly mutually inconsistent. The uncertainties are also tiny, since no $10^b$ error scaling can be used. The $b$ factor not only accounts for modeling uncertainties but also underestimated observational error bars. From the retrievals we find that especially \emph{MIRI MRS} uncertainties need to be scaled by about an order of magnitude. Since this is not possible for the grid fits here, this may be reflected in too narrow posteriors. Still, the spread of values across models can be informative about the likely range of values for the bulk parameters of \pso, incorporating model uncertainties. Consequently, we will below also not assess whether the \texttt{pRT} results are ``consistent'' with the grid fits. Consistency is commonly taken to mean that the 1-$\sigma$ credible regions of the posteriors of two different models overlap for a given parameter. Because the error bars of the \emph{JWST} observations are (too?) small, and because none of the models we explore is fully describing the true atmospheric state of \pso, we expect that parameter posteriors are going to be inconsistent. Instead, we test whether the constrained parameter values are broadly compatible, hence we are looking for more qualitative statements, for example: ``{\it most models point to a slight metal enrichment when compared to solar}'', if applicable.

For [M/H] the grid retrieval posteriors span ranges from 0 to 0.5, where the latter value comes from the Sonora Diamondback fit, that hits the upper grid boundary of 0.5. The \texttt{pRT} retrieval value of $0.31_{-0.01}^{+0.01}$ for the amorphous, spherical SiO particle cloud is compatible with this spread of retrieved metallicities, indicating a slightly enriched atmosphere when compared to solar. Only two models incorporate C/O as a free parameter. While the ATMO fit hits the lower grid boundary of 0.3, Exo-REM converges to 0.6 exactly, which is a coordinate point in its parameter grid (this likely means that without a $10^b$ treatment interpolation errors are non-negligible at the current grid spacing). The \texttt{pRT} retrieval constrains C/O to $0.789_{-0.003}^{+0.003}$ based on the gas phase absorber abundances alone. Applying a 25~\% oxygen sequestration due to silicate condensation, a value derived for atmospheres at solar composition \citep{sanchezlopezlandman2022}, leads to $0.592_{-0.002}^{+0.002}$ which is compatible with Exo-REM. Due to ATMO's behavior it is therefore a bit more difficult to judge in which value range \pso's C/O falls, but the retrieved \texttt{pRT} value, which is close to solar \citep[$\rm C/O=0.55$, see][]{asplund2009} is certainly not incompatible with the grid results. What is more, the low C/O value that ATMO finds could be related to the fact that the \ce{CH4} feature is still too strong in its best fitting model, when compared to the data.

The effective temperature posteriors obtained from the grids are 1109 (Sonora Diamondback), 1208 (ATMO), and 1250~K (Exo-REM, a grid point coordinate). \texttt{pRT}'s retrieved value of $1114_{-1}^{+1}$~K is compatible with these findings.

Comparing the atmospheric gravity ${\rm log}_{10}(g)$ between the \texttt{pRT} and the grid retrievals is less straightforward. For \texttt{pRT}, we adopted an evolutionary prior in our standard runs, otherwise the posteriors consistently ran into the ${\rm log}_{10}(g)=3$ lower prior boundary. The grid fits appear to suffer from the same problem, as they all run into their respective lower grid boundaries as well (Exo-REM: ${\rm log}_{10}(g)\rightarrow 3$, ATMO: ${\rm log}_{10}(g)\rightarrow 2.5$, Sonora Diamondback: ${\rm log}_{10}(g)\rightarrow 3.5$). Apparently there is not enough information on the gravity in the data as we consider them to constrain \pso's gravity, except for the fact that it is low. A triangular shape of the H-band, as well as the shape of the K-band have been cited as gravity indicators \citep[e.g.,][]{allersliu2013,faherty2018}, but the K-band's sole use as gravity indicator is not recommended, especially for dusty targets \citep{allersliu2013}. It therefore needs to be investigated whether near-infrared spectroscopy at medium resolution allows to constrain the gravity better, for example by using the potassium doublet at 1.25~µm. This data exists for \pso \ \citep[GNIRS data in][]{liumagnier2013}, but was not included in our study to keep the computational cost manageable. This problem should thus be revisited in future work. Applying the evolutionary prior from the retrievals to the grid models made them again converge to the lower grid boundaries for ${\rm log}(g)_{10}$ for ATMO and Sonora (while causing significantly longer run times, due to the strong prior violation). For Exo-REM the fit was worse than before and converged to ${\rm log}_{10}(g)=3.97$ with the evolutionary prior at ${\rm log}_{10}(g)=4$, but ran into the lower grid boundary for [M/H] ($-0.5$).

The radii constrained from the grids lie between 1.26 and 1.47~$R_{\rm Jup}$. This is roughly compatible with the value of $1.495_{-0.007}^{+0.007}$~$R_{\rm Jup}$ from the \texttt{pRT} retrieval, on which we put an evolutionary prior of $\mathcal{N}(1.36, 0.09)$~$R_{\rm Jup}$, where $\mathcal{N}(\mu,\sigma)$ is the normal distribution with mean value $\mu$ and standard deviation $\sigma$. However, even the radius retrieved in the case without evolutionary prior, $1.54_{-0.01}^{+0.02}$~$R_{\rm Jup}$, is not too far from the largest grid fit result.

Finally, we also investigated \pso's properties with the second approach presented in \cite{petruswhiteford2024}, that is, by splitting the data into 14 individual bands (corresponding to \emph{HST}, \emph{SpeX}, \emph{NIRSpec PRISM}, \emph{NIRSpec G395H NRS1} and \emph{NRS2}, \emph{MRS} channels 1A-3C) and running independent grid fits on them with the three self-consistent models. The resulting posteriors are presented in the three lowest rows of Fig.~\ref{fig:gridtrieval_vs_retrieval}. It can be seen that the posteriors tend to span the full parameter range accessible from the grid, at least when projected in one dimension, and generally lead to bounded constraints, except for in a few cases where a given spectral range is not sensitive to a parameter, even at the small error bars of the \emph{JWST} data. The most robust finding from this analysis is that Exo-REM and ATMO tend to prefer lower ${\rm log}_{10}(g)$ ($\lesssim 4$) solutions (Sonora Diamondback finds gravity solutions across the full grid range). For a number of models and parameters we observe a tendency for the fits to converge to grid coordinate values. This happens for [M/H] for all grids and many bands, for C/O for Exo-REM, for $T_{\rm eff}$ for most models and bands, and for ${\rm log}_{10}(g)$ for many bands in Exo-REM and Sonora Diamondback. Again, this is indicative of non-negligible model interpolation errors given the high data precision and likely too coarse grid spacing. As mentioned above, this might be absorbed by an $10^b$ error scaling, as could likely model biases. In \citet{petruswhiteford2024} individual band posteriors are also combined using the parameter sensitivity of a given band. We refrain from such an analysis here since our band posteriors appear to scatter more than what was reported for VHS~1256b in \citep{petruswhiteford2024}\rch{, with even tighter posteriors. A likely reason for this could be the smaller reported error bars for \pso's \emph{MRS} observations, for example, which result in fluxes that are about an order of magnitude more precise, but which is likely an overestimation.} A combination \rch{as in \citet{petruswhiteford2024}} would result in a multi-modal posterior here, for which a mean value and standard deviation would not add too much additional value when compared to the discussion above. \rch{For these fits in separate bands we note that we could not observe a clear correlation between posterior position and band wavelength.}

In summary, the free retrieval results with \texttt{pRT} are broadly compatible with the grid fits using self-consistent models, especially when the latter are run over the full spectral range. This is reassuring because of the aforementioned challenges encountered by \verb|PyMultiNest| in the era of precise \emph{JWST} data and complex many-parameter models. We note, however, that in the face of small \emph{JWST} uncertainties, better techniques for deriving model uncertainties should be developed: while the posteriors are broadly compatible, they are actually mutually inconsistent. Additionally, methods for deriving observational uncertainties, especially for \emph{MIRI MRS}, should be revisited.

Also the fact that running grid fits on individual bands leads to results spanning almost the full parameter range is noteworthy. Together with the residuals visible in the fits over the full spectral range, this means that self-consistent models may need to improve their description of disequilibrium chemistry (e.g., prescribe the mixing strength as a free parameter) and cloud modeling (this could potentially also require radiatively coupled multi-column treatments, such as presented in \citealt{morleymarley2014,lewapai2020}, as well as incorporating processes that lead to high-altitude clouds\rch{, see \citealt{lunamorley2021}}). If such improvements lead to better fits for \pso \ across the \emph{JWST} wavelength range, the comparison with the free retrievals should be revisited. We note, however, that the use of self-consistent models goes beyond the data-model comparison, and includes studying how the atmosphere reacts qualitatively to changes in parameter values describing a physical process, or changes in the process' description itself. For such purposes self-consistent models stay highly relevant even in their current condition.

\subsection{Free retrievals and the case for SiO nucleation}
\label{sect:bulk_retrievals}

\subsubsection{Pressure-temperature profile and cloud properties}

\begin{figure*}[h!]
    \centering
    \includegraphics[width=0.95\textwidth]{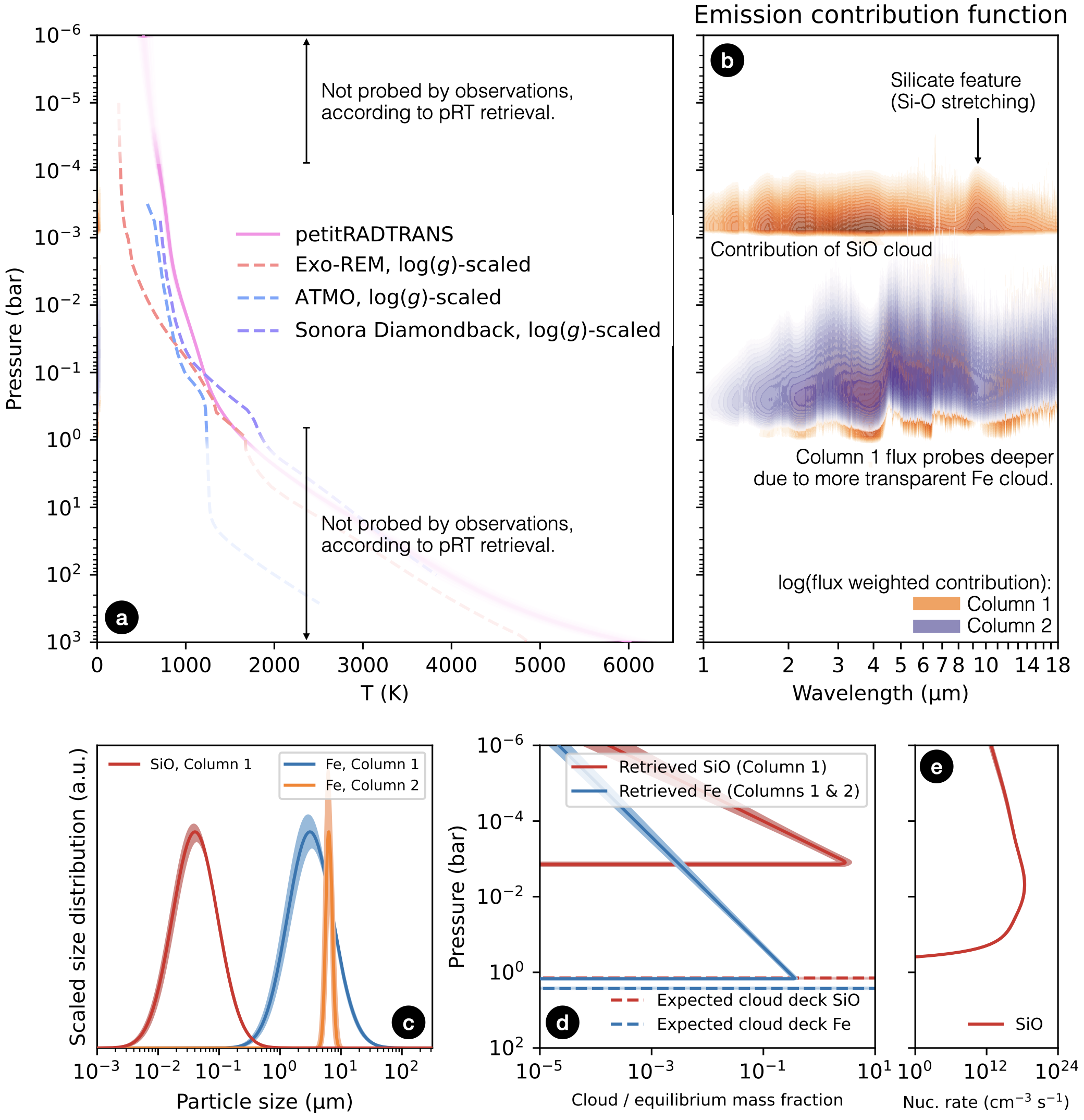}
    \caption{\rch{Overview plot of the temperature, emission  contribution, and cloud properties of the ``winning'' SiO cloud model with amorphous spherical particles, two atmospheric columns, and evolutionary priors.} {\it Panel a:} retrieved pressure-temperature profiles. The \texttt{pRT} retrieval profile is shown in magenta. The 1-3 $\sigma$ posterior regions of the temperature are indicated by progressively lighter magenta tones, but are difficult to distinguish because of the narrow posteriors. We also overplot the ${\rm log}_{10}(g)$-scaled $P$-$T$ profiles of the grid models that are closest to the best-fit values obtained with \texttt{species}. {\it Panel b} shows the log(emission contribution function)s of atmospheric Columns 1 and 2 as orange and purple contours. While nominally computed to sum to unity over all pressures at a given wavelength, the values have been scaled by the relative contribution of the two columns to the total atmospheric solution, and multiplied by the wavelength-dependent flux. {\it Panel c:} size distribution of SiO cloud particles in Column 1 (shown in red), and Fe cloud particles in Column 1 and 2 (shown in blue and orange, respectively). For better comparability the distributions' peak values have been normalized to the same y-axis values. The envelopes around the median distribution would correspond to the 1-$\sigma$ uncertainty if the posterior variations around the median followed a Gauss distribution. {\it Panel d:} altitude-dependent cloud mass fraction of the SiO cloud (present in Column 1, shown in red) and of the Fe cloud (present in both Columns, shown in blue). The colored envelopes again represent the 1-$\sigma$ uncertainty spread of the posterior distribution. The dashed lines correspond to the expected cloud deck position, obtained from intersecting the sampled P-T curves (see Panel a) with the saturation vapor pressure curves of SiO and Fe from \citet{gailwetzel2013,ackermanmarley2001}. {\it Panel e:} SiO nucleation rate computed according to \cite{gailwetzel2013}.}
    \label{fig:winning_model_cloud_characteristics}
\end{figure*}

In Panel a of Fig.~\ref{fig:winning_model_cloud_characteristics} we show the retrieved pressure-temperature ($P$-$T$) distribution, while in Panel~b we show the logarithmic emission contribution functions for both atmospheric columns of the winning retrieval model (i.e., spherical cloud particles of amorphous SiO). \rch{Logarithmic contribution functions are shown instead of linear ones to make contributions other than those of thick atmospheric clouds more visible.} The contribution in each column was calculated as reported in \citet{mollierewardenier2019} \rch{(replacing the Planck function with the full scattering+emission source function)}, and then weighted by the wavelength-dependent flux at the top of the atmosphere, as well as the respective column coverage fractions.
It can be seen that the P-T distribution shows a clear increase of temperature with pressure, such that clouds are necessary to redden the spectrum to the same degree as visible in the data (also see Fig.~\ref{fig:retrieval_summary} which exhibits significant brightening in the blue wavelengths if clouds are neglected in the best-fit model). A high SiO cloud with a cloud base at $\approx 10^{-3}$~bar and $\approx 800$~K in Column~1 leads to reddening of the spectra, but also produces the 10~µm absorption feature that is consistent with the \emph{MIRI MRS} data. The SiO cloud is not fully opaque, however, such that most molecular features form below it, between 0.01 and 1~bar, below which (towards lower altitudes and higher pressure) a deep iron cloud cuts off emission from deeper layers. In Column~2, which only includes a Fe cloud, the atmosphere is probed at similar pressures between 0.01 and 1~bar, but somewhat less deep; here a smaller width of the log-normal particle size distribution $\sigma$ leads to particles which are larger on average, leading to a stronger, approximately gray, absorption.

In Panel (a) we also show the $P$-$T$ curves derived from the self-consistent models, here the models closest to the reported best-fit values from the \texttt{species} analysis. Because the grid fits all hit the lower grid boundary, we scaled their pressures by $10^{4-{\rm log}(g_{\rm grid})}$ where ${\rm log}(g_{\rm grid})$ is the lower grid boundary. This was done to shift the photospheric location to similar pressures \cite[$P_{\rm phot}\propto g$, see, e.g.,][]{mollierevanboekel2015}, leading to better comparability with the \texttt{pRT} results which used the evolutionary prior at ${\rm log}(g)=4$. The grid model and the \texttt{pRT} temperature curves are broadly compatible, but a few noteworthy differences exist: Exo-REM is colder than all other models in the high atmosphere, for (scaled) pressures $<0.1$~bar. Exo-REM and Sonora Diamondback both exhibit greenhouse heating due to the cloud at approximately 1~bar, which is absent in \texttt{pRT}. Lastly, from pressures larger than 0.3~bar, ATMO becomes very isothermal, in order to mimic the cloud spectral reddening in a cloud-free model.

In the lower row of Fig.~\ref{fig:winning_model_cloud_characteristics}, we also show the retrieved particle size distributions (Panel c) and the mass fractions of the SiO and Fe clouds (Panel d). We find that the SiO cloud particle size distribution peaks at $\approx 40$~nm, such that together with the cloud deck at $10^{-3}$~bar the picture of a high-altitude, small-particle SiO haze arises. The particle size distributions of the iron clouds in Columns~1 and 2 peak around 3 and 6~µm, such that this cloud species appears to grow and rain out, with a deep cloud deck at around 1~bar. Comparing the freely retrieved cloud deck locations, that is, the pressure above which the cloud is expected to disperse, to the also plotted expected onset location of condensate stability \citep[taken from][accounting for their erratum]{ackermanmarley2001} reveals that the retrieved iron cloud is indeed consistent with iron condensates that have settled out to the bottom of their stability region.

\subsubsection{Evidence for SiO nucleation}
\label{sect:sio_evidence_discussion}
The location of the SiO cloud is surprising. Plotting the onset location for SiO condensate stability using the saturation vapor pressure from \citet{gailwetzel2013}, we find that the SiO cloud resides at pressures which are 3 dex lower than expected ($10^{-3}$ instead of 1~bar). In our implementation of the \citet{ackermanmarley2001} cloud model, the cloud mass fraction would have decayed by $10^{3f_{\rm sed}}$ from 1 to $10^{-3}$ bar, if the former truly was the cloud base pressure. Instead of assuming that cloud particles are mixed from a deep cloud base to such high altitudes, we suggest that \emph{JWST} probes the high-altitude formation of small \rch{SiO} nuclei \rch{that seed cloud condensation}. Our finding may therefore be consistent with the so-called ``top-down approach'' of cloud formation, beginning with nucleation in the upper reaches of the atmosphere \citep{hellingwoitke2008}. In general, cloud \rch{seeding} nuclei are always necessary for cloud formation to ensue. This can be because the species that are expected to dominate spectra have a too high surface tension. An example is iron, which cannot form directly from the gas phase, but needs a ``starting surface'' \citep{gaothorngren2020}. Another reason is that some species must form via grain-gas reactions, as in the case of the usual silicates (e.g., \ce{MgSiO3} and \ce{Mg2SiO4}). Indeed, homogeneous SiO nucleation from gas-phase SiO is discussed as a likely process for triggering condensate formation, since they may then provide a surface for the other condensates to grow on \citep{gailwetzel2013,leeblecic2018}. SiO is the most abundant Si-bearing gas phase species, and efficiently nucleates based on predictions from thermodynamic data at the temperatures expected across the L-T transition. However, usually \ce{TiO2} is treated as the more likely nucleation species. This is because 
\ce{TiO2} is more refractory\footnote{That is, \ce{TiO2} condensates are stable at higher temperatures than SiO.} and thus nucleates before SiO in gas upwelling from the deep atmosphere. In such a situation Si would condense into silicates on the \ce{TiO2} nuclei before being able to form SiO nuclei.

However, as discussed in \citet{leeblecic2018}, more refractory cloud nucleation species may be absent from the upper reaches of an atmosphere if the condensates in which they are incorporated rain out to low altitudes (so-called cold trapping). A likely species in the case of \ce{TiO2} would be the highly refractory calcium titanates (e.g., \ce{CaTiO3}, \ce{Ca3Ti2O7}, \ce{Ca4Ti3O10}, see \citealt{wakefordvisscher2017}). \citet{leeblecic2018} also argue that efficiently nucleating species may form a high-altitude haze (small particle) layer. Interestingly, when plotting the nucleation rate derived from classical nucleation theory in \citet{gailwetzel2013} (see lower Panel e in Fig.~\ref{fig:winning_model_cloud_characteristics}), we find that the peak nucleation rate is reached at $10^{-2}$~bar, so close to the retrieved SiO cloud base of $10^{-3}$~bar. For the high temperatures ($T\approx800>600$~K) and pressures ($10^{-3}$~bar) identified for the location of the SiO cloud here, a more complete nucleation theory finds that SiO nucleation is even more efficient than predicted by classical nucleation theory \citep{bromleygomez2016}.

In addition, nucleation is expected to form small particles which stay aloft more easily, while larger particles would settle out of the high atmosphere more quickly \citep[see][for a timescale estimate]{lunamorley2021}.  Our inferred mean particle size of 40~nm is indeed small, but it is larger than the typical condensation nuclei made from $({\rm SiO})_N$ clusters, which are expected to be even smaller ($\sim$nm sizes). We note, however, that constraining sizes for small particles from opacities becomes increasingly difficult, since the opacity becomes independent of particle size $r$ in the Rayleigh limit \citep[$\lambda\ll 2\pi r$, see, e.g.,][]{bohrenhuffman1983}. So it could be that our retrievals struggle to constrain the actual particle size, even though we do not find an upper limit (which could be related to our choice of how to parameterize the particle size distribution). In this case the remaining structure from 9.5-10~µm mentioned in Section~\ref{sect:brightness_temperature_method} could be caused by the actual ro-vibrational transitions of the smallest $({\rm SiO})_N$ clusters. Alternatively, the particles may already have grown to sizes of a few tens of nm when we observe them.

\rch{Lastly, we note that we do not see evidence for absorption of SiO in the gas phase \citep[gaseous SiO is the most important Si-bearing species prior to condensation of Si into silicates, see][]{visscherlodders2010}. In our retrievals it is only included as a trace species, with its abundance determined from chemical equilibrium. We do not see residuals around 8~µm, where the band head of SiO absorption lies, nor does turning the SiO gas opacity off in our best-fit model affect the fit quality noticeably.}

\subsubsection{Stability of atmospheric properties across retrievals}

In total we have run \rch{15} production retrievals for this study with a total computational cost of $\approx 10^6$ CPU core hours (many more retrievals were run during the model development process). Their derived atmospheric bulk properties are given in Table~\ref{tab:all_retrieval_posteriors}. In addition to testing \rch{11} different silicate species in our fiducial two-column model, we also tested departures from this model. Since spherical amorphous SiO particles were identified as most likely, this species was adopted for these additional retrievals. Specifically, we tested (i) turning off the evolutionary prior on the radius and ${\rm log}_{10}(g)$, (ii) making the second column completely cloud-free, (iii) keeping the SiO and Fe cloud parameters constant across both columns but varying the temperature profile in Column 2 with the temperature excursion method, (iv) using a single-column model with an Fe and SiO cloud. 

Here we discuss the stability of parameter posteriors across the different retrievals and, by extension, the likely robustness of the derived values. Such an analysis is particularly important because of the aforementioned issues with MultiNest in the present high-data-precision high-model-complexity situation. Again, robustness here is not defined as a significant overlap of retrieval posteriors for a given parameter; this is never the case, and would require that all models are sufficiently close to the optimal model or have sufficiently similar model properties. Analogous to our definition of compatibility in Section~\ref{sect:bulk_free_comparison}, it means that checking whether assertions such as ``{\it the metallicity is slightly super-solar, around a value of X}'' appear to be justified.

Before looking at the bulk parameters one-by-one, we note that the fiducial two-column model with a global iron and a patchy SiO cloud leads to the best fits consistently among all tested models (the best-fit spectra of all retrieval models are shown in Fig.~\ref{fig:best_fits_all_retrievals}). In particular, the single-column model, the two P-T model, and the clear-column cloudy-column model are not favored, and are rejected by $>21.8 \ \sigma$ (the rejection values are also given in Fig.~\ref{fig:best_fits_all_retrievals}).  Also the retrieval without the evolutionary prior is not favored from a BIC-based Bayes factor comparison. This is worrying as it is identical to the nominal two-column model otherwise, but more general, and should lead to a fit as least as good as the nominal model. We therefore see the aforementioned effect of MultiNest reaching its limits with the present data in classical retrieval analyses, and again caution that all results presented here should be interpreted with this finding in mind. We note that Multinest may fare better with more live points, but this is computationally unfeasible.

\noindent {\bf Effective temperature} $\bf T_{\rm eff}$: the winning model (spherical particles, amorphous SiO) constrains \pso's effective temperature to $1114_{-1}^{+1}$~K, while the range of values seen across all retrievals ranges from \rch{1073} to 1146~K. Given the breadth of assumptions tested in the retrievals, we may consider this parameter well constrained, with a range from \rch{1073}-1146~K.

\noindent {\bf Radius} $\bf R$: the winning model constrains the radius to ${1.495}_{-0.007}^{+0.007} \ R_{\rm Jup}$, with the retrievals ranging from \rch{1.454} to 1.516~$R_{\rm Jup}$ in the cases with the evolutionary prior, $\mathcal{N}(1.36 \ R_{\rm Jup},0.09 \ R_{\rm Jup})$. This is not too far from the result obtained when turning off the evolutionary prior, which is ${1.54}_{-0.01}^{+0.02} \ R_{\rm Jup}$. This parameter may therefore be considered robust, yielding a radius value of around 1.5~$R_{\rm Jup}$.

\begin{table*} 
\centering
\begin{threeparttable}

\caption{\pso \ retrieval results.}

\label{tab:all_retrieval_posteriors}

\centering
{\fontsize{8.6}{\baselineskip}\selectfont
\begin{tabular}{llccccccc}
\hline \hline
Silicate species & $T_{\rm eff}$ (K)$^{\rm (a)}$ & $R$ ($\rm R_{\rm Jup}$)$^{\rm (b)}$ & ${\rm{log}_{10}}(L/{\rm L_\odot})$$^{\rm (c)}$ & ${\rm log}_{10}(g)$ & $B^{\rm (d)}$ & $\rm [M/H]_{gas}^{\rm (e)}$ & $({\rm C/O})_{\rm gas}^{\rm (f)}$ & $\rm ^{12}C/^{13}C$ \\ \hline 
\multicolumn{9}{l}{{\it Two-column models with a global Fe cloud and a patchy silicate cloud, with evolutionary prior on ${\rm log}(g)$ and $R$} (41 free parameters):} \\ \hline
SiO, amorph., sph. & ${1114}_{-1}^{+1}$ & ${1.495}_{-0.007}^{+0.007}$ & ${-4.489}_{-0.005}^{+0.003}$ & ${4.009}_{-0.007}^{+0.007}$ & $0.695\pm 0.002$ & ${0.31}_{-0.01}^{+0.01}$ & ${0.789}_{-0.003}^{+0.003}$ & ${45}_{-4}^{+5}$  \\
SiO, amorph., irr. & ${1117}_{-2}^{+2}$ & ${1.49}_{-0.01}^{+0.01}$ & ${-4.486}_{-0.003}^{+0.003}$ & ${3.996}_{-0.009}^{+0.008}$ & $0.661\pm 0.003$ & ${0.43}_{-0.01}^{+0.01}$ & ${0.804}_{-0.004}^{+0.003}$ & ${36}_{-5}^{+5}$  \\
\rch{\ce{MgSiO3}, amorph.$^{\rm (g)}$, irr.} & ${1146}_{-1}^{+1}$ & ${1.485}_{-0.007}^{+0.010}$ & ${-4.441}_{-0.005}^{+0.007}$ & ${3.957}_{-0.006}^{+0.006}$ & $0.979 \pm 0.002$ & ${0.53}_{-0.01}^{+0.01}$ & ${0.820}_{-0.003}^{+0.003}$ & ${38}_{-4}^{+5}$ \\
\ce{MgSiO3}, cryst., sph. & ${1133}_{-1}^{+1}$ & ${1.513}_{-0.008}^{+0.008}$ & ${-4.444}_{-0.005}^{+0.005}$ & ${3.964}_{-0.008}^{+0.008}$ & $0.666\pm 0.004$ & ${0.39}_{-0.01}^{+0.01}$ & ${0.802}_{-0.002}^{+0.003}$ & ${65}_{-7}^{+7}$ \\
\rch{\ce{MgSiO3}, amorph.$^{\rm (g)}$, sph.} & ${1136}_{-1}^{+1}$ & ${1.466}_{-0.006}^{+0.006}$ & ${-4.468}_{-0.003}^{+0.004}$ & ${3.938}_{-0.009}^{+0.010}$ & $0.986\pm 0.003$ & ${0.60}_{-0.01}^{+0.01}$ & ${0.837}_{-0.003}^{+0.002}$ & ${57}_{-9}^{+9}$ \\
\rch{Mg$_{0.5}$Si$_{0.5}$O$_3$, amorph.$^{\rm (g)}$, irr.} & ${1132}_{-1}^{+1}$ & ${1.454}_{-0.005}^{+0.005}$ & ${-4.480}_{-0.004}^{+0.003}$ & ${3.927}_{-0.008}^{+0.009}$ & $0.966 \pm 0.003$ & ${0.49}_{-0.01}^{+0.01}$ & ${0.818}_{-0.003}^{+0.003}$ & ${35}_{-5}^{+4}$ \\
\ce{MgSiO3}, amorph.$^{\rm (s)}$, irr. & ${1102}_{-2}^{+1}$ & ${1.472}_{-0.004}^{+0.008}$ & ${-4.516}_{-0.003}^{+0.004}$ & ${3.949}_{-0.008}^{+0.008}$ & $0.965\pm 0.002$ & ${0.48}_{-0.01}^{+0.01}$ & ${0.809}_{-0.003}^{+0.003}$ & ${17}_{-2}^{+1}$ \\
\ce{Mg2SiO4}, amorph.$^{\rm (s)}$, irr. & ${1090}_{-2}^{+2}$ & ${1.479}_{-0.007}^{+0.006}$ & ${-4.532}_{-0.003}^{+0.004}$ & ${4.009}_{-0.007}^{+0.007}$ & $0.821\pm 0.003$ & ${0.44}_{-0.01}^{+0.01}$ & ${0.812}_{-0.002}^{+0.004}$ & ${21}_{-2}^{+3}$ \\
\ce{MgSiO3}, amorph.$^{\rm (s)}$, sph. & ${1098}_{-2}^{+2}$ & ${1.496}_{-0.005}^{+0.005}$ & ${-4.508}_{-0.003}^{+0.003}$ & ${3.968}_{-0.008}^{+0.007}$ & $0.963\pm 0.006$ & ${0.55}_{-0.01}^{+0.02}$ & ${0.834}_{-0.003}^{+0.003}$ & ${25}_{-3}^{+4}$ \\
\rch{\ce{Mg2SiO4}, amorph.$^{\rm (s)}$, sph.} & ${1073}_{-1}^{+1}$ & ${1.574}_{-0.003}^{+0.003}$ & ${-4.506}_{-0.002}^{+0.003}$ & ${4.002}_{-0.005}^{+0.005}$ & $0.567\pm0.003$ & ${0.24}_{-0.01}^{+0.01}$ & ${0.773}_{-0.005}^{+0.005}$ & ${18}_{-3}^{+3}$ \\
\ce{MgFeSiO4}, amorph.$^{\rm (g)}$, sph. & ${1102}_{-2}^{+2}$ & ${1.516}_{-0.006}^{+0.005}$ & ${-4.492}_{-0.003}^{+0.003}$ & ${3.95}_{-0.01}^{+0.01}$ & $0.844\pm 0.004$ & ${0.23}_{-0.01}^{+0.01}$ & ${0.793}_{-0.003}^{+0.003}$ & ${81}_{-12}^{+14}$ \\
\hline
\multicolumn{9}{l}{{\it Two-column models with a global Fe cloud and a patchy silicate cloud, without evolutionary prior} (41 free parameters):} \\ \hline
SiO, amorph., sph. & ${1106}_{-1}^{+1}$ & ${1.54}_{-0.01}^{+0.02}$ & ${-4.469}_{-0.007}^{+0.009}$ & $\rightarrow 3^{\rm {(i)}}$ & $0.66\pm 0.08$ & ${0.12}_{-0.01}^{+0.01}$ & ${0.872}_{-0.004}^{+0.003}$ & $> 50$ ($3\sigma$) \\ \hline
\multicolumn{9}{l}{{\it Two-column models where one column is clear and the other has Fe and silicate clouds, with evolutionary prior} (40 free parameters):} \\ \hline
SiO, amorph., sph. & ${1126.7}_{-0.7}^{+0.8}$ & ${1.489}_{-0.009}^{+0.009}$ & ${-4.468}_{-0.004}^{+0.004}$ & ${3.98}_{-0.01}^{+0.01}$ & $0.960\pm 0.006$ & ${0.373}_{-0.010}^{+0.009}$ & ${0.828}_{-0.003}^{+0.003}$ & ${140}_{-30}^{+130}$ \\
\hline
\multicolumn{9}{l}{{\it Two-column models with different pressure-temperature structures, and global Fe and silicate clouds, with evolutionary prior} (44 free parameters):} \\ \hline
SiO, amorph., sph. & ${1130}_{-1}^{+1}$ & ${1.464}_{-0.007}^{+0.006}$ & ${-4.478}_{-0.004}^{+0.004}$ & ${3.93}_{-0.01}^{+0.01}$ & unc.$\rm ^{(h)}$ & ${0.53}_{-0.01}^{+0.01}$ & ${0.823}_{-0.003}^{+0.003}$ & ${43}_{-4}^{+5}$ \\ \hline
\multicolumn{9}{l}{{\it Single-column models with global Fe and silicate clouds, with evolutionary prior on ${\rm log}(g)$ and $R$} (37 free parameters):} \\ \hline
SiO, amorph., sph. & ${1127}_{-2}^{+2}$ & ${1.456}_{-0.009}^{+0.006}$ & ${-4.487}_{-0.005}^{+0.004}$ & ${3.94}_{-0.01}^{+0.01}$ & -- & ${0.52}_{-0.01}^{+0.02}$ & ${0.823}_{-0.003}^{+0.003}$ & ${44}_{-5}^{+6}$ \\ \hline
\end{tabular}
}
\begin{tablenotes}
    \item Notes (a): defined as $\sigma T_{\rm eff}^4=\int [B F_{1}(\nu)+ (1-B) F_{2}(\nu)]d\nu$, where $\sigma$ is the Stefan-Boltzmann constant, $B$ is the coverage of atmospheric state 1, and $F_1$ and $F_2$ are the fluxes of atmospheric states 1 and 2 at the top of the atmosphere, respectively. (b): defined as $ R^2=\int [BR_1^2(\nu)F_{1}(\nu)+(1-B)R_2^2(\nu)F_{2}(\nu)]d\nu / (\sigma T_{\rm eff}^4)$, where $R_1(\nu)$ and $R_2(\nu)$ are the frequency-dependent photospheric radii of \pso\ for both atmospheric states, respectively, defined at a vertical optical depth of $\tau=2/3$. (c): defined as $L=4\pi \int [BR_1^2(\nu)F_{1}(\nu)+(1-B)R_2^2(\nu)F_{2}(\nu)]d\nu$. (d): simple average of the coverage of Column 1. Calculat\rch{ed} by taking the mean value of the coverages obtained from the \emph{HST}, \emph{SpEX} and \emph{JWST} datasets. (e): calculated by summing all metal-element-over-hydrogen ratios obtained from the gas phase absorbers in \pso\ and dividing this value by the corresponding solar value, taking ${\rm log}{10}$ of this value, \rch{and then the median across the atmosphere}. We note that elements that occur in the sun but are not constrained in \pso\ have been neglected. (f): obtained from the median of all vertical C/O values in the atmosphere, neglecting the O sequestered into silicates. (g): glassy material. (h): unconstrained, that is, posteriors on $B$ for the three observatories / epochs spanned ranges from 0 to 1, with uncertainties from 0.1-0.2. \rch{(i): converging towards a value of ${\rm log}(g)=3$, which is the lower prior boundary.} (s): sol-gel material.
\end{tablenotes}
\end{threeparttable}
\end{table*}

\noindent {\bf Luminosity} $\bf L$: the winning model constrains the bolometric luminosity to $\rm{log}(L/{\rm L_\odot})={-4.489}_{-0.005}^{+0.003}  $, with the retrievals ranging from $-4.532$ to $-4.441$~${\rm L_\odot}$, which makes for a robust determination of this parameter. This value range is somewhat lower than, but consistent with, the $\rm{log}(L/{\rm L_\odot})={-4.420}\pm0.060$ quoted in \citet{zhangliu2020}, which was derived based on near-infrared photometry.

\noindent {\bf Gravity} $\bf {\rm log}_{10}(g)$: the winning model constrains the atmospheric gravity to ${4.009}_{-0.007}^{+0.007}$, with most retrievals falling close to that range, which is determined by the adopted prior, $\mathcal{N}(4,0.04)$. If this prior is turned off and a uniform prior is adopted instead, the retrieval runs into the lower boundary of the prior (${\rm log}_{10}(g)=3$), similar to the grid fits with the self-consistent models. We therefore conclude that this parameter is not robust for the data considered here, and that medium- to high-resolution observations in the near-infrared may lead to a better constraint, as discussed in Section~\ref{sect:rce_gridtrievals}. Alternatively, running the \emph{JWST} retrievals at full, or at least higher, resolution may be beneficial; the \emph{JWST MIRI} studies of WISE~J1828 \citep{barradomolliere2023}, WISE~J0855 \citep{kuehnlepatapis2025}, and WISE~J0458 \citep{matthewsmolliere2025} all ran retrievals at $\lambda/\Delta\lambda=1000$ for \emph{MRS} and resulted in bounded constraints.  We again note that turning off the evolutionary prior led to a significantly worse fit. This demonstrates how the struggling nested sampling benefits from the use of informative priors, especially on ${\rm log}_{10}(g)$.

\noindent {\bf Column coverage} $\bf B$: while the winning model shows clear evidence for a two-column solution being favored ($B = 0.695 \pm 0.002$, where the $B$s of \emph{HST}, \emph{SpeX} and \emph{JWST} were averaged), some other models, for example the glassy \ce{MgSiO3} cases, do not require a 1+1D structure to result in a decent fit. We note, however, that these models still lead to a worse fit, and are disfavored \rch{with ${\rm log}_{10}(B)=32.18$}, using the BIC-to-evidence conversion method. We note that the glassy \ce{MgSiO3} cases in particular lead to residuals in the 4~µm emission peak, see Fig.~\ref{fig:best_fits_all_retrievals}.

\noindent {\bf Metallicity [M/H]}: the winning model constrains the atmospheric metallicity to be slightly super-solar, at ${0.31}_{-0.01}^{+0.01}$. The values observed for the various retrievals range from 0.12 to 0.6, such that the claim of a slightly super-solar metallicity appears to be justified, but the exact degree of enrichment is less straightforward to constrain.

\noindent {\bf Carbon-to-oxygen number ratio C/O}: the retrieved C/O values are largely consistent across the retrievals. The winning model constrains C/O to ${0.789}_{-0.003}^{+0.003}$, and the highest value seen across all runs goes up to 0.872. Applying the aforementioned 25~\% oxygen correction to account for its sequestration into silicates leads to a range from \rch{0.580} to 0.654, so slightly super-solar when compared to \citet{asplund2009}, who find ${\rm C/O}=0.55$ for the sun.

\begin{figure*}[t!]
    \centering
    \includegraphics[width=0.95\textwidth]{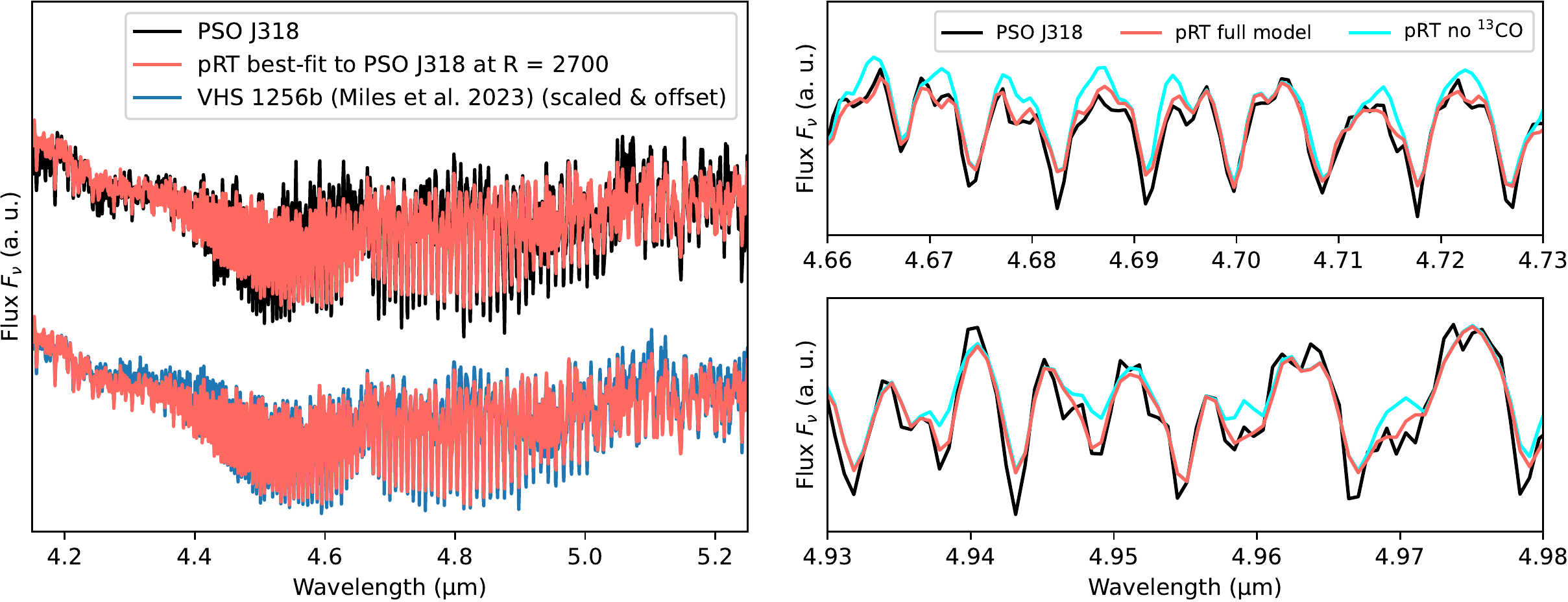}
    \caption{\rch{Evidence for $^{13}$CO absorption in \pso's atmosphere.} {\it Left panel:} the best-fit spectrum of the winning \texttt{pRT} retrieval model (amorphous spherical SiO particles) post-processed to high spectral resolution ($R=3200$, red solid line) and superimposed on the \emph{NIRSpec G395H NRS2} data of \pso \ at full spectral resolution (black solid line). For comparison, \texttt{pRT}'s best-fit model for \pso \ is also superimposed on a scaled version of VHS~1256b's \emph{G395H} spectrum from \citet{milesbiller2023} (blue solid line). {\it Right panel:} zoomed in version of the left panel, showing the data for \pso \ as well as the \texttt{pRT} best-fit model, where in one case (cyan solid line) the $^{13}$CO opacity has been turned off in post-processing.}
    \label{fig:13COinPSO}
\end{figure*}

\noindent {\bf Carbon isotope ratio $\bf ^{12}C/^{13}C$}: in the retrievals the abundances of $^{13}$CO and $^{12}$CO are treated independently, which allows to test for the presence of $^{13}$CO, and to measure its abundance in \pso. This is motivated by \citet{gandhideregt2023}, who identified $^{12}$CO, $^{13}$CO, C$^{17}$O and C$^{18}$O in the atmosphere of VHS~1256b, which is a similar object. $^{13}$CO is likely the easiest secondary isotopologue to detect in atmospheres of substellar objects hot enough to exhibit appreciable CO absorption \ \citep{mollieresnellen2019}. However, isotopologues are not the main focus of this study, and \emph{NIRSpec G395H NRS2} has been binned to $\lambda/\Delta\lambda=1000$ in our work (while its intrinsic resolution is $R\approx 3200 $ from 4-5~µm). This means that the minute changes of the individual lines of rarer isotopologues are harder to pick up. Therefore we did not search for the less abundant C$^{17}$O and C$^{18}$O isotopologues. The winning \texttt{pRT} model detects $^{13}$CO, and constrains $\rm {^{12}C/^{13}C}\approx {^{12}CO}/{^{13}CO}=45_{-4}^{+5}$. Across the retrievals this value varies quite substantially (from 17 to 81 for the retrievals with bounded constraints). These value are generally lower than the local value of 70 in the interstellar medium \citep{milramsavage2005}, which is surprising for an isolated object that may represent the low-mass end of star formation. To understand these constraints better we show a zoomed in version of \pso's spectrum in the 4-5~µm region, at the full resolution of \emph{G395H NRS2}, in Fig.~\ref{fig:13COinPSO}. We also show the best-fit model of \texttt{pRT}, post-processed to high spectral resolution $\lambda/\Delta\lambda=250,000$, convolved to $R=3200$, binned to the \emph{G395H NRS2} wavelength solution and then superimposed on the \pso \ data. The fit is decent, but residuals between the data and model persist. In particular, the \pso \ data appears to be noisier and still affected by systematic effects when compared to the corresponding spectrum of VHS~1256b \citep{milesbiller2023}, which we also show in the same panel. Even though the retrieval was run on the \pso \ and not the VHS~1256b data, the latter are better fit by the retrieval model. In the right panels we show a further zoomed-in view of the \pso \ data, together with the best-fit \texttt{pRT} model with and without considering the $^{13}$CO opacity. We conclude that while $^{13}$CO lines are clearly visible in the spectrum of \pso, noise and remaining systematic effects in the data might currently hinder an accurate determination of $\rm {^{12}C/^{13}C}$ (we note again that we use the standard reduction for \emph{G395H} from the MAST archive for \pso).

\section{Summary and discussion}
\label{sect:discussion}
In this work we analyzed the panchromatic emission spectrum of the young low-mass brown dwarf \pso, a ``rogue planet''. We considered data from 1-18~µm, taken with \emph{JWST NIRSpec} and \emph{MIRI} observations, together with archival \emph{HST} and \emph{IRTF SpeX} spectra. We summarize our findings below:

\begin{itemize}
    \item The emission of \pso \ is extremely red, and exhibits a broad and deep absorption feature at 10~µm, which we attribute to absorption by silicate clouds.
    \item We developed a technique to efficiently identify likely cloud species, based on the wavelength-dependent brightness temperature of \pso. This method suggests that the absorption is caused by spherical amorphous SiO grains. This prediction agrees with the subsequently run retrievals.
    \item Starting from the list of silicate species identified with the brightness temperature method we ran full retrievals on the spectra with \texttt{petitRADTRANS}. We find  satisfactory fits to the data across the full spectral range. These retrievals also identify SiO as the most likely species, \rch{strongly} favoring it over other candidates \rch{when using BIC-derived Bayes factors}.
    \item We hypothesize that our observations probe the homogeneous nucleation of SiO \rch{cloud seeds} from the gas phase, which is consistent with the high altitude retrieved for the cloud base ($\approx 1$~mbar), and the small particle sizes $\lesssim 0.1$~µm. These particles could then provide the surfaces required for cloud formation at lower altitudes.
    \item Our retrievals indicate that \pso \ may be best described by atmospheres more complex than what can be described by a 1D column. Our favored models describe the atmosphere in a 1+1D column approach. These two columns have a global Fe cloud, but only one column hosts a silicate cloud. While this need for a more than 1D solution is consistent with the reported variability of \pso, it does not perfectly reproduce its observed variability behavior \citep{billervos2018}. The silicate cloud coverage of the winning SiO model is 0.695 when averaged over the epochs of the three observatories (\emph{HST}, \emph{SpeX}, \emph{JWST}). The coverage of the three instruments, when compared to this mean, varies by $\sim\pm4$~\%. Taking this as a typical value of the expected coverage variability we can roughly reproduce the $\sim$~similar relative variability amplitudes observed between \emph{HST} and \emph{SpeX} wavelengths. However, while they are in phase for our models, they are phase shifted by~200$^\circ$ in \citet{billervos2018}. What is more, the variability reported in \citet{billervos2018} is gray across the \emph{HST} wavelengths, which is not what we find. This could indicate that more than two columns are necessary to explain \pso's variability and that more quantities than the silicate cloud cover could vary, consistent with the qualitative interpretation of \emph{JWST} variability data in \citet{billervos2024,mccarthyvos2025,chenbiller2025}.
    \item Our retrievals also indicate that solutions that vary the temperature structure across two columns, but assume identical cloud properties across the two columns, are not favored when compared to our fiducial two column model. So a temperature-profile induced reddening of the spectra as suggested by \citet{tremblinamundsen2015,tremblinamundsen2016,tremblinchabrier2017,tremblinpadioleau2019} is not supported by the results obtained with our current retrieval setup.
    \item We also run grid retrievals using various atmospheric models in radiative-convective equilibrium (Exo-REM, ATMO, Sonora Diamondback). The retrieved bulk properties between these self-consistent models and the \texttt{petitRADTRANS} retrievals are mutually compatible, while the small posterior uncertainties make them mutually inconsistent. Compatibility here means that we find that \pso \ is slightly enriched in metals when compared to solar across the various approaches, with potentially a solar to slightly super-solar C/O. The atmospheric gravity cannot be constrained from the data: all self-consistent model grids hit the lower grid boundary, and \texttt{petitRADTRANS} only achieves a bound value on ${\rm log}_{10}(g)$ if an evolutionary prior is used. This could potentially be alleviated by adding (existing) near-infrared observations at higher spectral resolution to the analysis, or by considering the \emph{JWST} data at full (or higher) resolution. The latter were binned down in the present analysis in order to make running the retrievals feasible (they still took $\sim 10^5$ CPU core hours per run).
    \item In general, the self-consistent models struggle to achieve a satisfactory fit across the full wavelength range, pointing to the need of an improved description (or parameterization) of atmospheric mixing in connection to disequilibrium chemistry, and an improved treatment of the deep iron cloud and the high-altitude silicate cloud that produces the 10~µm feature. Also exploring a $>$1D approach in self-consistent models may be beneficial.
    \item We observe that our retrievals struggle to converge to the best-fit parameter values, or converge at all, at times. For example, removing the evolutionary prior on ${\rm log}_{10}(g)$ and radius, in an otherwise identical retrieval, leads to a worse fit. This shows that techniques such as nested sampling are reaching their limit in the JWST era of high-precision data with wide wavelength coverage, which also require more complex ($\gtrsim$40 free parameters) models. This problem is also relevant with respect to the imminent first light of the ELT, slated for the early 2030s, from which we will get high-$S/N$ data at high spectral resolution and wide wavelength coverage.
\end{itemize}

\begin{figure}[t!]
    \centering
    \includegraphics[width=0.95\columnwidth]{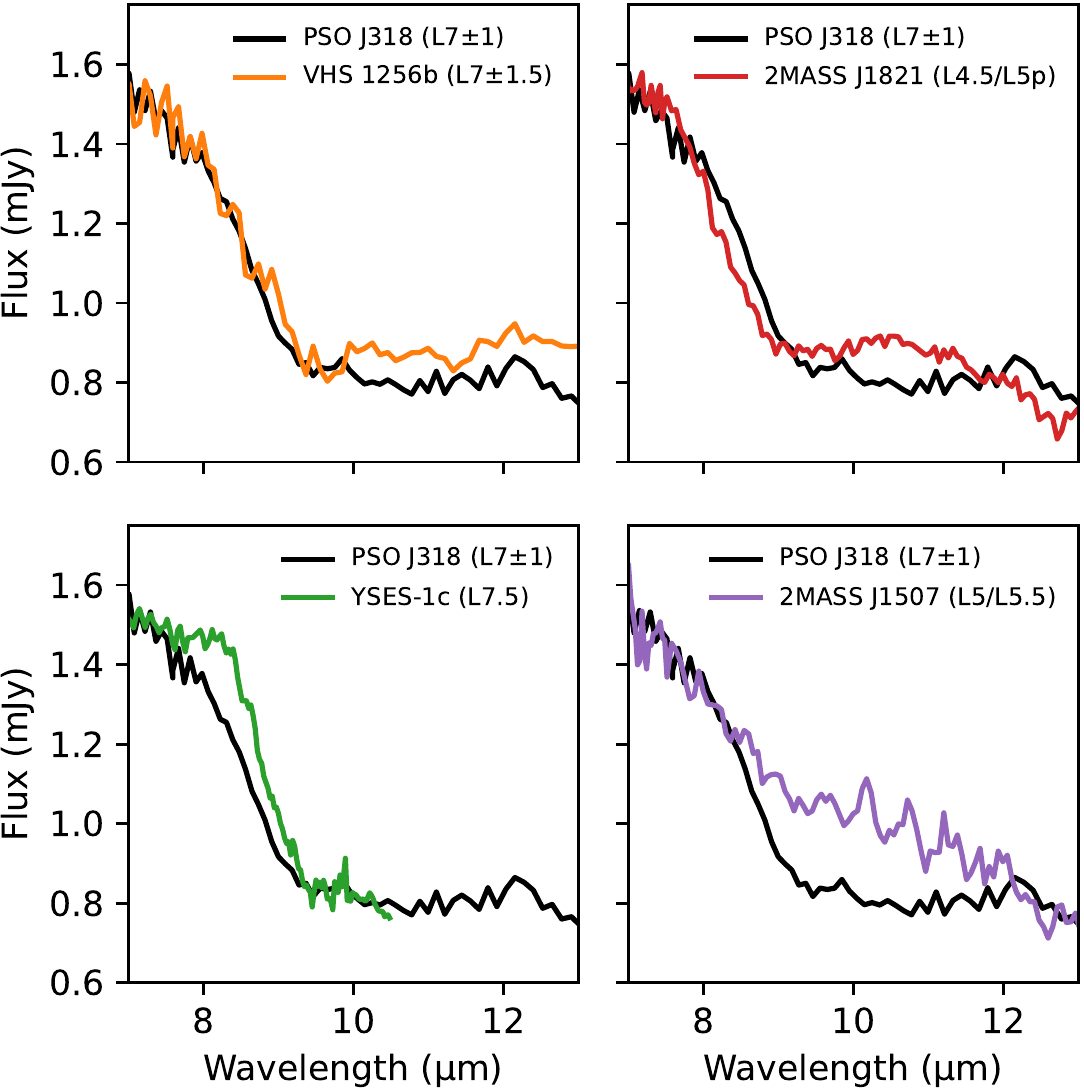}
    \caption{\rch{Comparison between the 10-µm flux of \pso \ and the (scaled) flux of brown dwarfs and low-mass companions with mid-L and L-T transition spectral types.}}
    \label{fig:ten_micron_target_comparison_study}
\end{figure}

We note here that the presence of high-altitude, small SiO cloud particles has also been inferred in the work of \citet{lunamorley2021,suarezmetchev2023b}. \citet{suarezmetchev2023b} reported that small SiO (or \ce{MgSiO3}) particles high-up in the atmosphere appear to be most prevalently occurring for high-gravity (${\rm log}_{10}(g)>5$) L-dwarfs, since small particles are the only ones that may stay aloft. In contrast, they found that larger ($\sim 1$~µm) pyroxene-type (Mg$_{x}$Fe$_{1-x}$\ce{SiO3}) grains may be present in the high atmosphere for lower gravity objects (${\rm log}{10}(g)\lesssim 4.5$). For \pso \ we find that SiO would still have to be mixed up across 3 dex in pressure from its thermo-chemically expected cloud base to be visible, which may be unlikely.  Additionally, if it is indeed SiO particles that are visible in the spectra, mixing from low altitudes to high up in the atmosphere would be challenging: SiO is not chemically favored if produced by condensation (rather than nucleation). For \pso \ we therefore argue that it is not mixing that keeps the small particles aloft in \pso. Instead we posit that the SiO particles nucleate locally, which is consistent with the so-called ``top-down'' cloud formation approach \citep{hellingwoitke2008}. Indeed, our actually inferred location of the SiO cloud coincides with the expected location of maximum SiO nucleation \citep{gailwetzel2013}. What is more, \pso \  is a known planetary-mass object, which is inconsistent with the ${\rm log}_{10}(g)$ trend identified in \citet{suarezmetchev2023b}. More retrieval work for a set of objects, spanning a large ${\rm log}_{10}(g)$ and temperature range is required to understand the behavior of these clouds better.

\rch{Some example spectra of brown dwarfs and low-mass companions are shown in Fig.~\ref{fig:ten_micron_target_comparison_study}, in comparison to the flux observed for \pso. Here we plot VHS 1256-1257b (spectral type L7$\pm$1.5, \citealt{milesbiller2023}), 2MASS J1821+1414  (L4.5/L5p, \citealt{suarezmetchev2022}), YSES-1c (L7.5, \citealt{hochrowland2025}), and 
2MASS J1507-1627 (L5/L5.5, \citealt{suarezmetchev2022}). Some objects appear to be consistent with the flux of \pso, especially VHS 1256b and 2MASS J1821+1414, but the properties of these objects, and also the ones that look different (here YSES-1c and 2MASS J1507-1627) should be studied more carefully. Interestingly, the potential preference for SiO in 2MASS J1821+1414 has been reported in the aforementioned \cite{lunamorley2021}, while they report preference for \ce{Mg2SiO4} and \ce{MgSiO3} for 2MASS J1507-1627. For YSES-1c combination of amorphous \ce{MgSiO3} and \ce{Mg2SiO4} is cited as explaining the feature well, or iron-enriched silicates \citep{hochrowland2025}, which would be consistent with the redder onset of the silicate feature. If SiO  nucleation is prevalent in at least some classes of directly imaged planets and brown dwarfs it may also explain why the 10~µm feature appears to be often best described by amorphous silicate absorption: while to this day the actual structure of SiO evades us \citep[see][for the still relevant models]{brady1959,philipp1975}, it is generally accepted that condensed SiO is highly amorphous \citep{hass1950,schnurregroebner2004}.}

Any retrieval analyses \rch{of such objects would benefit from being} backed up by dedicated microphysical cloud models that incorporate not only the nucleation of species such as SiO and \ce{TiO2}, but also model the condensation, settling and mixing of silicate species such as \ce{MgSiO3} or \ce{Mg2SiO4}, as well as gas-grain reactions (e.g., SiO should oxidize to \ce{SiO2} quite quickly). \rch{If SiO nuclei are present it} should also be investigated why only \rch{they are visible, and not the cloud species that are expected to condense on them}. Do \rch{SiO nuclei} form more efficiently than the expected condensation of the ``usual'' silicate species (\ce{MgSiO3}, \ce{Mg2SiO4}, ...) on their surfaces, or their expected oxidization into \ce{SiO2}?

We also find that the retrieval community is in need of improved inference techniques that both speed up and enhance the robustness of retrievals. We are facing the ``No free lunch'' theorem: we cannot significantly change the properties of our data (wider wavelength coverage, significantly better signal to noise) and the type of questions we ask (e.g., is the atmosphere multi-dimensional?) and expect our classical inference algorithms to still work. Separating the data into wavelength sections that allow us to ask questions that are independently answerable could be one approach, such as we did here with the brightness temperature method. Another example is the work by \citet{gandhideregt2023}, who only used \emph{NIRSpec G395H} data to measure CO isotopologue abundances in VHS~1256b. Alternatively, a promising avenue is exploring the usefulness of machine learning inference approaches, such as amortized or sequential simulation-based inference \citep{cranmerbrehmer2020}, for which early work for exoplanet and brown dwarf atmospheres exist \citep{vasistrozet2023,barradomolliere2023,ardevolmartinezmin2024,lueberkarchev2025}, but which have not yet been applied to models as complex (in terms of number of parameters) as the ones we employ here.

If such improved inference methods can be developed to run efficiently on the data investigated here, the suggested SiO detection from this work should be revisited, to explore the robustness of our finding. There are other runners-up for the most likely cloud species, such as glassy amorphous \ce{MgSiO3}, which lead to a good (but worse) fit. The properties of the silicate cloud should thus be studied by exploring a wider range of forward model setups. The current numerical cost of the retrievals makes this inefficient. Relatedly, such methods would potentially also enable to study the variability of objects such as \pso \ in more detail. Data sets that record the evolution of spectra over multiple rotational periods now exist \citep{billervos2024,mccarthyvos2025,chenbiller2025}, but it will be challenging to develop retrieval approaches that can take the full information content of these data into account.

\begin{acknowledgements}
P.M. thanks the referee, Caroline Morley, for a careful and constructive review of the paper. P.M. would also like to thank the editor Emmanuel Lellouch for additional comments on the manuscript. In addition, P.M. thanks Laura Kreidberg, Evert Nasedkin, and Johanna Vos for insightful discussions which improved the quality of the paper. This work is based (in part) on observations made with the NASA/ESA/CSA James Webb Space Telescope. The data were obtained from the Mikulski Archive for Space Telescopes at the Space Telescope Science Institute, which is operated by the Association of Universities for Research in Astronomy, Inc., under NASA contract NAS 5-03127 for JWST. These observations are associated with program \#1275.  The following National and Inter- national Funding Agencies funded and supported the development of JWST's MIRI instrument, which was used in this work: NASA; ESA; Belgian Science Policy Office (BELSPO); Centre Nationale d’Etudes Spatiales (CNES); Danish National Space Centre; Deutsches Zen- trum fur Luft und Raumfahrt (DLR); Enterprise Ireland; Ministerio De Economıa y Competividad; Netherlands Research School for Astronomy (NOVA); Netherlands Organisation for Scientific Research (NWO); Science and Technology Facilities Council; Swiss Space Office; Swedish National Space Agency; and UK Space Agency. P-O.L. acknowledges funding support by CNES. Z.Z. acknowledges the support of the NASA Hubble Fellowship grant HST-HF2-51522.001-A. P.P. thanks the Swiss National Science Foundation (SNSF) for financial support under grant number 200020\_200399. L.D. acknowledges funding support from the KU Leuven Interdisciplinary Grant (IDN/19/028), the European Union H2020-MSCA-ITN-2019 under Grant no. 860470 (CHAMELEON) and the FWO research grant G086217N. B.A.B. acknowledges funding by the UK Science and Technology Facilities Council (STFC) grant no. ST/V000594/1. O.A. is a Senior Research Associate of the Fonds de la Recherche Scientifique -- FNRS. This project has received funding from the European Research Council (ERC) under the European Union's Horizon 2020 research and innovation programme (grant agreement No 819155). I.A. would like to thank the European Space Agency (ESA) and the Belgian Federal Science Policy Office (BELSPO) for their support in the framework of the PRODEX Programme. DB is funded by grant PID2023-150468NB-I00 by the Spanish Ministry of Science and Innovation/State Agency of Research MCIN/AEI/10.13039/501100011033. JPP acknowledges financial support from the UK Science and Technology Facilities Council, and the UK Space Agency. M.S. acknowledges support from the European Research Council under the Horizon 2020 Framework Program via the ERC Advanced Grant Origins 83 24 28. N.W. acknowledges support from NSF award \#2238468, \#1909776, and NASA Award \#80NSSC22K0142. EvD acknowledges support from A-ERC grant 101019751 MOLDISK. G.\"O. acknowledges support from the Swedish National Space Agency (SNSA).
\end{acknowledgements}

%
%

\bibliographystyle{aa}
\bibliography{references}{}

\newcommand{\noop}[1]{}
\begin{thebibliography}{168}
\expandafter\ifx\csname natexlab\endcsname\relax\def\natexlab#1{#1}\fi

\bibitem[{{Ackerman} \& {Marley}(2001)}]{ackermanmarley2001}
{Ackerman}, A.~S. \& {Marley}, M.~S. 2001, \apj, 556, 872

\bibitem[{{Allard} {et~al.}(2001){Allard}, {Hauschildt}, {Alexander},
  {Tamanai}, \& {Schweitzer}}]{allardhauschildt2001}
{Allard}, F., {Hauschildt}, P.~H., {Alexander}, D.~R., {Tamanai}, A., \&
  {Schweitzer}, A. 2001, \apj, 556, 357

\bibitem[{{Allard} {et~al.}(2019){Allard}, {Spiegelman}, {Leininger}, \&
  {Molliere}}]{allardspiegelman2019}
{Allard}, N.~F., {Spiegelman}, F., {Leininger}, T., \& {Molliere}, P. 2019,
  \aap, 628, A120

\bibitem[{{Allers} {et~al.}(2016){Allers}, {Gallimore}, {Liu}, \&
  {Dupuy}}]{allersgallimore2016}
{Allers}, K.~N., {Gallimore}, J.~F., {Liu}, M.~C., \& {Dupuy}, T.~J. 2016,
  \apj, 819, 133

\bibitem[{{Allers} \& {Liu}(2013)}]{allersliu2013}
{Allers}, K.~N. \& {Liu}, M.~C. 2013, \apj, 772, 79

\bibitem[{{Apai} {et~al.}(2013){Apai}, {Radigan}, {Buenzli}, {Burrows}, {Reid},
  \& {Jayawardhana}}]{apairadigan2013}
{Apai}, D., {Radigan}, J., {Buenzli}, E., {et~al.} 2013, \apj, 768, 121

\bibitem[{{Ard{\'e}vol Mart{\'\i}nez} {et~al.}(2024){Ard{\'e}vol
  Mart{\'\i}nez}, {Min}, {Huppenkothen}, {Kamp}, \&
  {Palmer}}]{ardevolmartinezmin2024}
{Ard{\'e}vol Mart{\'\i}nez}, F., {Min}, M., {Huppenkothen}, D., {Kamp}, I., \&
  {Palmer}, P.~I. 2024, \aap, 681, L14

\bibitem[{{Asplund} {et~al.}(2009){Asplund}, {Grevesse}, {Sauval}, \&
  {Scott}}]{asplund2009}
{Asplund}, M., {Grevesse}, N., {Sauval}, A.~J., \& {Scott}, P. 2009, \araa, 47,
  481

\bibitem[{{Azzam} {et~al.}(2016){Azzam}, {Tennyson}, {Yurchenko}, \&
  {Naumenko}}]{azzamtennyson2016}
{Azzam}, A. A.~A., {Tennyson}, J., {Yurchenko}, S.~N., \& {Naumenko}, O.~V.
  2016, \mnras, 460, 4063

\bibitem[{Barber {et~al.}(2013)Barber, Strange, Hill, Polyansky, Mellau,
  Yurchenko, \& Tennyson}]{barberstrange2013}
Barber, R.~J., Strange, J.~K., Hill, C., {et~al.} 2013, Monthly Notices of the
  Royal Astronomical Society, 437, 1828

\bibitem[{{Barrado} {et~al.}(2023){Barrado}, {Molli{\`e}re}, {Patapis}, {Min},
  {Tremblin}, {Ardevol Martinez}, {Whiteford}, {Vasist}, {Argyriou}, {Samland},
  {Lagage}, {Decin}, {Waters}, {Henning}, {Morales-Calder{\'o}n}, {Guedel},
  {Vandenbussche}, {Absil}, {Baudoz}, {Boccaletti}, {Bouwman}, {Cossou},
  {Coulais}, {Crouzet}, {Gastaud}, {Glasse}, {Glauser}, {Kamp}, {Kendrew},
  {Krause}, {Lahuis}, {Mueller}, {Olofsson}, {Pye}, {Rouan}, {Royer},
  {Scheithauer}, {Waldmann}, {Colina}, {van Dishoeck}, {Ray}, {{\"O}stlin}, \&
  {Wright}}]{barradomolliere2023}
{Barrado}, D., {Molli{\`e}re}, P., {Patapis}, P., {et~al.} 2023, \nat, 624, 263

\bibitem[{{Barrado y Navascu{\'e}s} {et~al.}(1999){Barrado y Navascu{\'e}s},
  {Stauffer}, {Song}, \& {Caillault}}]{Barrado1999_BPMG}
{Barrado y Navascu{\'e}s}, D., {Stauffer}, J.~R., {Song}, I., \& {Caillault},
  J.~P. 1999, \apjl, 520, L123

\bibitem[{{Baudino} {et~al.}(2015){Baudino}, {B{\'e}zard}, {Boccaletti},
  {Bonnefoy}, {Lagrange}, \& {Galicher}}]{baudinobezard2015}
{Baudino}, J.-L., {B{\'e}zard}, B., {Boccaletti}, A., {et~al.} 2015, \aap, 582,
  A83

\bibitem[{{Benneke} \& {Seager}(2013)}]{bennekeseager2013}
{Benneke}, B. \& {Seager}, S. 2013, \apj, 778, 153

\bibitem[{{Biller}(2017)}]{biller2017}
{Biller}, B. 2017, The Astronomical Review, 13, 1

\bibitem[{{Biller} {et~al.}(2015){Biller}, {Vos}, {Bonavita}, {Buenzli},
  {Baxter}, {Crossfield}, {Allers}, {Liu}, {Bonnefoy}, {Deacon}, {Brandner},
  {Schlieder}, {Dupuy}, {Kopytova}, {Manjavacas}, {Allard}, {Homeier}, \&
  {Henning}}]{billervos2015}
{Biller}, B.~A., {Vos}, J., {Bonavita}, M., {et~al.} 2015, \apjl, 813, L23

\bibitem[{{Biller} {et~al.}(2018){Biller}, {Vos}, {Buenzli}, {Allers},
  {Bonnefoy}, {Charnay}, {B{\'e}zard}, {Allard}, {Homeier}, {Bonavita},
  {Brandner}, {Crossfield}, {Dupuy}, {Henning}, {Kopytova}, {Liu},
  {Manjavacas}, \& {Schlieder}}]{billervos2018}
{Biller}, B.~A., {Vos}, J., {Buenzli}, E., {et~al.} 2018, \aj, 155, 95

\bibitem[{{Biller} {et~al.}(2024){Biller}, {Vos}, {Zhou}, {McCarthy}, {Tan},
  {Crossfield}, {Whiteford}, {Suarez}, {Faherty}, {Manjavacas}, {Chen}, {Liu},
  {Sutlieff}, {Limbach}, {Molliere}, {Dupuy}, {Oliveros-Gomez}, {Muirhead},
  {Henning}, {Mace}, {Crouzet}, {Karalidi}, {Morley}, {Tremblin}, \&
  {Kataria}}]{billervos2024}
{Biller}, B.~A., {Vos}, J.~M., {Zhou}, Y., {et~al.} 2024, \mnras, 532, 2207

\bibitem[{{Blain} {et~al.}(2021){Blain}, {Charnay}, \&
  {B{\'e}zard}}]{blaincharnay2021}
{Blain}, D., {Charnay}, B., \& {B{\'e}zard}, B. 2021, \aap, 646, A15

\bibitem[{{Blain} {et~al.}(2024){Blain}, {Molli{\`e}re}, \&
  {Nasedkin}}]{blainmolliereJOSS2024}
{Blain}, D., {Molli{\`e}re}, P., \& {Nasedkin}, E. 2024, The Journal of Open
  Source Software, 9, 7028

\bibitem[{{Bodenheimer} {et~al.}(2013){Bodenheimer}, {D'Angelo}, {Lissauer},
  {Fortney}, \& {Saumon}}]{bodenheimerdangelo2013}
{Bodenheimer}, P., {D'Angelo}, G., {Lissauer}, J.~J., {Fortney}, J.~J., \&
  {Saumon}, D. 2013, \apj, 770, 120

\bibitem[{{Bohren} \& {Huffman}(1983)}]{bohrenhuffman1983}
{Bohren}, C.~F. \& {Huffman}, D.~R. 1983, {Absorption and scattering of light
  by small particles}

\bibitem[{{Bouwman} {et~al.}(2001){Bouwman}, {Meeus}, {de Koter}, {Hony},
  {Dominik}, \& {Waters}}]{bouwman2001}
{Bouwman}, J., {Meeus}, G., {de Koter}, A., {et~al.} 2001, \aap, 375, 950

\bibitem[{Brady(1959)}]{brady1959}
Brady, G.~W. 1959, The Journal of Physical Chemistry, 63, 1119

\bibitem[{Bromley {et~al.}(2016)Bromley, Gómez~Martín, \&
  Plane}]{bromleygomez2016}
Bromley, S.~T., Gómez~Martín, J.~C., \& Plane, J. M.~C. 2016, Phys. Chem.
  Chem. Phys., 18, 26913

\bibitem[{{Buchner}(2021)}]{buchner2021}
{Buchner}, J. 2021, arXiv e-prints, arXiv:2101.09675

\bibitem[{{Buchner} {et~al.}(2014){Buchner}, {Georgakakis}, {Nandra}, {Hsu},
  {Rangel}, {Brightman}, {Merloni}, {Salvato}, {Donley}, \&
  {Kocevski}}]{buchnergeorgakakis2014}
{Buchner}, J., {Georgakakis}, A., {Nandra}, K., {et~al.} 2014, \aap, 564, A125

\bibitem[{{Burningham} {et~al.}(2021){Burningham}, {Faherty}, {Gonzales},
  {Marley}, {Visscher}, {Lupu}, {Gaarn}, {Fabienne Bieger}, {Freedman}, \&
  {Saumon}}]{burninghamfaherty2021}
{Burningham}, B., {Faherty}, J.~K., {Gonzales}, E.~C., {et~al.} 2021, \mnras,
  506, 1944

\bibitem[{{Burrows} {et~al.}(1997){Burrows}, {Marley}, {Hubbard}, {Lunine},
  {Guillot}, {Saumon}, {Freedman}, {Sudarsky}, \& {Sharp}}]{burrows1997}
{Burrows}, A., {Marley}, M., {Hubbard}, W.~B., {et~al.} 1997, \apj, 491, 856

\bibitem[{{Burrows} {et~al.}(2002){Burrows}, {Ram}, {Bernath}, {Sharp}, \&
  {Milsom}}]{burrowsram2002}
{Burrows}, A., {Ram}, R.~S., {Bernath}, P., {Sharp}, C.~M., \& {Milsom}, J.~A.
  2002, \apj, 577, 986

\bibitem[{{Bushouse} {et~al.}(2023){Bushouse}, {Eisenhamer}, {Dencheva},
  {Davies}, {Greenfield}, {Morrison}, {Hodge}, {Simon}, {Grumm}, {Droettboom},
  {Slavich}, {Sosey}, {Pauly}, {Miller}, {Jedrzejewski}, {Hack}, {Davis},
  {Crawford}, {Law}, {Gordon}, {Regan}, {Cara}, {MacDonald}, {Bradley},
  {Shanahan}, {Jamieson}, {Teodoro}, {Williams}, \&
  {Pena-Guerrero}}]{Bushouse2023}
{Bushouse}, H., {Eisenhamer}, J., {Dencheva}, N., {et~al.} 2023, {JWST
  Calibration Pipeline}

\bibitem[{{Calamari} {et~al.}(2024){Calamari}, {Faherty}, {Visscher}, {Gemma},
  {Burningham}, \& {Rothermich}}]{calamarifaherty2024}
{Calamari}, E., {Faherty}, J.~K., {Visscher}, C., {et~al.} 2024, \apj, 963, 67

\bibitem[{{Campos Estrada} {et~al.}(2025){Campos Estrada}, {Lewis}, {Helling},
  {Booth}, {Ard{\'e}vol Mart{\'\i}nez}, \& {J{\o}rgensen}}]{camposestrada2025}
{Campos Estrada}, B., {Lewis}, D.~A., {Helling}, C., {et~al.} 2025, \aap, 694,
  A275

\bibitem[{{Chabrier} {et~al.}(2014){Chabrier}, {Johansen}, {Janson}, \&
  {Rafikov}}]{chabrierjohansen2014}
{Chabrier}, G., {Johansen}, A., {Janson}, M., \& {Rafikov}, R. 2014, in
  Protostars and Planets VI, ed. H.~{Beuther}, R.~S. {Klessen}, C.~P.
  {Dullemond}, \& T.~{Henning}, 619--642

\bibitem[{{Charnay} {et~al.}(2018){Charnay}, {B{\'e}zard}, {Baudino},
  {Bonnefoy}, {Boccaletti}, \& {Galicher}}]{charnaybezard2018}
{Charnay}, B., {B{\'e}zard}, B., {Baudino}, J.~L., {et~al.} 2018, \apj, 854,
  172

\bibitem[{{Chen} {et~al.}(2025){Chen}, {Biller}, {Tan}, {Vos}, {Zhou},
  {Su{\'a}rez}, {McCarthy}, {Morley}, {Whiteford}, {Dupuy}, {Faherty},
  {Sutlieff}, {Oliveros-Gomez}, {Manjavacas}, {Limbach}, {Lee}, {Karalidi},
  {Crossfield}, {Liu}, {Molliere}, {Muirhead}, {Henning}, {Mace}, {Crouzet}, \&
  {Kataria}}]{chenbiller2025}
{Chen}, X., {Biller}, B.~A., {Tan}, X., {et~al.} 2025, arXiv e-prints,
  arXiv:2505.00794

\bibitem[{{Chubb} \& {Min}(2022)}]{chubbmin2022}
{Chubb}, K.~L. \& {Min}, M. 2022, \aap, 665, A2

\bibitem[{Chubb {et~al.}(2021)Chubb, Rocchetto, Yurchenko, Min, Waldmann,
  Barstow, Mollière, Al-Refaie, Phillips, \& Tennyson}]{chubb2021exomolop-1d1}
Chubb, K.~L., Rocchetto, M., Yurchenko, S.~N., {et~al.} 2021, Astronomy \&
  Astrophysics, 646, A21

\bibitem[{Coles {et~al.}(2019)Coles, Yurchenko, \&
  Tennyson}]{colesyurchenko2019}
Coles, P.~A., Yurchenko, S.~N., \& Tennyson, J. 2019, Monthly Notices of the
  Royal Astronomical Society, 490, 4638

\bibitem[{{Cranmer} {et~al.}(2020){Cranmer}, {Brehmer}, \&
  {Louppe}}]{cranmerbrehmer2020}
{Cranmer}, K., {Brehmer}, J., \& {Louppe}, G. 2020, Proceedings of the National
  Academy of Science, 117, 30055

\bibitem[{{Cushing} {et~al.}(2006){Cushing}, {Roellig}, {Marley}, {Saumon},
  {Leggett}, {Kirkpatrick}, {Wilson}, {Sloan}, {Mainzer}, {Van Cleve}, \&
  {Houck}}]{cushing2006}
{Cushing}, M.~C., {Roellig}, T.~L., {Marley}, M.~S., {et~al.} 2006, \apj, 648,
  614

\bibitem[{{Dittmann}(2024)}]{dittmann2024}
{Dittmann}, A. 2024, The Open Journal of Astrophysics, 7, 79

\bibitem[{{Dorschner} {et~al.}(1995){Dorschner}, {Begemann}, {Henning},
  {Jaeger}, \& {Mutschke}}]{dorschnerbegemann1995}
{Dorschner}, J., {Begemann}, B., {Henning}, T., {Jaeger}, C., \& {Mutschke}, H.
  1995, \aap, 300, 503

\bibitem[{{Draine}(1988)}]{draine1988}
{Draine}, B.~T. 1988, \apj, 333, 848

\bibitem[{{Dyrek} {et~al.}(2024){Dyrek}, {Min}, {Decin}, {Bouwman}, {Crouzet},
  {Molli{\`e}re}, {Lagage}, {Konings}, {Tremblin}, {G{\"u}del}, {Pye},
  {Waters}, {Henning}, {Vandenbussche}, {Ardevol Martinez}, {Argyriou},
  {Ducrot}, {Heinke}, {van Looveren}, {Absil}, {Barrado}, {Baudoz},
  {Boccaletti}, {Cossou}, {Coulais}, {Edwards}, {Gastaud}, {Glasse}, {Glauser},
  {Greene}, {Kendrew}, {Krause}, {Lahuis}, {Mueller}, {Olofsson}, {Patapis},
  {Rouan}, {Royer}, {Scheithauer}, {Waldmann}, {Whiteford}, {Colina}, {van
  Dishoeck}, {{\"O}stlin}, {Ray}, \& {Wright}}]{dyrekmin2024}
{Dyrek}, A., {Min}, M., {Decin}, L., {et~al.} 2024, \nat, 625, 51

\bibitem[{{Fabian} {et~al.}(2001){Fabian}, {Henning}, {J{\"a}ger}, {Mutschke},
  {Dorschner}, \& {Wehrhan}}]{fabianhenning2001}
{Fabian}, D., {Henning}, T., {J{\"a}ger}, C., {et~al.} 2001, \aap, 378, 228

\bibitem[{{Fabian} {et~al.}(2000){Fabian}, {J{\"a}ger}, {Henning}, {Dorschner},
  \& {Mutschke}}]{fabianjaeger2000}
{Fabian}, D., {J{\"a}ger}, C., {Henning}, T., {Dorschner}, J., \& {Mutschke},
  H. 2000, \aap, 364, 282

\bibitem[{{Faherty}(2018)}]{faherty2018}
{Faherty}, J.~K. 2018, {Spectral Properties of Brown Dwarfs and Unbound
  Planetary Mass Objects}, 188

\bibitem[{{Feroz} \& {Hobson}(2008)}]{ferrozhobson2008}
{Feroz}, F. \& {Hobson}, M.~P. 2008, \mnras, 384, 449

\bibitem[{{Feroz} {et~al.}(2009){Feroz}, {Hobson}, \&
  {Bridges}}]{ferrozhobson2009}
{Feroz}, F., {Hobson}, M.~P., \& {Bridges}, M. 2009, \mnras, 398, 1601

\bibitem[{{Feroz} {et~al.}(2019){Feroz}, {Hobson}, {Cameron}, \&
  {Pettitt}}]{ferozhobson2019}
{Feroz}, F., {Hobson}, M.~P., {Cameron}, E., \& {Pettitt}, A.~N. 2019, The Open
  Journal of Astrophysics, 2, 10

\bibitem[{{Gail}(2001)}]{gail2001}
{Gail}, H.-P. 2001, \aap, 378, 192

\bibitem[{{Gail}(2004)}]{gail2004}
{Gail}, H.-P. 2004, \aap, 413, 571

\bibitem[{{Gail}(2010)}]{gail2010}
{Gail}, H.-P. 2010, {Astromineralogy}, ed. T.~{Henning}, Vol. 815 (Springer)

\bibitem[{{Gail} {et~al.}(2013){Gail}, {Wetzel}, {Pucci}, \&
  {Tamanai}}]{gailwetzel2013}
{Gail}, H.~P., {Wetzel}, S., {Pucci}, A., \& {Tamanai}, A. 2013, \aap, 555,
  A119

\bibitem[{{Gandhi} {et~al.}(2023){Gandhi}, {de Regt}, {Snellen}, {Zhang},
  {Rugers}, {van Leur}, \& {Bosschaart}}]{gandhideregt2023}
{Gandhi}, S., {de Regt}, S., {Snellen}, I., {et~al.} 2023, \apjl, 957, L36

\bibitem[{{Gao} {et~al.}(2018){Gao}, {Marley}, \& {Ackerman}}]{gaomarley2018}
{Gao}, P., {Marley}, M.~S., \& {Ackerman}, A.~S. 2018, \apj, 855, 86

\bibitem[{{Gao} {et~al.}(2020){Gao}, {Thorngren}, {Lee}, {Fortney}, {Morley},
  {Wakeford}, {Powell}, {Stevenson}, \& {Zhang}}]{gaothorngren2020}
{Gao}, P., {Thorngren}, D.~P., {Lee}, G. K.~H., {et~al.} 2020, Nature
  Astronomy, 4, 951

\bibitem[{{Gardner} {et~al.}(2023){Gardner}, {Mather}, {Abbott}, {Abell},
  {Abernathy}, {Abney}, {Abraham}, {Abraham}, {Abul-Huda}, {Acton}, {Adams},
  {Adams}, {Adler}, {Adriaensen}, {Aguilar}, {Ahmed}, {Ahmed}, {Ahmed},
  {Albat}, {Albert}, {Alberts}, {Aldridge}, {Allen}, {Allen}, {Altenburg},
  {Altunc}, {Alvarez}, {{\'A}lvarez-M{\'a}rquez}, {Alves de Oliveira},
  {Ambrose}, {Anandakrishnan}, {Andersen}, {Anderson}, {Anderson}, {Anderson},
  {Anderson}, {Aprea}, {Archer}, {Arenberg}, {Argyriou}, {Arribas}, {Artigau},
  {Arvai}, {Atcheson}, {Atkinson}, {Averbukh}, {Aymergen}, {Bacinski},
  {Baggett}, {Bagnasco}, {Baker}, {Balzano}, {Banks}, {Baran}, {Barker},
  {Barrett}, {Barringer}, {Barto}, {Bast}, {Baudoz}, {Baum}, {Beatty},
  {Beaulieu}, {Bechtold}, {Beck}, {Beddard}, {Beichman}, {Bellagama}, {Bely},
  {Berger}, {Bergeron}, {Bernier}, {Bertch}, {Beskow}, {Betz}, {Biagetti},
  {Birkmann}, {Bjorklund}, {Blackwood}, {Blazek}, {Blossfeld}, {Bluth},
  {Boccaletti}, {Boegner}, {Bohlin}, {Boia}, {B{\"o}ker}, {Bonaventura},
  {Bond}, {Bosley}, {Boucarut}, {Bouchet}, {Bouwman}, {Bower}, {Bowers},
  {Bowers}, {Boyce}, {Boyer}, {Boyer}, {Boyer}, {Boyer}, {Bradley}, {Brady},
  {Brandl}, {Brannen}, {Breda}, {Bremmer}, {Brennan}, {Bresnahan}, {Bright},
  {Broiles}, {Bromenschenkel}, {Brooks}, {Brooks}, {Brown}, {Brown}, {Brown},
  {Bruce}, {Bryson}, {Bujanda}, {Bullock}, {Bunker}, {Bureo}, {Burt}, {Bush},
  {Bushouse}, {Bussman}, {Cabaud}, {Cale}, {Calhoon}, {Calvani}, {Canipe},
  {Caputo}, {Cara}, {Carey}, {Case}, {Cesari}, {Cetorelli}, {Chance},
  {Chandler}, {Chaney}, {Chapman}, {Charlot}, {Chayer}, {Cheezum}, {Chen},
  {Chen}, {Cherinka}, {Chichester}, {Chilton}, {Chittiraibalan}, {Clampin},
  {Clark}, {Clark}, {Clark}, {Claybrooks}, {Cleveland}, {Cohen}, {Cohen},
  {Col{\'o}n}, {Coleman}, {Colina}, {Comber}, {Comeau}, {Comer}, {Conde Reis},
  {Connolly}, {Conroy}, {Contos}, {Contreras}, {Cook}, {Cooper}, {Cooper},
  {Correia}, {Correnti}, {Cossou}, {Costanza}, {Coulais}, {Cox}, {Coyle},
  {Cracraft}, {Crew}, {Curtis}, {Cusveller}, {Da Costa Maciel}, {Dailey},
  {Daugeron}, {Davidson}, {Davies}, {Davis}, {Davis}, {Day}, {de Chambure}, {de
  Jong}, {De Marchi}, {Dean}, {Decker}, {Delisa}, {Dell}, \&
  {Dellagatta}}]{gardnermather2023}
{Gardner}, J.~P., {Mather}, J.~C., {Abbott}, R., {et~al.} 2023, \pasp, 135,
  068001

\bibitem[{{Grant} {et~al.}(2023){Grant}, {Lewis}, {Wakeford}, {Batalha},
  {Glidden}, {Goyal}, {Mullens}, {MacDonald}, {May}, {Seager}, {Stevenson},
  {Valenti}, {Visscher}, {Alderson}, {Allen}, {Ca{\~n}as}, {Col{\'o}n},
  {Clampin}, {Espinoza}, {Gressier}, {Huang}, {Lin}, {Long}, {Louie},
  {Pe{\~n}a-Guerrero}, {Ranjan}, {Sotzen}, {Valentine}, {Anderson}, {Balmer},
  {Bellini}, {Hoch}, {Kammerer}, {Libralato}, {Mountain}, {Perrin}, {Pueyo},
  {Rickman}, {Rebollido}, {Sohn}, {van der Marel}, \&
  {Watkins}}]{grantlewis2023}
{Grant}, D., {Lewis}, N.~K., {Wakeford}, H.~R., {et~al.} 2023, \apjl, 956, L32

\bibitem[{{Hansen}(1971)}]{hansen1971}
{Hansen}, J.~E. 1971, Journal of the Atmospheric Sciences, 28, 1400

\bibitem[{{Hargreaves} {et~al.}(2020){Hargreaves}, {Gordon}, {Rey}, {Nikitin},
  {Tyuterev}, {Kochanov}, \& {Rothman}}]{hargreaves2020}
{Hargreaves}, R.~J., {Gordon}, I.~E., {Rey}, M., {et~al.} 2020, \apjs, 247, 55

\bibitem[{{Harker} \& {Desch}(2002)}]{harkerdesch2002}
{Harker}, D.~E. \& {Desch}, S.~J. 2002, \apjl, 565, L109

\bibitem[{{Hass}(1950)}]{hass1950}
{Hass}, G. 1950, Journal of the American Ceramic Society, 33, 353

\bibitem[{{Helling} {et~al.}(2008){Helling}, {Woitke}, \&
  {Thi}}]{hellingwoitke2008}
{Helling}, C., {Woitke}, P., \& {Thi}, W.~F. 2008, \aap, 485, 547

\bibitem[{{Henning}(2010)}]{henning2010}
{Henning}, T. 2010, \araa, 48, 21

\bibitem[{{Henning} \& {Mutschke}(1997)}]{henningmutschke1997}
{Henning}, T. \& {Mutschke}, H. 1997, \aap, 327, 743

\bibitem[{{Henning} \& {Stognienko}(1993)}]{henningsognienko1993}
{Henning}, T. \& {Stognienko}, R. 1993, \aap, 280, 609

\bibitem[{Himes(2022)}]{himes2022retrievalfail}
Himes, M. 2022, PhD thesis, University of Central Florida, doctoral
  Dissertation (Open Access)

\bibitem[{Hoch {et~al.}(2025)Hoch, Rowland, Petrus, Nasedkin, Ingebretsen,
  Kammerer, Perrin, D’Orazi, Balmer, Barman, Bonnefoy, Chauvin, Chen,
  De~Rosa, Girard, Gonzales, Kenworthy, Konopacky, Macintosh, Moran, Morley,
  Palma-Bifani, Pueyo, Ren, Rickman, Ruffio, Theissen, Ward-Duong, \&
  Zhang}]{hochrowland2025}
Hoch, K. K.~W., Rowland, M., Petrus, S., {et~al.} 2025, Nature

\bibitem[{{Hubeny}(2017)}]{hubeny2017}
{Hubeny}, I. 2017, \mnras, 469, 841

\bibitem[{{Inglis} {et~al.}(2024){Inglis}, {Batalha}, {Lewis}, {Kataria},
  {Knutson}, {Kilpatrick}, {Gagnebin}, {Mukherjee}, {Pettyjohn}, {Crossfield},
  {Foote}, {Grant}, {Henry}, {Lally}, {McKemmish}, {Sing}, {Wakeford}, {Zapata
  Trujillo}, \& {Zellem}}]{inglisbatalha2024}
{Inglis}, J., {Batalha}, N.~E., {Lewis}, N.~K., {et~al.} 2024, arXiv e-prints,
  arXiv:2409.11395

\bibitem[{{Jaeger} {et~al.}(1998){Jaeger}, {Molster}, {Dorschner}, {Henning},
  {Mutschke}, \& {Waters}}]{jaeger1998}
{Jaeger}, C., {Molster}, F.~J., {Dorschner}, J., {et~al.} 1998, \aap, 339, 904

\bibitem[{{J{\"a}ger} {et~al.}(2003{\natexlab{a}}){J{\"a}ger}, {Dorschner},
  {Mutschke}, {Posch}, \& {Henning}}]{jaegerdorschner2003}
{J{\"a}ger}, C., {Dorschner}, J., {Mutschke}, H., {Posch}, T., \& {Henning}, T.
  2003{\natexlab{a}}, \aap, 408, 193

\bibitem[{{J{\"a}ger} {et~al.}(2003{\natexlab{b}}){J{\"a}ger}, {Il'in},
  {Henning}, {Mutschke}, {Fabian}, {Semenov}, \&
  {Voshchinnikov}}]{jaegerilin2003}
{J{\"a}ger}, C., {Il'in}, V.~B., {Henning}, T., {et~al.} 2003{\natexlab{b}},
  \jqsrt, 79-80, 765

\bibitem[{{J\"ager} {et~al.}(1994){J\"ager}, {Mutschke}, {Begemann},
  {Dorschner}, \& {Henning}}]{jaegermutschke1994}
{J\"ager}, C., {Mutschke}, H., {Begemann}, B., {Dorschner}, J., \& {Henning},
  T. 1994, \aap, 292, 641

\bibitem[{{Jakobsen} {et~al.}(2022){Jakobsen}, {Ferruit}, {Alves de Oliveira},
  {Arribas}, {Bagnasco}, {Barho}, {Beck}, {Birkmann}, {B{\"o}ker}, {Bunker},
  {Charlot}, {de Jong}, {de Marchi}, {Ehrenwinkler}, {Falcolini}, {Fels},
  {Franx}, {Franz}, {Funke}, {Giardino}, {Gnata}, {Holota}, {Honnen}, {Jensen},
  {Jentsch}, {Johnson}, {Jollet}, {Karl}, {Kling}, {K{\"o}hler}, {Kolm},
  {Kumari}, {Lander}, {Lemke}, {L{\'o}pez-Caniego}, {L{\"u}tzgendorf},
  {Maiolino}, {Manjavacas}, {Marston}, {Maschmann}, {Maurer}, {Messerschmidt},
  {Moseley}, {Mosner}, {Mott}, {Muzerolle}, {Pirzkal}, {Pittet}, {Plitzke},
  {Posselt}, {Rapp}, {Rauscher}, {Rawle}, {Rix}, {R{\"o}del}, {Rumler},
  {Sabbi}, {Salvignol}, {Schmid}, {Sirianni}, {Smith}, {Strada}, {te Plate},
  {Valenti}, {Wettemann}, {Wiehe}, {Wiesmayer}, {Willott}, {Wright}, {Zeidler},
  \& {Zincke}}]{jakobsenferruit2022}
{Jakobsen}, P., {Ferruit}, P., {Alves de Oliveira}, C., {et~al.} 2022, \aap,
  661, A80

\bibitem[{{Juh{\'a}sz} {et~al.}(2010){Juh{\'a}sz}, {Bouwman}, {Henning},
  {Acke}, {van den Ancker}, {Meeus}, {Dominik}, {Min}, {Tielens}, \&
  {Waters}}]{juhaszbouwman2010}
{Juh{\'a}sz}, A., {Bouwman}, J., {Henning}, T., {et~al.} 2010, \apj, 721, 431

\bibitem[{{Kiefer} {et~al.}(2024){Kiefer}, {Samra}, {Lewis}, {Schneider},
  {Min}, {Carone}, {Decin}, \& {Helling}}]{kiefersamra2024}
{Kiefer}, S., {Samra}, D., {Lewis}, D.~A., {et~al.} 2024, arXiv e-prints,
  arXiv:2409.01121

\bibitem[{{Kipping} \& {Benneke}(2025)}]{kippingbenneke2025}
{Kipping}, D. \& {Benneke}, B. 2025, arXiv e-prints, arXiv:2506.05392

\bibitem[{{Kitzmann} \& {Heng}(2018)}]{kitzmannheng2018b}
{Kitzmann}, D. \& {Heng}, K. 2018, \mnras, 475, 94

\bibitem[{{Koike} {et~al.}(1995){Koike}, {Kaito}, {Yamamoto}, {Shibai},
  {Kimura}, \& {Suto}}]{koikekaito1995}
{Koike}, C., {Kaito}, C., {Yamamoto}, T., {et~al.} 1995, \icarus, 114, 203

\bibitem[{{Koike} {et~al.}(2006){Koike}, {Mutschke}, {Suto}, {Naoi}, {Chihara},
  {Henning}, {J{\"a}ger}, {Tsuchiyama}, {Dorschner}, \&
  {Okuda}}]{koikemutschke2006}
{Koike}, C., {Mutschke}, H., {Suto}, H., {et~al.} 2006, \aap, 449, 583

\bibitem[{{K{\"u}hnle} {et~al.}(2025){K{\"u}hnle}, {Patapis}, {Molli{\`e}re},
  {Tremblin}, {Matthews}, {Glauser}, {Whiteford}, {Vasist}, {Absil}, {Barrado},
  {Min}, {Lagage}, {Waters}, {Guedel}, {Henning}, {Vandenbussche}, {Baudoz},
  {Decin}, {Pye}, {Royer}, {van Dishoeck}, {{\"O}stlin}, {Ray}, \&
  {Wright}}]{kuehnlepatapis2025}
{K{\"u}hnle}, H., {Patapis}, P., {Molli{\`e}re}, P., {et~al.} 2025, \aap, 695,
  A224

\bibitem[{{Law} {et~al.}(2023){Law}, {E. Morrison}, {Argyriou}, {Patapis},
  {{\'A}lvarez-M{\'a}rquez}, {Labiano}, \& {Vandenbussche}}]{Law2023}
{Law}, D.~R., {E. Morrison}, J., {Argyriou}, I., {et~al.} 2023, \aj, 166, 45

\bibitem[{{Leconte}(2021)}]{leconte2021}
{Leconte}, J. 2021, \aap, 645, A20

\bibitem[{{Lee} {et~al.}(2018){Lee}, {Blecic}, \& {Helling}}]{leeblecic2018}
{Lee}, E.~K.~H., {Blecic}, J., \& {Helling}, C. 2018, \aap, 614, A126

\bibitem[{{Lei} \& {Molli{\`e}re}(2024)}]{leimolliere2024}
{Lei}, E. \& {Molli{\`e}re}, P. 2024, arXiv e-prints, arXiv:2410.21364

\bibitem[{{Lew} {et~al.}(2020){Lew}, {Apai}, {Marley}, {Saumon}, {Schneider},
  {Zhou}, {Cowan}, {Karalidi}, {Manjavacas}, {Bedin}, \&
  {Miles-P{\'a}ez}}]{lewapai2020}
{Lew}, B. W.~P., {Apai}, D., {Marley}, M., {et~al.} 2020, \apj, 903, 15

\bibitem[{{Line} {et~al.}(2017){Line}, {Marley}, {Liu}, {Burningham}, {Morley},
  {Hinkel}, {Teske}, {Fortney}, {Freedman}, \& {Lupu}}]{linemarley2017}
{Line}, M.~R., {Marley}, M.~S., {Liu}, M.~C., {et~al.} 2017, \apj, 848, 83

\bibitem[{{Line} {et~al.}(2015){Line}, {Teske}, {Burningham}, {Fortney}, \&
  {Marley}}]{lineteske2015}
{Line}, M.~R., {Teske}, J., {Burningham}, B., {Fortney}, J.~J., \& {Marley},
  M.~S. 2015, \apj, 807, 183

\bibitem[{{Liu} {et~al.}(2013){Liu}, {Magnier}, {Deacon}, {Allers}, {Dupuy},
  {Kotson}, {Aller}, {Burgett}, {Chambers}, {Draper}, {Hodapp}, {Jedicke},
  {Kaiser}, {Kudritzki}, {Metcalfe}, {Morgan}, {Price}, {Tonry}, \&
  {Wainscoat}}]{liumagnier2013}
{Liu}, M.~C., {Magnier}, E.~A., {Deacon}, N.~R., {et~al.} 2013, \apjl, 777, L20

\bibitem[{{Lueber} {et~al.}(2025){Lueber}, {Karchev}, {Fisher}, {Heim},
  {Trotta}, \& {Heng}}]{lueberkarchev2025}
{Lueber}, A., {Karchev}, K., {Fisher}, C., {et~al.} 2025, arXiv e-prints,
  arXiv:2502.18045

\bibitem[{{Luna} \& {Morley}(2021)}]{lunamorley2021}
{Luna}, J.~L. \& {Morley}, C.~V. 2021, \apj, 920, 146

\bibitem[{{MacDonald} \& {Batalha}(2023)}]{macdonaldbatalha2023}
{MacDonald}, R.~J. \& {Batalha}, N.~E. 2023, Research Notes of the American
  Astronomical Society, 7, 54

\bibitem[{Madhusudhan(2018)}]{madhusudhan2018b}
Madhusudhan, N. 2018, Atmospheric Retrieval of Exoplanets, ed. H.~J. Deeg \&
  J.~A. Belmonte (Cham: Springer International Publishing), 2153--2182

\bibitem[{{Madhusudhan} {et~al.}(2014){Madhusudhan}, {Amin}, \&
  {Kennedy}}]{madhusudhan2014}
{Madhusudhan}, N., {Amin}, M.~A., \& {Kennedy}, G.~M. 2014, \apjl, 794, L12

\bibitem[{Mamajek \& Bell(2014)}]{mamajekbell2014}
Mamajek, E.~E. \& Bell, C. P.~M. 2014, Monthly Notices of the Royal
  Astronomical Society, 445, 2169

\bibitem[{{Marley} \& {Robinson}(2015)}]{marleyrobinson2015}
{Marley}, M.~S. \& {Robinson}, T.~D. 2015, \araa, 53, 279

\bibitem[{{Matthews} {et~al.}(2025){Matthews}, {Molli{\`e}re}, {K{\"u}hnle},
  {Patapis}, {Whiteford}, {Samland}, {Lagage}, {Waters}, {Tsai}, {Zahnle},
  {Guedel}, {Henning}, {Vandenbussche}, {Absil}, {Argyriou}, {Barrado},
  {Coulais}, {Glauser}, {Olofsson}, {Pye}, {Rouan}, {Royer}, {van Dishoeck},
  {Ray}, \& {{\"O}stlin}}]{matthewsmolliere2025}
{Matthews}, E.~C., {Molli{\`e}re}, P., {K{\"u}hnle}, H., {et~al.} 2025, \apjl,
  981, L31

\bibitem[{{McCarthy} {et~al.}(2025){McCarthy}, {Vos}, {Muirhead}, {Biller},
  {Morley}, {Faherty}, {Burningham}, {Calamari}, {Cowan}, {Cruz}, {Gonzales},
  {Limbach}, {Liu}, {Nasedkin}, {Su{\'a}rez}, {Tan}, {O'Toole}, {Visscher},
  {Whiteford}, \& {Zhou}}]{mccarthyvos2025}
{McCarthy}, A.~M., {Vos}, J.~M., {Muirhead}, P.~S., {et~al.} 2025, \apjl, 981,
  L22

\bibitem[{{McKemmish} {et~al.}(2019){McKemmish}, {Masseron}, {Hoeijmakers},
  {P{\'e}rez-Mesa}, {Grimm}, {Yurchenko}, \&
  {Tennyson}}]{mckemmishmasseron2019}
{McKemmish}, L.~K., {Masseron}, T., {Hoeijmakers}, H.~J., {et~al.} 2019,
  \mnras, 488, 2836

\bibitem[{{Milam} {et~al.}(2005){Milam}, {Savage}, {Brewster}, {Ziurys}, \&
  {Wyckoff}}]{milramsavage2005}
{Milam}, S.~N., {Savage}, C., {Brewster}, M.~A., {Ziurys}, L.~M., \& {Wyckoff},
  S. 2005, \apj, 634, 1126

\bibitem[{{Miles} {et~al.}(2023){Miles}, {Biller}, {Patapis}, {Worthen},
  {Rickman}, {Hoch}, {Skemer}, {Perrin}, {Whiteford}, {Chen}, {Sargent},
  {Mukherjee}, {Morley}, {Moran}, {Bonnefoy}, {Petrus}, {Carter}, {Choquet},
  {Hinkley}, {Ward-Duong}, {Leisenring}, {Millar-Blanchaer}, {Pueyo}, {Ray},
  {Sallum}, {Stapelfeldt}, {Stone}, {Wang}, {Absil}, {Balmer}, {Boccaletti},
  {Bonavita}, {Booth}, {Bowler}, {Chauvin}, {Christiaens}, {Currie},
  {Danielski}, {Fortney}, {Girard}, {Grady}, {Greenbaum}, {Henning}, {Hines},
  {Janson}, {Kalas}, {Kammerer}, {Kennedy}, {Kenworthy}, {Kervella}, {Lagage},
  {Lew}, {Liu}, {Macintosh}, {Marino}, {Marley}, {Marois}, {Matthews},
  {Matthews}, {Mawet}, {McElwain}, {Metchev}, {Meyer}, {Molliere}, {Pantin},
  {Quirrenbach}, {Rebollido}, {Ren}, {Schneider}, {Vasist}, {Wyatt}, {Zhou},
  {Briesemeister}, {Bryan}, {Calissendorff}, {Cantalloube}, {Cugno}, {De
  Furio}, {Dupuy}, {Factor}, {Faherty}, {Fitzgerald}, {Franson}, {Gonzales},
  {Hood}, {Howe}, {Kraus}, {Kuzuhara}, {Lagrange}, {Lawson}, {Lazzoni}, {Liu},
  {Llop-Sayson}, {Lloyd}, {Martinez}, {Mazoyer}, {Quanz}, {Redai}, {Samland},
  {Schlieder}, {Tamura}, {Tan}, {Uyama}, {Vigan}, {Vos}, {Wagner}, {Wolff},
  {Ygouf}, {Zhang}, {Zhang}, \& {Zhang}}]{milesbiller2023}
{Miles}, B.~E., {Biller}, B.~A., {Patapis}, P., {et~al.} 2023, \apjl, 946, L6

\bibitem[{{Min}(2015)}]{Min2015}
{Min}, M. 2015, in European Physical Journal Web of Conferences, Vol. 102,
  European Physical Journal Web of Conferences, 00005

\bibitem[{{Min} {et~al.}(2005){Min}, {Hovenier}, \& {de
  Koter}}]{minhovenier2005}
{Min}, M., {Hovenier}, J.~W., \& {de Koter}, A. 2005, \aap, 432, 909

\bibitem[{{Molli{\`e}re} {et~al.}(2022){Molli{\`e}re}, {Molyarova}, {Bitsch},
  {Henning}, {Schneider}, {Kreidberg}, {Eistrup}, {Burn}, {Nasedkin},
  {Semenov}, {Mordasini}, {Schlecker}, {Schwarz}, {Lacour}, {Nowak}, \&
  {Schulik}}]{mollieremolyarova2022}
{Molli{\`e}re}, P., {Molyarova}, T., {Bitsch}, B., {et~al.} 2022, \apj, 934, 74

\bibitem[{{Molli{\`e}re} \& {Mordasini}(2012)}]{mollieremordasini2012}
{Molli{\`e}re}, P. \& {Mordasini}, C. 2012, \aap, 547, A105

\bibitem[{{Molli{\`e}re} \& {Snellen}(2019)}]{mollieresnellen2019}
{Molli{\`e}re}, P. \& {Snellen}, I.~A.~G. 2019, \aap, 622, A139

\bibitem[{{Molli{\`e}re} {et~al.}(2020){Molli{\`e}re}, {Stolker}, {Lacour},
  {Otten}, {Shangguan}, {Charnay}, {Molyarova}, {Nowak}, {Henning}, {Marleau},
  {Semenov}, {van Dishoeck}, {Eisenhauer}, {Garcia}, {Garcia Lopez}, {Girard},
  {Greenbaum}, {Hinkley}, {Kervella}, {Kreidberg}, {Maire}, {Nasedkin},
  {Pueyo}, {Snellen}, {Vigan}, {Wang}, {de Zeeuw}, \&
  {Zurlo}}]{mollierestolker2020}
{Molli{\`e}re}, P., {Stolker}, T., {Lacour}, S., {et~al.} 2020, \aap, 640, A131

\bibitem[{{Molli{\`e}re} {et~al.}(2017){Molli{\`e}re}, {van Boekel}, {Bouwman},
  {Henning}, {Lagage}, \& {Min}}]{mollierevanboekel2016}
{Molli{\`e}re}, P., {van Boekel}, R., {Bouwman}, J., {et~al.} 2017, \aap, 600,
  A10

\bibitem[{{Molli{\`e}re} {et~al.}(2015){Molli{\`e}re}, {van Boekel},
  {Dullemond}, {Henning}, \& {Mordasini}}]{mollierevanboekel2015}
{Molli{\`e}re}, P., {van Boekel}, R., {Dullemond}, C., {Henning}, T., \&
  {Mordasini}, C. 2015, \apj, 813, 47

\bibitem[{{Molli{\`e}re} {et~al.}(2019){Molli{\`e}re}, {Wardenier}, {van
  Boekel}, {Henning}, {Molaverdikhani}, \& {Snellen}}]{mollierewardenier2019}
{Molli{\`e}re}, P., {Wardenier}, J.~P., {van Boekel}, R., {et~al.} 2019, \aap,
  627, A67

\bibitem[{{Moran} {et~al.}(2024){Moran}, {Marley}, \&
  {Crossley}}]{moranmarley2024}
{Moran}, S.~E., {Marley}, M.~S., \& {Crossley}, S.~D. 2024, \apjl, 973, L3

\bibitem[{{Mordasini} {et~al.}(2016){Mordasini}, {van Boekel}, {Molli{\`e}re},
  {Henning}, \& {Benneke}}]{mordasinivanboekel2016}
{Mordasini}, C., {van Boekel}, R., {Molli{\`e}re}, P., {Henning}, T., \&
  {Benneke}, B. 2016, \apj, 832, 41

\bibitem[{{Morley} {et~al.}(2012){Morley}, {Fortney}, {Marley}, {Visscher},
  {Saumon}, \& {Leggett}}]{morleyfortney2012}
{Morley}, C.~V., {Fortney}, J.~J., {Marley}, M.~S., {et~al.} 2012, \apj, 756,
  172

\bibitem[{{Morley} {et~al.}(2014){Morley}, {Marley}, {Fortney}, {Lupu},
  {Saumon}, {Greene}, \& {Lodders}}]{morleymarley2014}
{Morley}, C.~V., {Marley}, M.~S., {Fortney}, J.~J., {et~al.} 2014, \apj, 787,
  78

\bibitem[{{Morley} {et~al.}(2024){Morley}, {Mukherjee}, {Marley}, {Fortney},
  {Visscher}, {Lupu}, {Gharib-Nezhad}, {Thorngren}, {Freedman}, \& {Batalha
  7}}]{morleymukherjee2024}
{Morley}, C.~V., {Mukherjee}, S., {Marley}, M.~S., {et~al.} 2024, arXiv
  e-prints, arXiv:2402.00758

\bibitem[{{Nasedkin} {et~al.}(2024){Nasedkin}, {Molli{\`e}re}, \&
  {Blain}}]{nasedkinmolliere2024}
{Nasedkin}, E., {Molli{\`e}re}, P., \& {Blain}, D. 2024, The Journal of Open
  Source Software, 9, 5875

\bibitem[{{Nasedkin} {et~al.}(2025){Nasedkin}, {Schrader}, {Vos}, {Biller},
  {Burningham}, {Cowan}, {Faherty}, {Gonzales}, {Lam}, {McCarthy}, {Muirhead},
  {O'Toole}, {Plummer}, {Su{\'a}rez}, {Tan}, {Visscher}, {Whiteford}, \&
  {Zhou}}]{nasedkinschrader2025}
{Nasedkin}, E., {Schrader}, M., {Vos}, J.~M., {et~al.} 2025, arXiv e-prints,
  arXiv:2507.07772

\bibitem[{{{\"O}berg} {et~al.}(2011){{\"O}berg}, {Murray-Clay}, \&
  {Bergin}}]{oeberg2011}
{{\"O}berg}, K.~I., {Murray-Clay}, R., \& {Bergin}, E.~A. 2011, \apjl, 743, L16

\bibitem[{{Olofsson} {et~al.}(2009){Olofsson}, {Augereau}, {van Dishoeck},
  {Mer{\'\i}n}, {Lahuis}, {Kessler-Silacci}, {Dullemond}, {Oliveira}, {Blake},
  {Boogert}, {Brown}, {Evans}, {Geers}, {Knez}, {Monin}, \&
  {Pontoppidan}}]{olofssonaugereau2009}
{Olofsson}, J., {Augereau}, J.~C., {van Dishoeck}, E.~F., {et~al.} 2009, \aap,
  507, 327

\bibitem[{{Palik}(1985)}]{palik1985}
{Palik}, E.~D. 1985, {Handbook of optical constants of solids}

\bibitem[{{Pegourie}(1988)}]{pegourie1988}
{Pegourie}, B. 1988, \aap, 194, 335

\bibitem[{{Petrus} {et~al.}(2024){Petrus}, {Whiteford}, {Patapis}, {Biller},
  {Skemer}, {Hinkley}, {Su{\'a}rez}, {Palma-Bifani}, {Morley}, {Tremblin},
  {Charnay}, {Vos}, {Wang}, {Stone}, {Bonnefoy}, {Chauvin}, {Miles}, {Carter},
  {Lueber}, {Helling}, {Sutlieff}, {Janson}, {Gonzales}, {Hoch}, {Absil},
  {Balmer}, {Boccaletti}, {Bonavita}, {Booth}, {Bowler}, {Briesemeister},
  {Bryan}, {Calissendorff}, {Cantalloube}, {Chen}, {Choquet}, {Christiaens},
  {Cugno}, {Currie}, {Danielski}, {De Furio}, {Dupuy}, {Factor}, {Faherty},
  {Fitzgerald}, {Fortney}, {Franson}, {Girard}, {Grady}, {Henning}, {Hines},
  {Hood}, {Howe}, {Kalas}, {Kammerer}, {Kennedy}, {Kenworthy}, {Kervella},
  {Kim}, {Kitzmann}, {Kraus}, {Kuzuhara}, {Lagage}, {Lagrange}, {Lawson},
  {Lazzoni}, {Leisenring}, {Lew}, {Liu}, {Liu}, {Llop-Sayson}, {Lloyd},
  {Macintosh}, {M{\^a}lin}, {Manjavacas}, {Marino}, {Marley}, {Marois},
  {Martinez}, {Matthews}, {Matthews}, {Mawet}, {Mazoyer}, {McElwain},
  {Metchev}, {Meyer}, {Millar-Blanchaer}, {Molli{\`e}re}, {Moran}, {Mukherjee},
  {Pantin}, {Perrin}, {Pueyo}, {Quanz}, {Quirrenbach}, {Ray}, {Rebollido},
  {Adams Redai}, {Ren}, {Rickman}, {Sallum}, {Samland}, {Sargent}, {Schlieder},
  {Stapelfeldt}, {Tamura}, {Tan}, {Theissen}, {Uyama}, {Vasist}, {Vigan},
  {Wagner}, {Ward-Duong}, {Wolff}, {Worthen}, {Wyatt}, {Ygouf}, {Zurlo},
  {Zhang}, {Zhang}, {Zhang}, \& {Zhou}}]{petruswhiteford2024}
{Petrus}, S., {Whiteford}, N., {Patapis}, P., {et~al.} 2024, \apjl, 966, L11

\bibitem[{Philipp(1971)}]{philipp1975}
Philipp, H. 1971, Journal of Physics and Chemistry of Solids, 32, 1935

\bibitem[{Polyansky {et~al.}(2018)Polyansky, Kyuberis, Zobov, Tennyson,
  Yurchenko, \& Lodi}]{polanskykyuberis2018}
Polyansky, O.~L., Kyuberis, A.~A., Zobov, N.~F., {et~al.} 2018, Monthly Notices
  of the Royal Astronomical Society, 480, 2597

\bibitem[{{Powell} {et~al.}(2018){Powell}, {Zhang}, {Gao}, \&
  {Parmentier}}]{powellzhang2018}
{Powell}, D., {Zhang}, X., {Gao}, P., \& {Parmentier}, V. 2018, \apj, 860, 18

\bibitem[{{Purcell} \& {Pennypacker}(1973)}]{purcell1973}
{Purcell}, E.~M. \& {Pennypacker}, C.~R. 1973, \apj, 186, 705

\bibitem[{{Radigan} {et~al.}(2014){Radigan}, {Lafreni{\`e}re}, {Jayawardhana},
  \& {Artigau}}]{radiganlafreniere2014}
{Radigan}, J., {Lafreni{\`e}re}, D., {Jayawardhana}, R., \& {Artigau}, E. 2014,
  \apj, 793, 75

\bibitem[{Raftery(1995)}]{raftery1995}
Raftery, A.~E. 1995, Sociological Methodology, 25, 111

\bibitem[{{Rieke} {et~al.}(2015){Rieke}, {Wright}, {B{\"o}ker}, {Bouwman},
  {Colina}, {Glasse}, {Gordon}, {Greene}, {G{\"u}del}, {Henning}, {Justtanont},
  {Lagage}, {Meixner}, {N{\o}rgaard-Nielsen}, {Ray}, {Ressler}, {van Dishoeck},
  \& {Waelkens}}]{riekewright2015}
{Rieke}, G.~H., {Wright}, G.~S., {B{\"o}ker}, T., {et~al.} 2015, \pasp, 127,
  584

\bibitem[{{Robinson} \& {Marley}(2014)}]{robinsonmarley2014}
{Robinson}, T.~D. \& {Marley}, M.~S. 2014, \apj, 785, 158

\bibitem[{{Rossow}(1978)}]{rossow1978}
{Rossow}, W.~B. 1978, \icarus, 36, 1

\bibitem[{{Rothman} {et~al.}(2010){Rothman}, {Gordon}, {Barber}, {Dothe},
  {Gamache}, {Goldman}, {Perevalov}, {Tashkun}, \& {Tennyson}}]{rothman2010}
{Rothman}, L.~S., {Gordon}, I.~E., {Barber}, R.~J., {et~al.} 2010, \jqsrt, 111,
  2139

\bibitem[{{S{\'a}nchez-L{\'o}pez} {et~al.}(2022){S{\'a}nchez-L{\'o}pez},
  {Landman}, {Molli{\`e}re}, {Casasayas-Barris}, {Kesseli}, \&
  {Snellen}}]{sanchezlopezlandman2022}
{S{\'a}nchez-L{\'o}pez}, A., {Landman}, R., {Molli{\`e}re}, P., {et~al.} 2022,
  \aap, 661, A78

\bibitem[{{Saumon} \& {Marley}(2008)}]{saumonmarley2008}
{Saumon}, D. \& {Marley}, M.~S. 2008, \apj, 689, 1327

\bibitem[{Schnurre {et~al.}(2004)Schnurre, Gröbner, \&
  Schmid-Fetzer}]{schnurregroebner2004}
Schnurre, S., Gröbner, J., \& Schmid-Fetzer, R. 2004, Journal of
  Non-Crystalline Solids, 336, 1

\bibitem[{Servoin \& Piriou(1973)}]{servoinpiriou1973}
Servoin, J.~L. \& Piriou, B. 1973, physica status solidi (b), 55, 677

\bibitem[{Sousa-Silva {et~al.}(2014)Sousa-Silva, Al-Refaie, Tennyson, \&
  Yurchenko}]{sousasilva2014}
Sousa-Silva, C., Al-Refaie, A.~F., Tennyson, J., \& Yurchenko, S.~N. 2014,
  Monthly Notices of the Royal Astronomical Society, 446, 2337

\bibitem[{{Spiegel} {et~al.}(2011){Spiegel}, {Burrows}, \&
  {Milsom}}]{spiegelburrows2011}
{Spiegel}, D.~S., {Burrows}, A., \& {Milsom}, J.~A. 2011, \apj, 727, 57

\bibitem[{{Stolker} {et~al.}(2020){Stolker}, {Quanz}, {Todorov}, {K{\"u}hn},
  {Molli{\`e}re}, {Meyer}, {Currie}, {Daemgen}, \& {Lavie}}]{stolkerquanz2020}
{Stolker}, T., {Quanz}, S.~P., {Todorov}, K.~O., {et~al.} 2020, \aap, 635, A182

\bibitem[{{Su{\'a}rez} \& {Metchev}(2022)}]{suarezmetchev2022}
{Su{\'a}rez}, G. \& {Metchev}, S. 2022, \mnras, 513, 5701

\bibitem[{{Su{\'a}rez} \& {Metchev}(2023)}]{suarezmetchev2023b}
{Su{\'a}rez}, G. \& {Metchev}, S. 2023, \mnras, 523, 4739

\bibitem[{{Su{\'a}rez} {et~al.}(2023){Su{\'a}rez}, {Vos}, {Metchev}, {Faherty},
  \& {Cruz}}]{suarezvos2023}
{Su{\'a}rez}, G., {Vos}, J.~M., {Metchev}, S., {Faherty}, J.~K., \& {Cruz}, K.
  2023, \apjl, 954, L6

\bibitem[{{Tremblin} {et~al.}(2016){Tremblin}, {Amundsen}, {Chabrier},
  {Baraffe}, {Drummond}, {Hinkley}, {Mourier}, \&
  {Venot}}]{tremblinamundsen2016}
{Tremblin}, P., {Amundsen}, D.~S., {Chabrier}, G., {et~al.} 2016, \apjl, 817,
  L19

\bibitem[{{Tremblin} {et~al.}(2015){Tremblin}, {Amundsen}, {Mourier},
  {Baraffe}, {Chabrier}, {Drummond}, {Homeier}, \&
  {Venot}}]{tremblinamundsen2015}
{Tremblin}, P., {Amundsen}, D.~S., {Mourier}, P., {et~al.} 2015, \apjl, 804,
  L17

\bibitem[{{Tremblin} {et~al.}(2017){Tremblin}, {Chabrier}, {Baraffe}, {Liu},
  {Magnier}, {Lagage}, {Alves de Oliveira}, {Burgasser}, {Amundsen}, \&
  {Drummond}}]{tremblinchabrier2017}
{Tremblin}, P., {Chabrier}, G., {Baraffe}, I., {et~al.} 2017, \apj, 850, 46

\bibitem[{{Tremblin} {et~al.}(2019){Tremblin}, {Padioleau}, {Phillips},
  {Chabrier}, {Baraffe}, {Fromang}, {Audit}, {Bloch}, {Burgasser}, {Drummond},
  {Gonz{\'a}lez}, {Kestener}, {Kokh}, {Lagage}, \&
  {Stauffert}}]{tremblinpadioleau2019}
{Tremblin}, P., {Padioleau}, T., {Phillips}, M.~W., {et~al.} 2019, \apj, 876,
  144

\bibitem[{{Tremblin} {et~al.}(2020){Tremblin}, {Phillips}, {Emery}, {Baraffe},
  {Lew}, {Apai}, {Biller}, \& {Bonnefoy}}]{tremblinphillips2020}
{Tremblin}, P., {Phillips}, M.~W., {Emery}, A., {et~al.} 2020, \aap, 643, A23

\bibitem[{{Tsuji} {et~al.}(1996){Tsuji}, {Ohnaka}, \& {Aoki}}]{tsuji2996}
{Tsuji}, T., {Ohnaka}, K., \& {Aoki}, W. 1996, \aap, 305, L1

\bibitem[{{van Boekel} {et~al.}(2005){van Boekel}, {Min}, {Waters}, {de Koter},
  {Dominik}, {van den Ancker}, \& {Bouwman}}]{boekelmin2005}
{van Boekel}, R., {Min}, M., {Waters}, L.~B.~F.~M., {et~al.} 2005, \aap, 437,
  189

\bibitem[{{Vasist} {et~al.}(2023){Vasist}, {Rozet}, {Absil}, {{\bf
  \color{orange} Molli\`ere}}, {Nasedkin}, \& {Louppe}}]{vasistrozet2023}
{Vasist}, M., {Rozet}, F., {Absil}, O., {et~al.} 2023, \aap, 672, A147

\bibitem[{{Visscher} {et~al.}(2010){Visscher}, {Lodders}, \&
  {Fegley}}]{visscherlodders2010}
{Visscher}, C., {Lodders}, K., \& {Fegley}, Jr., B. 2010, \apj, 716, 1060

\bibitem[{{Vos} {et~al.}(2017){Vos}, {Allers}, \& {Biller}}]{vosallers2017}
{Vos}, J.~M., {Allers}, K.~N., \& {Biller}, B.~A. 2017, \apj, 842, 78

\bibitem[{{Vos} {et~al.}(2023){Vos}, {Burningham}, {Faherty}, {Alejandro},
  {Gonzales}, {Calamari}, {Bardalez Gagliuffi}, {Visscher}, {Tan}, {Morley},
  {Marley}, {Gemma}, {Whiteford}, {Gaarn}, \& {Park}}]{vosburningham2023}
{Vos}, J.~M., {Burningham}, B., {Faherty}, J.~K., {et~al.} 2023, \apj, 944, 138

\bibitem[{{Wakeford} {et~al.}(2017){Wakeford}, {Visscher}, {Lewis}, {Kataria},
  {Marley}, {Fortney}, \& {Mandell}}]{wakefordvisscher2017}
{Wakeford}, H.~R., {Visscher}, C., {Lewis}, N.~K., {et~al.} 2017, \mnras, 464,
  4247

\bibitem[{{Wende} {et~al.}(2010){Wende}, {Reiners}, {Seifahrt}, \&
  {Bernath}}]{wendereiners2010}
{Wende}, S., {Reiners}, A., {Seifahrt}, A., \& {Bernath}, P.~F. 2010, \aap,
  523, A58

\bibitem[{{Wetzel} {et~al.}(2013){Wetzel}, {Klevenz}, {Gail}, {Pucci}, \&
  {Trieloff}}]{wetzel2013}
{Wetzel}, S., {Klevenz}, M., {Gail}, H.~P., {Pucci}, A., \& {Trieloff}, M.
  2013, \aap, 553, A92

\bibitem[{Yurchenko {et~al.}(2020)Yurchenko, Mellor, Freedman, \&
  Tennyson}]{yurchenkomellor2020}
Yurchenko, S.~N., Mellor, T.~M., Freedman, R.~S., \& Tennyson, J. 2020, Monthly
  Notices of the Royal Astronomical Society, 496, 5282

\bibitem[{Yurchenko {et~al.}(2021)Yurchenko, Tennyson, Syme, Adam, Clark,
  Cooper, Dobney, Donnelly, Gorman, Lynas-Gray, Meltzer, Owens, Qu, Semenov,
  Somogyi, Upadhyay, Wright, \& Zapata Trujillo}]{yurchenkotennyson2021}
Yurchenko, S.~N., Tennyson, J., Syme, A.-M., {et~al.} 2021, Monthly Notices of
  the Royal Astronomical Society, 510, 903

\bibitem[{{Zahnle} \& {Marley}(2014)}]{zahnlemarley2014}
{Zahnle}, K.~J. \& {Marley}, M.~S. 2014, \apj, 797, 41

\bibitem[{{Zalesky} {et~al.}(2019){Zalesky}, {Line}, {Schneider}, \&
  {Patience}}]{zaleskyline2019}
{Zalesky}, J.~A., {Line}, M.~R., {Schneider}, A.~C., \& {Patience}, J. 2019,
  \apj, 877, 24

\bibitem[{{Zeidler} {et~al.}(2015){Zeidler}, {Mutschke}, \&
  {Posch}}]{zeidlermutschke2015}
{Zeidler}, S., {Mutschke}, H., \& {Posch}, T. 2015, \apj, 798, 125

\bibitem[{{Zeidler} {et~al.}(2013){Zeidler}, {Posch}, \&
  {Mutschke}}]{zeidlerposch2013}
{Zeidler}, S., {Posch}, T., \& {Mutschke}, H. 2013, \aap, 553, A81

\bibitem[{{Zhang} {et~al.}(2020){Zhang}, {Liu}, {Hermes}, {Magnier}, {Marley},
  {Tremblay}, {Tucker}, {Do}, {Payne}, \& {Shappee}}]{zhangliu2020}
{Zhang}, Z., {Liu}, M.~C., {Hermes}, J.~J., {et~al.} 2020, \apj, 891, 171

\bibitem[{{Zhang} {et~al.}(2025){Zhang}, {Molli{\`e}re}, {Fortney}, \&
  {Marley}}]{zhangmolliere2025}
{Zhang}, Z., {Molli{\`e}re}, P., {Fortney}, J.~J., \& {Marley}, M.~S. 2025,
  arXiv e-prints, arXiv:2502.18559

\bibitem[{{Zhang} {et~al.}(2023){Zhang}, {Molli{\`e}re}, {Hawkins}, {Manea},
  {Fortney}, {Morley}, {Skemer}, {Marley}, {Bowler}, {Carter}, {Franson},
  {Maas}, \& {Sneden}}]{zhangmolliere2023}
{Zhang}, Z., {Molli{\`e}re}, P., {Hawkins}, K., {et~al.} 2023, \aj, 166, 198

\end{thebibliography}

\begin{appendix}

\section{Extracting cloud opacities from brightness temperatures}
\label{app:opacity_properties}
\subsection{Assumptions \rch{and their validity}}
In extracting cloud opacities from brightness temperatures as derived below, we make the following assummptions:
\begin{enumerate}
    \item The cloud must be the dominating opacity source over the considered wavelength range. Here we focus on the silicate absorption feature around 10~µm. \rch{Studying emission contribution functions from our full retrievals (see Section~\ref{sect:10mu_retrievals}) with different silicate cloud species we found that the wavelength range from 8-12.5~µm was optimal for our purposes. Choosing a wider range resulted in gas phase species such as \ce{H2O} becoming dominant. We note that our assumption breaks down in cases where the silicate cloud is optically thin and only slightly affects the flux in the 10~µm region.}
    \item The cloud opacity is approximately constant over the pressure range where the cloud becomes optically thick. \rch{From studying emission contribution functions of various retrievals we find that the cloud becomes optically thick over a small pressure range at any given wavelength, in which the cloud mass fraction changes by less than an order of magnitude. We note that in our models the mass fraction is the only way in which the cloud opacity may change, since the mean radius and width of the particle size distribution are assumed to be vertically constant.}
    \item The cloud must become optically thick quickly. Alternatively put, the width of the peak of the emission contribution function must be narrow, such that only a small pressure range contributes most of the visible emission at any given wavelength. This also means that, at any given wavelength, the emission stems from a small temperature range and its intensity is well described by the Planck function at that temperature. \rch{From spot checking some of the species considered in our retrievals we found that the atmospheric temperature varied between 3-6~\% across the pressures where the clouds became optically thick.}
    \item In the wavelength range of interest absorption (and not scattering) is dominant. \rch{This is a good assumption because for silicates the scattering opacity in the 10~µm wavelength region is much smaller than the absorption opacity for particles with sizes $\lesssim$ 1~µm (but scattering can be important or dominant at shorter wavelengths). For larger particles, for which scattering becomes important also at longer wavelengths, the 10~µm feature vanishes. As a compromise, while neglecting scattering during the derivation of the brightness temperature method, we use the sum of the absorption and scattering opacity when fitting. In this way we are able to at least somewhat handle the transition regime when the 10~µm feature vanishes.}
    \item If the spectrum contains contributions from a cloudy and a cloud-free column, or a column with and without a silicate cloud, then the flux arising from the latter column has a constant brightness temperature and can be described using Planck's law. \rch{Testing this based on the retrievals from our full analysis, the column without the silicate cloud has a much smaller brightness temperature variation across our adopted wavelength range for the brightness temperature method. For example, while in the SiO column the brightness temperature varies by 250~K across the 10~µm feature, the variation is in the 50~K range for the column without the SiO cloud. This is not surprising, since this wavelength range is typically free of other strong absorbers for objects around the L-T transition and because the deep iron loud, present in both columns in our retrievals, tends to flatten the relevant brightness temperature range in the column without the SiO cloud even more.}
\end{enumerate}

\subsection{The opacity -- brightness temperature relation}
The goal of the derivation presented below is to match observed brightness temperatures with the temperature at which the cloud becomes optically thick, which we define as the location where its vertical optical depth $\tau=2/3$.

We begin by deriving the brightness temperature of the cloudy column of the atmosphere. For this we express the observed flux as $F(\nu)=fF_{\rm cloudy}+(1-f)F_{\rm clear}$, where $f$ is the coverage fraction of the cloudy column on the planet, and $F_{\rm cloudy}(\nu)$ and $F_{\rm clear}(\nu)$ are the fluxes of the cloudy and clear columns, respectively. We then equate $F_{\rm cloudy}(\nu)$ with the flux of a blackbody, which is simply Planck's law $B(\nu)$ as a function of frequency $\nu$ multiplied by $\pi$:
\begin{equation}
    F(\nu) - (1-f)F_{\rm clear}(\nu) = \pi B(\nu)f\frac{R^2}{d^2}.
\end{equation}
The planet's radius is denoted by $R$, while the distance between the observer and the planet is denoted by $d$. Solving for the temperature in Planck's law $B(\nu,T) = (2h\nu^3/c^2)({\rm exp}(h\nu/k_{\rm B}T)-1)^{-1}$ one finds that
\begin{equation}
    T_{\rm bright}(\nu) = \frac{h\nu}{k_{\rm B}}\left[{\rm ln}\left(1+\pi f\frac{R^2}{d^2}\frac{2h\nu^3}{c^2}\frac{1}{F(\nu) - (1-f)F_{\rm clear}(\nu)}\right)\right]^{-1},
\end{equation}
where $h$, $k_{\rm B}$ and $c$ are the Planck constant, the Boltzmann constant and the speed of light, respectively.

Next we derive the temperature at which the cloud becomes optically thick. We use the standard definition of the atmosphere's vertical optical depth $\tau$, namely $d\tau = \rho \kappa dr$, where $\rho$ is the local mass density, $\kappa$ the opacity in units of cross-section per mass, and $r$ is the atmosphere's radial coordinate. Using the equation of hydrostatic equilibrium, $dP/dr = -\rho g$, where $P$ is the pressure and $g$ is the gravitational acceleration, we can write that $d\tau = (\kappa/g)dP$. The photospheric pressure (where $\tau=2/3$) is thus $P_{\rm phot}=(2g)/(3\kappa)$, where we assumed $\kappa$ and $g$ to be constant. Now we can consider two different types of local pressure-temperature profiles in the photospheric region of the atmosphere, either
\begin{equation}
    T_{\alpha}(P) = T_{\rm deep} + \alpha {\rm ln}(P/P_{\rm deep}),
\end{equation}
where $T_{\rm deep}$, $P_{\rm deep}$ and $\alpha$ are free parameters. Alternatively, we can assume a power law dependence:
\begin{equation}
    T_{\nabla}(P) = T_{\rm deep} (P/P_{\rm deep})^\nabla,
\end{equation}
where $T_{\rm deep}$, $P_{\rm deep}$ and $\nabla$ are free parameters. We can now solve $T_{\alpha}(P_{\rm phot})=T_{\rm bright}(\nu)$ or $T_{\nabla}(P_{\rm phot})=T_{\rm bright}(\nu)$ for $\kappa$ to extract the cloud opacity as a function of frequency (or wavelength, by setting $\lambda=c/\nu$).

For $T_{\alpha}$ one finds that
\begin{equation}
    T_{\rm bright}(\nu) = T_{\rm deep} - \alpha {\rm ln}(\kappa) + \alpha {\rm ln}(2g/3P_{\rm deep}) 
\end{equation}
and thus
\begin{equation}
    T_{\rm bright}(\nu) = - \alpha {\rm ln}(\kappa) + {\rm cst},
\end{equation}
such that
\begin{equation}
    \kappa(\nu) \propto {\rm exp}\left(-T_{\rm bright}/\alpha\right).
    \label{equ:kappa_alpha}
\end{equation}

For $T_{\nabla}$ one finds that
\begin{equation}
    T_{\rm bright}(\nu) = T_{\rm deep} (2g/3\kappa P_{\rm deep})^\nabla,
\end{equation}
such that
\begin{equation}
    \kappa(\nu) \propto T_{\rm bright}^{-1/\nabla}.
    \label{equ:kappa_nabla}
\end{equation}

\subsection{Fitting and observational uncertainties}
\label{appendix:kappa_uncertainties}
In order to fit opacities derived from brightness temperatures, $\kappa(\nu)$, with opacities derived from optical constants, $\kappa_{\rm opt}(\nu)$, the former need observational error bars. Since the fundamental measurement is the observed flux $F$, we will derive the uncertainty $\Delta \kappa$ as a function of the flux uncertainty $\Delta F$ for the $T_\alpha$ and $T_\nabla$ temperature-pressure relations.

We begin by calculating $\Delta T_{\rm bright}=(\partial T_{\rm bright}/\partial F) \Delta F$:
\begin{equation}
    \Delta T_{\rm bright} = \frac{T_{\rm bright}}{{\rm ln}(\mathcal{M})}\frac{\mathcal{M}-1}{\mathcal{M}}\frac{\Delta F}{F(\nu) - (1-f)F_{\rm clear}(\nu)},
\end{equation}
where
\begin{equation}
    \mathcal{M}= 1+\pi f\frac{R^2}{d^2}\frac{2h\nu^3}{c^2}\frac{1}{F(\nu) - (1-f)F_{\rm clear}(\nu)}.
\end{equation}

We set the proportionality constant to unity in Equations \ref{equ:kappa_alpha} and \ref{equ:kappa_nabla}; during the fitting we will then simply apply a multiplicative scaling parameter to the opacities predicted from optical constants. In addition, the fit is carried out in ${\rm log}(\kappa)$ space. If this is not done, small $\nabla$ values can lead to very small numerical values and thus underflow errors. The multiplicative scaling applied to the opacities derived from optical constants opacity thus becomes an additive term.
We then find
\begin{equation}
    {\rm ln}(\kappa) = -T_{\rm bright}/\alpha
\end{equation}
and
\begin{equation}
    \Delta {\rm ln}(\kappa) = \Delta T_{\rm bright}/|\alpha|
\end{equation}
for the $T_\alpha$ profile.
For the $T_\nabla$ profile we find
\begin{equation}
    {\rm ln}(\kappa) = -{\rm ln}(T_{\rm bright})/\nabla
\end{equation}
and
\begin{equation}
    \Delta {\rm ln}(\kappa) = \frac{1}{|\nabla|}\frac{\Delta T_{\rm bright}}{T_{\rm bright}}.
\end{equation}

We note that, due to our assumptions made above, the derived $\Delta{\rm log}(\kappa)$ uncertainties are lower limits because the derived ${\rm log}(\kappa)$ values may be subject to additional systematic modeling errors.

\section{Considered silicate species for brightness temperature fitting}

The references for the optical constants  of the silicon-bearing condensates considered in this work are listed in Table~\ref{tab:si_cloud_species_for_brigthness_temp_fitting}.

\begin{table*}[t!]
    \centering
    \begin{threeparttable}
    \caption{Silicate cloud species considered for brightness temperature fitting}
    \label{tab:si_cloud_species_for_brigthness_temp_fitting}
    \begin{tabular}{l l l l}
        \hline\hline
        Species & Morphology & Shape & Reference  \\
        \hline
\ce{Fe2SiO4} & crystalline & Sphere/Irregular & \citet{fabianhenning2001} \\
\ce{Mg$_{0.5}$Fe$_{0.5}$SiO3} & amorphous (glassy) & Sphere/Irregular & \citet{jaegermutschke1994} \\
\ce{Mg2SiO4} & amorphous (sol-gel) & Sphere/Irregular & \citet{jaegerilin2003} \\
\ce{Mg2SiO4} & crystalline & Sphere/Irregular & \citet{servoinpiriou1973} \\
\ce{MgFeSiO4} & amorphous (glassy) & Sphere/Irregular & \citet{dorschnerbegemann1995} \\
\ce{MgSiO3} & amorphous (glassy) & Sphere/Irregular & \citet{jaegermutschke1994} \\
\ce{MgSiO3} & amorphous (sol-gel) & Sphere/Irregular & \citet{jaegerilin2003} \\
\ce{MgSiO3} & crystalline & Sphere/Irregular & \citet{jaeger1998} \\
\ce{SiC} & crystalline & Sphere/Irregular & \citet{pegourie1988} \\
\ce{SiO} & amorphous & Sphere/Irregular & \citet{palik1985,wetzel2013} \\
\ce{SiO2} & amorphous (glassy) & Sphere/Irregular & \citet{henningmutschke1997, palik1985} \\
\ce{SiO2} & crystalline & Sphere/Irregular & \citet{zeidlerposch2013, palik1985} \\
        \hline
    \end{tabular}
    \end{threeparttable}

\end{table*}

\section{Best-fit spectra for all retrieval models}

\begin{figure*}
    \centering
    \includegraphics[width=0.90\textwidth]{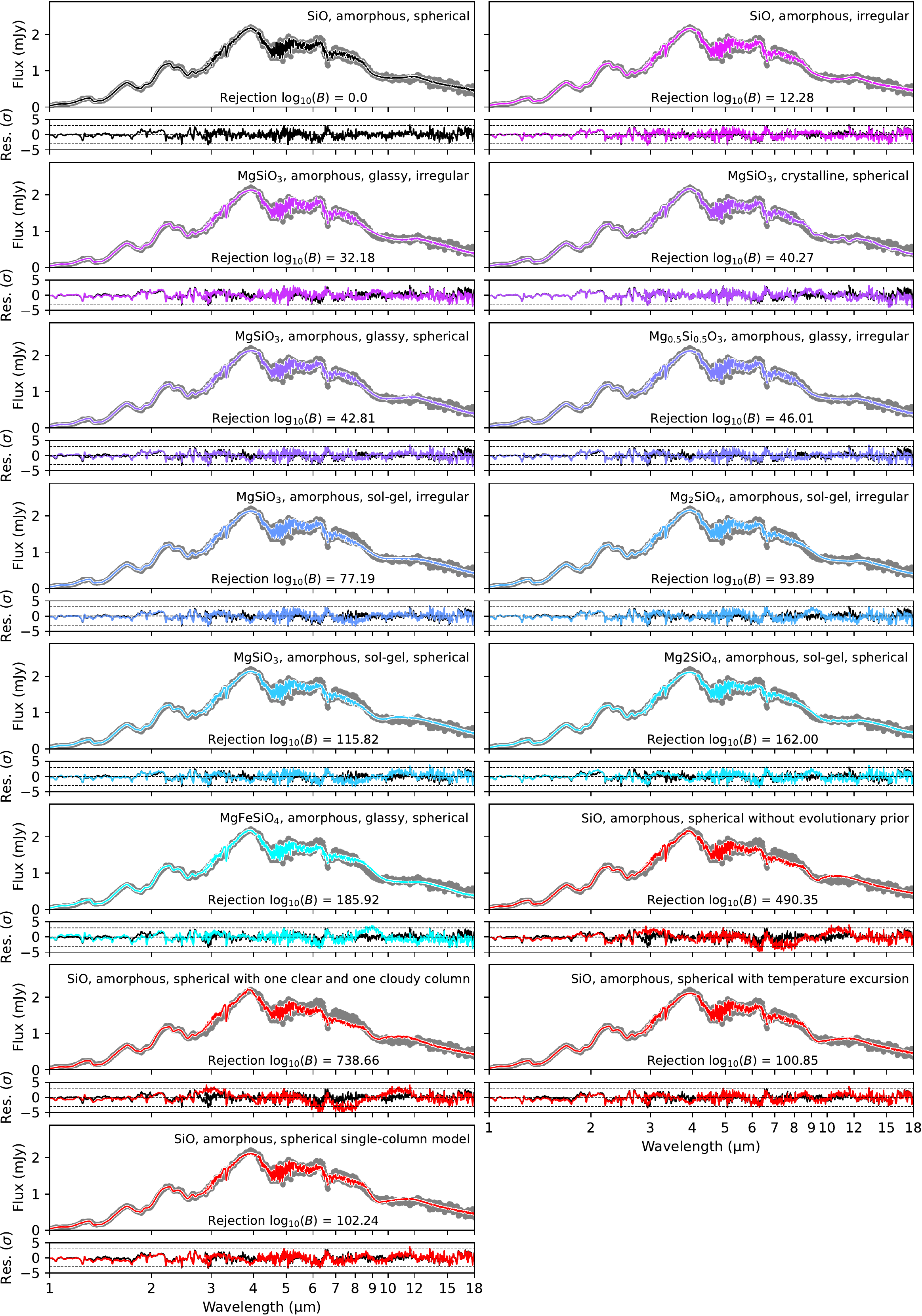}
    \caption{Best-fit spectra for all retrieval models listed in Table~\ref{tab:all_retrieval_posteriors}. From left to right, top to bottom, the ordering is identical to the model ordering of Table~\ref{tab:all_retrieval_posteriors}. The residuals of the winning model (SiO cloud with amorphous spherical particles in a two-column model with a global Fe cloud) are plotted in black below the respective residuals of all other models.}
    \label{fig:best_fits_all_retrievals}
\end{figure*}

The best-fit spectra of all retrieval models listed in Table~\ref{tab:all_retrieval_posteriors} are shown in Fig.~\ref{fig:best_fits_all_retrievals}.

\section{Best-fit opacities for all silicon-bearing species}
\begin{figure*}
    \centering
    \includegraphics[width=0.95\textwidth]{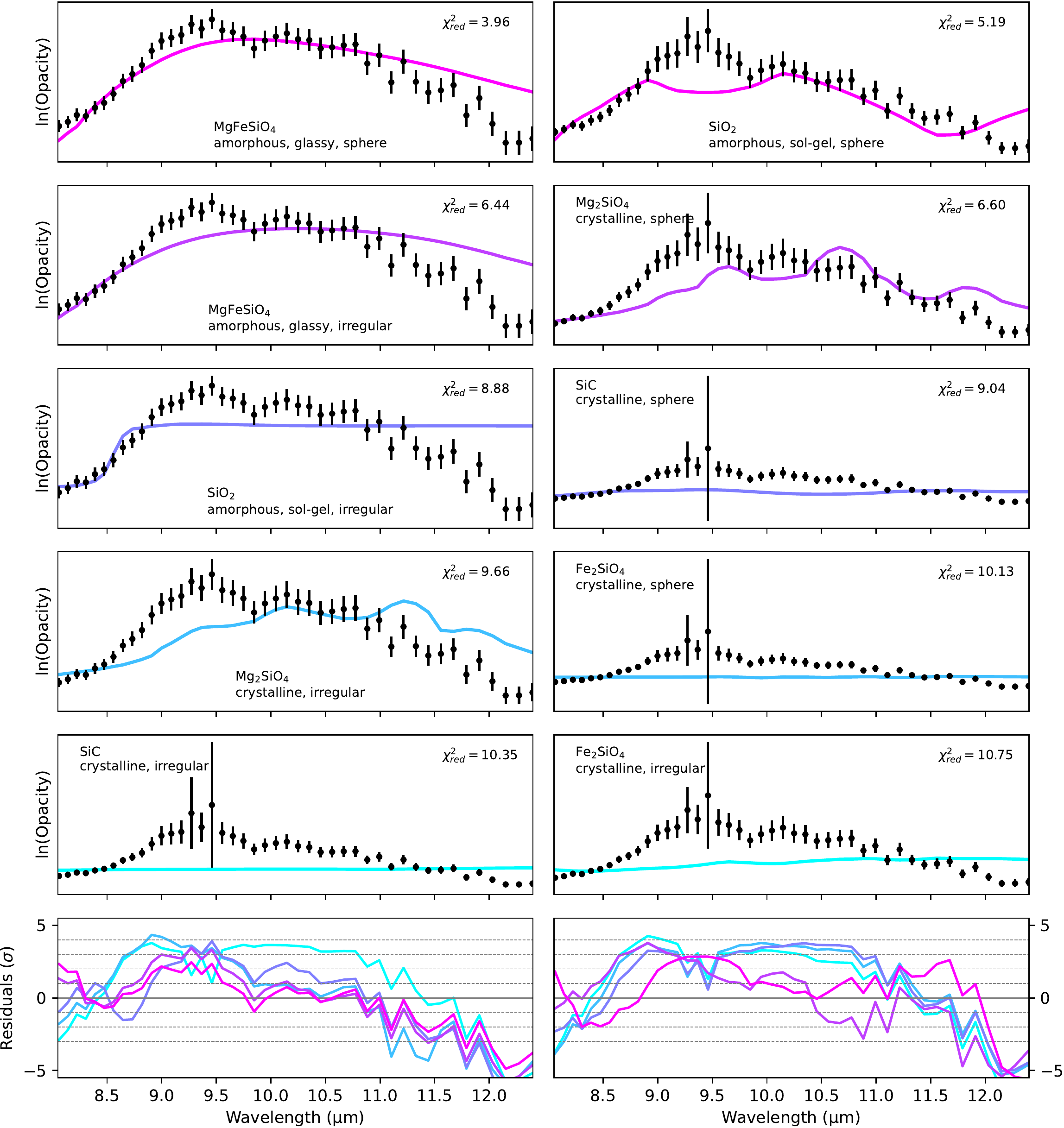}
    \caption{Like Fig.~\ref{fig:cloud_shape_fits}, but showing results for the remaining silicate species listed in Table~\ref{table:cloud_fit_full}.}
    \label{fig:brightness_temperature_all_appendix_1}
\end{figure*}

\rch{In Fig.~\ref{fig:cloud_shape_fits} we already showed the brightness temperature fits for the best 14 candidate silicate species. The fits for the remaining species considered in this work (see Table~\ref{table:cloud_fit_full}) are shown in Fig.~\ref{fig:brightness_temperature_all_appendix_1}.}

\end{appendix}

\end{document}